\documentclass[pra,%
 reprint,%
 twocolumn,
superscriptaddress,
nofootinbib,
longbibliography]{revtex4-2}

\usepackage{graphicx}% Include figure files
\usepackage{dcolumn}% Align table columns on decimal point
\usepackage{amsfonts}
\usepackage{amssymb}
\usepackage{amsmath}
\usepackage{amsthm}

\usepackage[colorlinks=true,linkcolor=blue,citecolor=blue,urlcolor=blue,plainpages=false,pdfpagelabels]{hyperref}
\usepackage{color}
\usepackage[dvipsnames]{xcolor}
\usepackage{mathtools}
\usepackage{bbm}

\usepackage{float}
\usepackage{array}
\usepackage{longtable}

\usepackage{verbatim}

\usepackage[utf8]{inputenc}
\usepackage[T1]{fontenc}
\usepackage{etoolbox}

\usepackage{anyfontsize}  % Can help to remove certain font size warnings
\usepackage{microtype} % Helps eliminate protrusion into the margins

\usepackage{newtxtext}
\usepackage{newtxmath}

\makeatletter
\g@addto@macro\bfseries{\boldmath}
\makeatother

\newcommand{\bra}[1]{\langle #1|}
\newcommand{\ket}[1]{|#1\rangle}

\newcommand{\ketbra}[2]{\ket{#1}\!\bra{#2}}

\newcommand{\e}{\mathrm{e}}
\newcommand{\I}{\mathrm{i}}

\renewcommand{\t}{{\scriptscriptstyle\mathsf{T}}}
\newcommand{\id}{\operatorname{id}}

\theoremstyle{definition}

\renewcommand{\qedsymbol}{$\blacksquare$}

\renewcommand{\qedsymbol}{\unskip\nobreak\quad\qedsymbol}
\renewcommand{\qedsymbol}{$\blacksquare$}

\DeclarePairedDelimiter{\ceil}{\lceil}{\rceil}

\begin{document}

\title{Fast and reliable entanglement distribution with quantum repeaters: principles for improving protocols using reinforcement learning}

\author{Stav Haldar}\email{hstav1@lsu.edu}
\affiliation{Hearne Institute for Theoretical Physics, Department of Physics and Astronomy,
Louisiana State University, Baton Rouge, Louisiana 70803, USA}

\author{Pratik J. Barge}\email{pbarge1@lsu.edu}
\affiliation{Hearne Institute for Theoretical Physics, Department of Physics and Astronomy,
Louisiana State University, Baton Rouge, Louisiana 70803, USA}

\author{Sumeet Khatri}\email{sumeet.khatri@fu-berlin.de}
\affiliation{Dahlem Center for Complex Quantum Systems, Freie Universit\"{a}t Berlin, 14195 Berlin, Germany}

\author{Hwang Lee}\email{hwlee@lsu.edu}
\affiliation{Hearne Institute for Theoretical Physics, Department of Physics and Astronomy,
Louisiana State University, Baton Rouge, Louisiana 70803, USA}

\date{\today}

\begin{abstract}

Future quantum technologies such as quantum communication, quantum sensing, and distributed quantum computation, will rely on networks of shared entanglement between spatially separated nodes.
%Distributing entanglement between these nodes, especially over long distances, currently remains a challenge, due to limitations resulting from the fragility of quantum systems, such as photon losses, non-ideal measurements, and quantum memories with short coherence times. In the absence of full-scale fault-tolerant quantum error correction, which can in principle overcome these limitations, we should understand the extent to which we can circumvent these limitations.
In this work, we provide improved protocols/policies for entanglement distribution along a linear chain of nodes, both homogeneous and inhomogeneous, that take practical limitations into account. For a wide range of parameters, our policies improve upon previously known policies, such as the ``swap-as-soon-as-possible'' policy, with respect to both the waiting time and the fidelity of the end-to-end entanglement. This improvement is greatest for the most practically relevant cases, namely, for short coherence times, high link losses, and highly asymmetric links. To obtain our results, we model entanglement distribution using a Markov decision process, and then we use the Q-learning reinforcement learning (RL) algorithm to discover new policies. These new policies are characterized by dynamic, state-dependent memory cutoffs and collaboration between the nodes. In particular, we quantify this collaboration between the nodes. Our quantifiers tell us how much ``global'' knowledge of the network every node has, specifically, how much knowledge two distant nodes have of each other's states. In addition to the usual figures of merit, these quantifiers add an extra important dimension to the performance analysis and practical implementation of quantum repeaters. Finally, our understanding of the performance of large quantum networks is currently limited by the computational inefficiency of simulating them using RL or other optimization methods. The other main contribution of our work is to address this limitation. We present a method for nesting policies in order to obtain policies for large repeater chains. By nesting our RL-based policies for small repeater chains, we obtain policies for large repeater chains that improve upon the swap-as-soon-as-possible policy, and thus we pave the way for a scalable method for obtaining policies for long-distance entanglement distribution under practical constraints.

\end{abstract}

\maketitle

%\tableofcontents

\section{Introduction}

The development of advanced and practical quantum technologies is a hallmark of the \textit{second quantum revolution}~\cite{dowling2003quantum}. One of the frontiers of this revolution is a global-scale quantum internet~\cite{VanMeter_book,Dowling_book2,kimble2008quantuminternet,simon2017towards,WEH18,CCT+20,ICM+22,munro2022tomorrowquantuminternet}, which promises the realization of a plethora of quantum technologies, such as distributed quantum computing~\cite{cirac1999distributed,barz2012demonstration}, distributed quantum sensing~\cite{ge2018distributed, proctor2018multiparameter, komar2014quantum, simon2017towards, gottesman2012longer}, and quantum key distribution~\cite{bennett2020quantum, ekert1991quantum,xu2020QKDrealistic,pirandola2020advancescrypto}. A critical milestone on the road to a quantum internet is the ability to distribute quantum entanglement over long distances.

One of the main obstacles to achieving long-distance entanglement distribution, and consequently many of the above promises of quantum technologies, is noise, i.e., errors caused by the difficulty of maintaining good control over qubits and their environment. Noise arises in entanglement distribution due to loss in the quantum channels used to send qubits between spatially-separated nodes, and at every node noise arises due to short-lived quantum memories and imperfect entanglement swapping~\cite{heshami2016quantum,awschalom2021interconnects}. These sources of noise ultimately limit the rate, the distance, and the quality of entanglement distribution. Quantum error correction~\cite{Shor95errorcorrection,shor96faulttolerant,gottesman2009qecreview,roffe2019qecreview}, which includes entanglement distillation~\cite{bennett1996concentrating,bennett1996purification,bennett1996distillQEC}, has been understood for almost two decades to be the primary method to combat noise in order to achieve long-distance entanglement distribution, as well as full-scale, fault-tolerant quantum computation more generally. However, building devices with several thousands of qubits and then implementing error correction is currently a major technological and engineering challenge. 

Motivated by this current state of affairs, our work is inspired by the following ensuing idea: instead of viewing noise as something that should be fought, let us take noise as a \textit{given} and then see what protocols we can design, and what performance and potential advantages we can achieve. 
This idea lies at the intersection of theory and experiment, and our goal is to prove theoretical statements that provide a guide to researchers in the lab on how to design their devices in order to achieve the best performance. 
We note that this type of question, on making the best use of noisy quantum resources, has already been the focus of recent theoretical and experimental research on noisy intermediate-scale quantum computing~\cite{Preskill2018quantumcomputingin,bharti2021NISQreview,cerezo2021VQAreview}, particularly in the context of noise resilience~\cite{sharma2020noiseresilience,gentini2020noiseresilient,fontana2021noiseresilience}, quantum error mitigation~\cite{cai2022errormitigationreview}, and quantum advantage~\cite{bravyi2020quantumadvantagenoisy}.

\begin{figure*}
    \centering
    \includegraphics[width=0.95\textwidth]{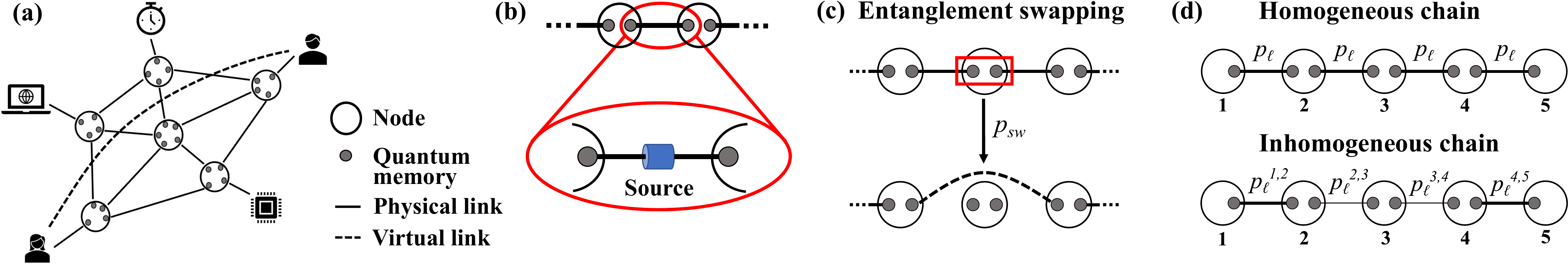}
    \caption{(a) A schematic of a general quantum network. The intermediate nodes represent quantum repeaters, and the symbols at the end nodes indicate possible applications. (b) A closer look at the elementary links and nodes of a linear network topology. Every node has two quantum memories, and every pair of neighboring quantum memories has an associated entanglement source. (c) Once both memories at a node are active (each storing one-half of a two-qubit entangled state), entanglement swapping can be performed to extend the entanglement to non-neighboring nodes, with success probability $p_{sw}$. (d) A linear quantum repeater chain. In a homogeneous chain, all elementary links have the same elementary link success probability $p_{\ell}$. For an inhomogeneous chain, these success probabilities can be different. We use $p_{\ell}^{i,j}$ to denote the elementary link success probability for nodes $i$ and $j$.}
    \label{fig:q_net}
\end{figure*}

Long-distance entanglement distribution typically proceeds by breaking up a quantum communication channel between a sender and a receiver into segments with intermediate nodes called ``quantum repeaters''~\cite{briegel1998quantum, duer1999repeaterspurification,muralidharan2016optimal}. The quantum repeaters (at least those of the first generation; see Refs.~\cite{MATN15,muralidharan2016optimal}) perform entanglement distillation and entanglement swapping~\cite{bennett1993teleporting,zukowski93swapping} in order to increase the quality and distance, respectively, of an entangled quantum state; see Fig.~\ref{fig:q_net}. Once entanglement between the sender and the receiver has been established, they can use the quantum teleportation protocol~\cite{bennett1993teleporting} in order to send arbitrary quantum information to each other. The advantage of using quantum repeaters is that the rates for the end-to-end quantum information transmission can be improved compared to directly sending the quantum information. Entanglement distribution has been experimentally demonstrated on numerous platforms for neighboring nodes~\cite{olmschenk2009quantum, nolleke2013efficient, langenfeld2021quantum, humphreys2018deterministic, pfaff2014unconditional}, and recently for non-neighboring nodes in a quantum network of solid-state spin qubits~\cite{hermans2022qubit}. 

Understanding what parameter regimes of today's noisy devices correspond to high performance (high rates and high fidelities)---or, conversely, what performance can be achieved with such devices---is crucial in order to realize long-distance entanglement distribution with currently available quantum devices. One challenge is to determine what protocols achieve \textit{optimal} performance in the presence of device imperfections. For example, a simple strategy one can always do is perform entanglement swapping as soon as both elementary links are active at a particular node~\cite{coopmans2021thesis,kamin2022exactrateswapasap}, the so-called ``swap-as-soon-as-possible'' (\textsc{swap-asap}) policy---but is this the \textit{best} strategy, especially in the presence of device imperfections, and particularly probabilistic elementary link generation and entanglement swapping? This is the primary question we are concerned with in this work. In more detail, we are concerned with the following questions:
\begin{itemize}
    \item Can we find optimal entanglement distribution policies, or at least policies better than \textsc{swap-asap}, for small, inhomogeneous, resource-constrained quantum networks, which provide faster connectivity as well as high end-to-end fidelities?
    \item Can general unifying features be identified for such improved policies?
    \item Can these unifying features help in devising improved policies for large networks with many nodes, for which a direct optimization is not practical?
\end{itemize}
Under certain conditions, which we outline below, we answer \textit{``yes''} to all of the above questions. 

\subsection{Summary of our work}

In this work, we provide a theoretical framework for both modeling protocols/policies for entanglement distribution and for optimizing policies with respect to the performance parameters of average waiting time and fidelity. We do so while taking specific aspects of physical implementations into account, but the framework is abstract enough that our results are agnostic to any particular implementation. We consider specifically the task of distributing entanglement between two distant nodes separated by a linear chain of quantum repeaters; see Fig.~\ref{fig:q_net}. We consider probabilistic elementary link generation, probabilistic entanglement swapping, and quantum memories with finite coherence time at the nodes. The policies that we obtain are thus functions of these high-level parameters that characterize the noise and imperfections of the quantum devices that are available today. We note that in our policy optimization, our notion of waiting time is based on the number of rounds of local operations and classical communication (LOCC), rather than the actual classical communication time. We also assume that all nodes have full, global knowledge of the state of the repeater chain; see Sec.~\ref{sec:MDP_model} for more details. While having full global knowledge is challenging in practice, especially for large repeater chains, we also in this work present a method for nesting policies that are ``global'' only for small sections of a full repeater chain; see Sec.~\ref{sec:nested} for details. Our theoretical framework is based on Markov decision processes (MDPs), which we use to model states of the repeater chain and the actions that can be taken at the nodes; see Sec.~\ref{sec:MDP_model}.

In principle, MDPs can be solved exactly using dynamic and linear programming~\cite{Puterman2014book}. However, because the state and action space grows exponentially with the input size (in our case, the input is the number of nodes), other techniques are needed. In this work, we use reinforcement learning (RL)~\cite{SuttonBarto2018book}, and we use it to obtain policies that, while not necessarily globally optimal, are better in certain parameter regimes than human-developed policies such as the \textsc{swap-asap} policy referred to above.

The main results of our work are as follows.
\begin{itemize}
    \item Using Q-learning, a model-free RL technique, we find improved policies for entanglement distribution in a linear repeater chain with up to five nodes. Our policies yield faster communication (i.e., lower waiting time) (Sec.~\ref{sec:opt_pol_waiting_time}) and higher fidelity (Sec.~\ref{sec-opt_pol_fidelity}) compared to the \textsc{swap-asap} policy. Furthermore, our fidelity-based results apply to arbitrary Pauli noise models, such as dephasing, for the decoherence of the quantum memories. The improvement of our policies over the \textsc{swap-asap} policy is the largest in the most non-ideal cases, and thus the most practically relevant cases: when the memory coherence times are low, the elementary link success probabilities are low, and the asymmetries are high. Furthermore, the policies obtained using Q-learning are known to be optimal, in principle, as long as enough training is done. In this sense, our policies are near-optimal. We refer to Appendix~\ref{sec-Q_learning} for more information about the Q-learning algorithm and the specifics of our implementation.
    
    \item Another important practical consideration is inhomogeneous device parameters. In this work, we also find improved policies for small inhomogeneous repeater chains of up to five nodes; see Sec.~\ref{sec-waiting_time_inhomogeneous}.
    
    \item Apart from discovering policies that outperform \textsc{swap-asap} over a wide range of parameters, in Sec.~\ref{sec:advantage} we also quantitatively address the question: \emph{Where do improved policies derive their advantage from?} We find that our improved policies have the following key features:
        \begin{enumerate}
            \item Using global knowledge of the network, i.e., using the state of the entire repeater chain when executing actions at a particular node.
            \item Collaboration between nodes of the network, i.e., correlations between the states and actions of links and nodes.
            \item Dynamic, state-dependent cutoffs for the links. This means that links are kept only for a limited amount of time before being discarded due to too much decoherence. Compared to the \textsc{swap-asap} policy, these cutoffs improve the end-to-end fidelity, as one would expect, but surprisingly they also improve the expected waiting time. 
        \end{enumerate}
        We provide in this work concrete quantifiers for the aforementioned ``global knowledge'', ``collaboration'', and ``dynamic cutoffs''; see Sec.~\ref{sec:advantage}. These results, and in particular the quantifiers that we use, provide new conceptual insights into the functioning of quantum repeaters. In particular, because our policies provide the most improvement in highly non-ideal parameter regimes, as described above, we see that collaboration between the nodes, one of the central features of our policies, is crucial when dealing with realistic, small-to-medium scale quantum networks with noisy and imperfect quantum devices.
    
    \item Finally, practical quantum networks require policies for distributing entanglement over very long distances. Finding an optimal policy directly for large repeater chains under general assumptions is neither analytically tractable nor computationally efficient. Global policies for such repeater chains, i.e., those in which the actions are based on the entire state of the repeater chain, would also not necessarily be practically useful, because classical communication times would adversely effect the average waiting time and end-to-end fidelity. Another main contribution of this work is to address this challenge. We introduce a nesting method that takes our improved RL-based policies for small sections of a long repeater chain and combines them with a local policy for connecting the sections amongst themselves; see Sec.~\ref{sec:nested}. We demonstrate this method for up to 13 nodes with two levels of nesting. In this case, we show that the nested policies substantially outperform fully local policies, and thus we demonstrate the utility of our RL-based approach even for very large quantum repeater chains.
    
\end{itemize}
It is worth mentioning that the \textsc{swap-asap} policy is a ``system-agnostic'' policy, not depending explicitly on the device parameters of the repeater chain. In contrast, our RL-based policies do in general depend on the input device parameters and on the entanglement swapping success probability. The advantage of obtaining these ``system-dependent'' policies is that it allows practitioners to use policies that are tailored to their specific device parameters. Furthermore, while it might not be surprising that we would obtain waiting time and fidelity improvements by using RL (or any other optimization technique that takes information about the system into account), understanding \textit{how} these improvements come about is certainly of interest, even for small repeater chains, and it is this question that we primarily address in this work. Notably, just by extracting the general principles of improved policies for small repeater chains in Sec.~\ref{sec-discussion} and Sec.~\ref{sec:advantage}, we are able to construct improved policies for large repeater chains in Sec.~\ref{sec:nested}.

\subsection{Summary of prior work}

Due to the complexity of the problem that we consider in this work, analytical treatments of performance and policy optimization~\cite{khatri2019practical, vinay2019statistical, shchukin2019waiting, VGNT20, PGGT20, DT21, Khatri2021policieselementary,coopmans2022improved,Kha22} are limited to simple geometries (e.g., linear or star networks) of only a handful of nodes, often with the simplifying assumption of infinite quantum memory coherence time. Hence, numerical optimization seems to be a more accessible route to analyzing realistic quantum networks. For example, Ref.~\cite{li2020efficient} numerically optimizes memory cutoffs, Ref.~\cite{jiang2007optimal} uses dynamic programming to optimize quantum repeater protocols, and Ref.~\cite{daSilva2021genetic} uses genetic algorithms. There also exist several numerical simulation platforms for quantum networks~\cite{MMG+15,DW18,Bart18,Mat19,DNZB20,WKC+20,CKD+20,wallnofer2022ReQuSim}.

Recently, in Ref.~\cite{wallnofer2020MLQComm}, the idea of using machine learning, in particular reinforcement learning, for quantum communication tasks was introduced. Then, in Ref.~\cite{Khatri2021policieselementary} (see also Ref.~\cite{Kha22}), a particular kind of MDP for elementary links in a quantum network was introduced, taking memory decoherence into account, with an exact solution using dynamic~\cite{Khatri2021policieselementary} and linear~\cite{Kha22} programming. Thereafter, in Refs.~\cite{shchukin2022optimal,inesta2022optimal,reiss2022deep}, an analogous MDP formulation for repeater chains was presented, with exact solutions based on linear programming in Ref.~\cite{shchukin2022optimal} and RL-based solutions in Refs.~\cite{inesta2022optimal,reiss2022deep}. This work is most similar in spirit to Refs.~\cite{inesta2022optimal,reiss2022deep}; however, there are notable differences in the modeling of the repeater chain using MDPs between those two works and ours, and we defer a detailed outline of these and other differences between our work and Refs.~\cite{inesta2022optimal,reiss2022deep} to Sec.~\ref{sec-compare_prior_work}.

Finally, it is important to emphasize that this line of work departs from the usual information-theoretic analysis of entanglement distribution and quantum communication~\cite{Hol12_book,NC00_book,KW20_book}, which is concerned with what rates, in principle, can be achieved when constrained by the laws of quantum mechanics alone. (See, in particular, Refs.~\cite{BCHW15,Pir16,AML16,LP17,CM17,AK17,BA17,RKB+18,Pir19,Pir19b,DBWH19,BAKE20,HP22} for information-theoretic analyses for quantum communication networks.) From a practical perspective, the fundamental rate limitations derived in those works should be thought of as informing us about what performance rates \textit{cannot at all} be achieved in practice. On the other hand, while this work still uses tools from quantum information theory, it also dives deeper into the physical implementation, thus resulting in a more realistic performance analysis, telling us what rates and performance criteria \textit{can} be achieved with quantum devices available to us today, and provides us with a mathematical framework for obtaining optimal protocols in realistic settings. Prior works along similar lines include Refs.~\cite{VLMN09,AME11,JKR+16,MDV19,rozpedek2018practicalentdistill,krastanov2019optentpurif,CCvM20,DPW20,CERW20,GEW20,bugalho2021multipartitenoisy,ramiro2021optimized,PGGT20,VGNT20,DT21,VanMeter_book,daSilva2021genetic,azuma2021tools}. We emphasize that while the framework and results presented in this work take the specifics of physical implementations into account, they are agnostic to any particular implementation.

\section{Theoretical Model} \label{sec:MDP_model}

In this section, we outline our theoretical model for entanglement distribution in a quantum repeater chain. We start with the model for the creation of elementary and virtual links, and then proceed to the definition of the two types of rewards that are provided to the learning agent in our Q-learning algorithm for determining policies for creating end-to-end entanglement in the quantum repeater chain. Further mathematical details on the model presented here can be found in Ref.~\cite{Kha22}.

In summary, our model consists of the following elements.
\begin{itemize}
    \item The quantum repeater chain consists of $n$ nodes. Every node has at most two quantum memories; see Fig.~\ref{fig:q_net}. All of these memories have a maximum cutoff time $m^{\star}\in\{0,1,2,\dotsc\}$, which can be thought of in terms of the coherence time of the quantum memory. If a pair of memories has been holding an entangled state for $m^{\star}$ time steps, then the state is discarded and the entanglement is regenerated.
    
    \item Elementary link generation succeeds with probability $p_{\ell}\in[0,1]$. Different elementary links can have different success probabilities, in which case we refer to the repeater chain as ``inhomogeneous''. Otherwise, when all of the elementary links have the same value of $p_{\ell}$, we refer to the repeater chain as ``homogeneous''.
    
    \item Entanglement swapping is assumed to be error-free but non-deterministic in general, succeeding with probability $p_{sw}\in[0,1]$.
    
    \item We consider discrete time steps. In every time step, nodes, or pairs of nodes, can perform the following actions:
        \begin{itemize}
            \item A pair of neighboring nodes can attempt to generate an elementary link between them, discard the link that might already exist between them, or wait (do nothing).
            
            \item A node can attempt entanglement swapping on its two quantum memories.
        \end{itemize}
        
\end{itemize}
It is important to note that, in this work, we consider the abstract setting in which the size of the repeater chain is given simply by the total number $n$ of nodes, and neither the physical distance between the end points of the chain nor the distances between the intermediate nodes is explicitly specified. Also, classical communication---specifically, LOCC---is assumed to be free in our framework. This abstract setting is entirely analogous to the information-theoretic setting of quantum communication, in which only the number of channel uses (or the number of ``network uses''---see, e.g., Refs.~\cite{Pir16,Pir19}) is the resourceful quantity used to determine the waiting time/rate, and LOCC is assumed to be free. Accordingly, in this work, a time step is synonymous with a single use of the repeater chain, and thus entanglement swapping operations (because they are LOCC operations) do not add a time step, and thus do not contribute to the waiting time; see Sec.~\ref{sec:opt_pol_waiting_time} below for more information. A detailed calculation of waiting times, taking into account classical communication in both the uses of the repeater chain and of the entanglement swapping operations, is left for future work. 

We now describe our model for the creation of elementary and virtual links.

\subsection{Creation of elementary links}

As shown in Fig.~\ref{fig:q_net}(b), an \emph{elementary link} exists between two neighboring nodes when one part of a bipartite entangled state is successfully delivered to each of them from a common source. Elementary link creation succeeds with probability $p_{\ell} \in [0,1]$. This success probability includes the effects of channel loss, detector inefficiencies, noise, finite absorption cross-sections of quantum memories, etc.; see Refs.~\cite{Khatri2021policieselementary,Kha22} for details and examples. We call the elementary link ``active'' if an entangled state is shared between the two nodes.

Assume that the quantum state shared by the quantum memories of two nodes, call them $R_1$ and $R_2$, after successful creation of the elementary link, is the two-qubit state $\sigma_{R_1R_2}^0$. Let us also assume that the qubit memories $R_1$ and $R_2$ undergo decoherence according to quantum channels $\mathcal{N}^1$ and $\mathcal{N}^2$, respectively. Then, let
\begin{equation}\label{eq-elem_link_state}
    \sigma_{R_1R_2}(m)\coloneqq ((\mathcal{N}_{R_1}^{1})^{\circ m}\otimes(\mathcal{N}_{R_2}^2)^{\circ m})(\sigma_{R_1R_2}^0)
\end{equation}
be the quantum state of the two memories after $m\in\{0,1,2,\dotsc\}$ discrete time steps, where $(\mathcal{N}^1)^{\circ m}\equiv \mathcal{N}^1\circ\mathcal{N}^1\circ\dotsb\circ\mathcal{N}^1$ ($m$ times), and similarly for $(\mathcal{N}^2)^{\circ m}$. We refer to $m$ as the \textit{age} of the elementary link. In Sec.~\ref{sec-practical}, we give an example of a noise model and an explicit form for the states $\sigma_{R_1R_2}(m)$.

As time progresses, the fidelity of the quantum state in \eqref{eq-elem_link_state} deteriorates until it is considered too noisy to be useful after a given number $m^{\star}$ of time steps, at which point the state is discarded and the elementary link is created again. We let
\begin{equation}\label{eq-elem_link_fidelity}
    f(m)\coloneqq \bra{\Phi^+}\sigma(m)\ket{\Phi^+}
\end{equation}
be the fidelity of the state of the elementary link after $m$ time steps, where $\ket{\Phi^+}=\frac{1}{\sqrt{2}}(\ket{0,0}+\ket{1,1})$ is the two-qubit maximally-entangled Bell state. In Sec.~\ref{sec-practical}, we give an example of a noise model for which we can analytically determine this fidelity function.

\subsection{Creation of virtual links}\label{sec-virtual_links}

As shown in Fig.~\ref{fig:q_net}(c), a \emph{virtual link} can be established between two distant nodes with intermediate nodes acting as quantum repeaters, which execute the entanglement swapping protocol with success probability $p_{sw} \in [0,1]$. With reference to Fig.~\ref{fig:q_net}(c), consider three nodes, corresponding to two elementary links, such that the quantum state of the two elementary links is $\sigma_{AR_1}(m_1)\otimes\sigma_{R_2B}(m_2)$. By performing the entanglement swapping protocol on the two qubits $R_1,R_2$, the resulting quantum state shared by $A$ and $B$ is $\rho_{AB}(m_1,m_2)$; we refer to Ref.~\cite{Kha22} for an explicit expression for this state, which is not required for our purposes in what follows. Now, we would like to determine the fidelity
\begin{equation}\label{eq-virtual_link_fidelity}
    f_{\textsc{swap}}(m_1,m_2)\coloneqq\bra{\Phi^+}\rho(m_1,m_2)\ket{\Phi^+}
\end{equation}
of the swapped state without keeping track of the full quantum state of the elementary links. Depending on the noise model of the individual quantum memories, there exists an $m'\in\{0,1,\dotsc\}$ such that $f_{\textsc{swap}}(m_1,m_2)=f(m')$. The age $m'$ of the swapped state can thus be calculated as $m'=f^{-1}(f_{\textsc{swap}}(m_1,m_2))$, as long as $f$ is invertible. For example, suppose that the initial state $\sigma^0$ in \eqref{eq-elem_link_state} is the perfect maximally-entangled state $\Phi^+$. Then,  for Pauli noise, because it commutes with the entanglement swapping operation~\cite{bowen2001teleportation,schmidt2020memoryassisted}, we have that $f_{\textsc{swap}}(m_1,m_2)=f(m_1+m_2)$, so that $m'=m_1+m_2$. This update rule can be viewed as a consequence of the developments in Ref.~\cite[Appendix~D]{schmidt2020memoryassisted}; however, for convenience, we provide a short proof in Appendix~\ref{sec:ent_swap_Pauli_update}. For non-Pauli, in particular non-unital noise models, the function $f_{\textsc{swap}}$ will not necessarily have a simple form, and we leave the analysis of entanglement swapping under non-unital noise as an interesting direction for future work.

With the update rule $f_{\textsc{swap}}(m_1,m_2)=f(m_1+m_2)$ for entanglement swapping of two elementary links under Pauli noise, we have that for a repeater chain with $n$ nodes, i.e., $n-1$ elementary links, the end-to-end fidelity after entanglement swapping is given simply by $f(m_1+m_2+\dotsb+m_{n-1})$, where $m_1,m_2,\dotsc,m_{n-1}$ are the ages of the elementary links. A basic requirement for our end-to-end link is then that $m_1+m_2+\dotsb+m_{n-1}\leq m^{\star}$, so that the fidelity of the end-to-end link will be at least equal to $f(m^{\star})$. This requirement can be achieved if $(n-1)t^{\star}\leq m^{\star}$, where $t^{\star}=\max\{m_1,m_2,\dotsc,m_{n-1}\}$. Alternatively, if we have a more stringent requirement on the fidelity of the end-to-end link, say $F_{\min}\in(f(m^{\star}),1)$, then the condition becomes
\begin{equation}\label{eq-n_tStar_Fmin_tradeoff_basic}
    (n-1)t^{\star}\leq f^{-1}(F_{\min}).
\end{equation}
The quantity $t^{\star}\in\{0,1,\dotsc,m^{\star}\}$ can be thought of as a \textit{cutoff time} for the elementary links, a time that could be less than $m^{\star}$ at which the elementary links are discarded and regenerated. The relation in \eqref{eq-n_tStar_Fmin_tradeoff_basic} can then be interpreted as telling us the maximum number $n-1$ of elementary links that are allowed in a repeater chain with a given end-to-end fidelity requirement $F_{\min}\in[f(m^{\star}),1)$ and a given cutoff time $t^{\star}$ for the elementary links. In this work, using the reinforcement learning model that we introduce below, in Sec.~\ref{sec-practical} we obtain values for the minimum fidelity $F_{\min}$ that can be achieved for a given number $n$ of nodes, through optimization of the cutoff $t^{\star}$, while allowing this cutoff to be different for each of the elementary links.

\subsection{Policies via reinforcement learning}

We are interested in optimizing the waiting time and fidelity in quantum repeater chains. To do this, in this work we resort to using reinforcement learning, specifically, the Q-learning reinforcement learning algorithm~\cite{SuttonBarto2018book}. Q-learning is a model-free reinforcement learning algorithm, in the sense that the learning agent is not given a description of its environment. In principle, the Q-learning algorithm provides us with an optimal policy, in the limit that all states and actions are visited by the agent sufficiently often. The mathematical backbone of reinforcement learning is a model of the agent's environment as a Markov decision process (MDP). Our MDP for a quantum repeater chain is summarized in Fig.~\ref{fig:mdp}, and we provide a detailed description of the MDP in Appendix~\ref{sec-MDP_details}. In Appendix~\ref{sec-Q_learning}, we provide a few details on the Q-learning algorithm and our implementation of it.

\begin{figure}
    \centering
    \includegraphics[width=0.70\columnwidth]{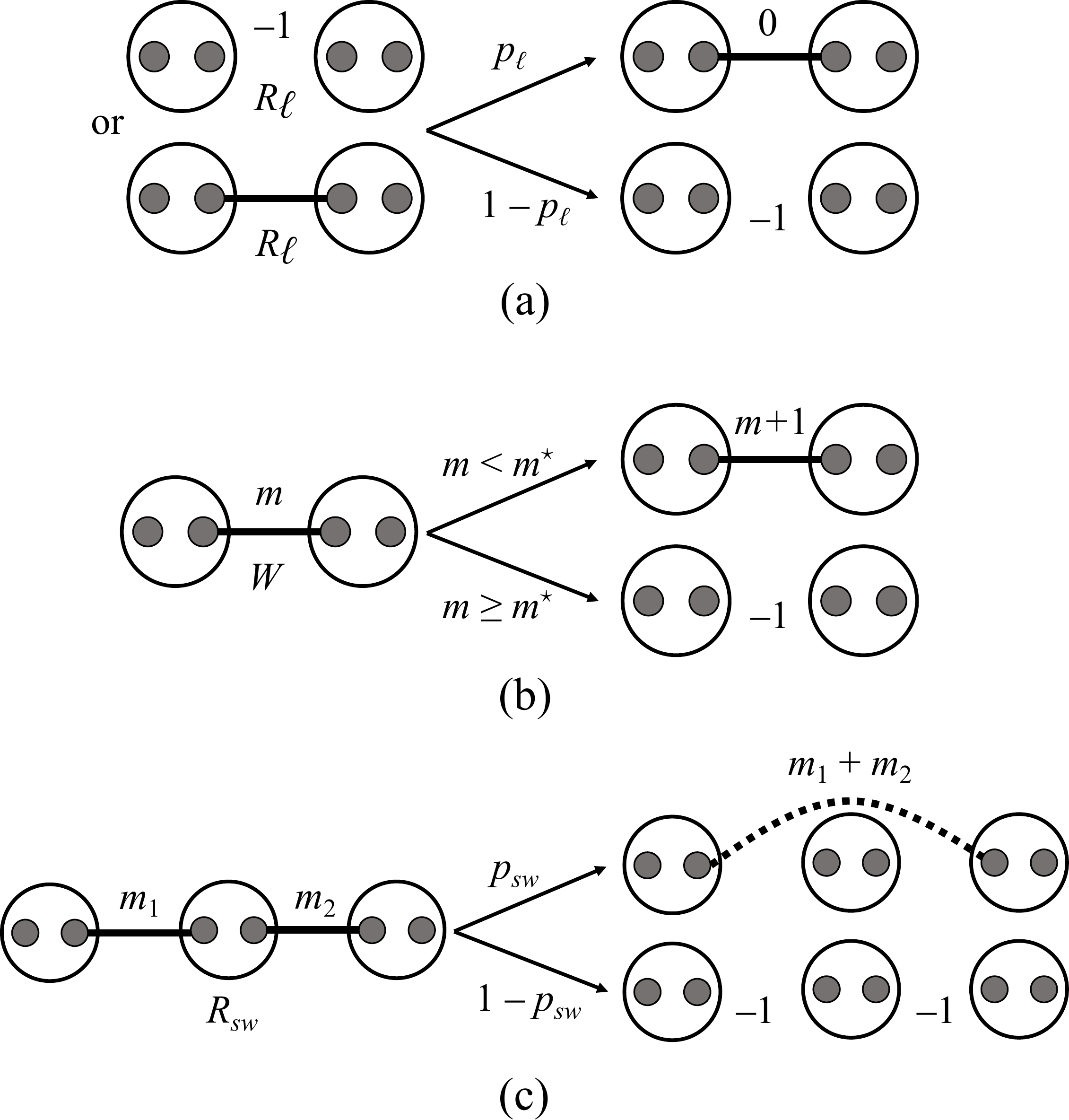}
    \caption{Illustration of the states, actions, and transition rules of our MDP for entanglement distribution in a quantum repeater chain; see Appendix~\ref{sec-MDP_details} for details. (a)~The effect of the elementary link request action ($R_{\ell}$) between two nodes is to discard the entangled state, if there is one stored in their respective quantum memories, and attempt to generate an elementary link between the two nodes, which succeeds with probability $p_{\ell}$.  We note that the agent can choose to discard a link at any time $t^{\star}\leq m^{\star}$, thereby allowing it to choose a cutoff that is different from the maximum cutoff $m^{\star}$. (b)~The effect of the wait action ($W$) is to increase the age $m$ of a link by one time unit, if $m$ is strictly less than the maximum cutoff time $m^{\star}$; otherwise, if $m=m^{\star}$, then the link is discarded. (c)~Given two links with ages $m_1$ and $m_2$, the entanglement swapping action ($R_{sw}$) succeeds with probability $p_{sw}$, in which case the age of the new, virtual link is given by the sum $m_1+m_2$ of the ages of the links. We emphasize that all virtual links are discarded at the max cutoff time $m^\star$, i.e., if $m_1+m_2 > m^\star$, then the virtual link creation is considered a failure, even if the entanglement swapping operation succeeds.} %the age of the oldest elementary link, i.e., $\max\{m_1,m_2\}$.}
    \label{fig:mdp}
\end{figure}

Our implementation of the Q-learning algorithm requires the learning agent to receive a reward in a manner such that maximizing the expected total reward over time corresponds to optimizing either the waiting time or the fidelity. We now provide our definitions of these rewards.

\paragraph*{Reward for the waiting time.} When optimizing for the expected waiting time, we want the agent to receive high rewards for actions that bring it to the terminal state, namely, the state in which the two end nodes are connected. Hence, we define the reward $R$ as follows:
\begin{equation}\label{eq-opt_waiting_time_reward}
     R(S,A) = \left\{
     \begin{array}{l l}
       100 & \quad \text{if } S \text{ is a terminal state}, \\
       -1 & \quad \text{otherwise}, \\
     \end{array}\right.
\end{equation}
for every state $S$ and action $A$. Note that with this definition, the agent, in order to achieve the highest expected total reward, is incentivized to take the least number of steps to reach the terminal state, and this is why it can be used to minimize the expected waiting time for an end-to-end link.

\paragraph*{Reward for the fidelity.} When maximizing the fidelity, we want to give rewards to the Q-learning agent based on minimizing the age of the end-to-end entangled state. Note that when optimizing for the waiting time, the reward in \eqref{eq-opt_waiting_time_reward} only depends on whether the state is a terminal state. Thus, in that case, the total reward collected is the highest if the number of intermediate states required to reach the terminal states from a fully disconnected initial state is minimized (least number of negative rewards collected). Now, we are not only interested in taking a small number of steps, but also at the end we want to reach a terminal state with a small value of the age $S_{1,n}$ of the end-to-end entangled state; we refer to Appendix~\ref{sec-MDP_details} for an explanation of the notation $S_{1,n}$. The reward should therefore incentivize the agent to not only reduce the waiting time, but also reduce the age of the end-to-end link. We therefore define the reward as follows:
\begin{equation}\label{eq-reward_Q_learning_fidelity}
    R(S,A) = \left\{
     \begin{array}{l l}
       \frac{100}{S_{1,n}} & \quad \text{if } S \text{ is a terminal state}, \\
       -1 &\quad \text{otherwise}, \\
     \end{array}\right.
\end{equation}
for every state $S$ and action $A$. In other words, higher positive rewards are awarded to the agent if the age of the end-to-end link is low. We also emphasize that by defining the reward in terms of the age of the entangled state, we circumvent the need to keep track of quantum states during our simulations.

Let us emphasize a subtlety in the definition of the reward in Eq.~\eqref{eq-reward_Q_learning_fidelity}. This reward does not maximize the fidelity alone. In fact, for optimizing the fidelity alone, the optimal policy is clear: just wait for all of the elementary links to become newly active \textit{at the same time}, and then perform entanglement swapping. This policy maximizes the end-to-end fidelity, but the waiting time is exceedingly large. On the other hand, with the reward function in \eqref{eq-reward_Q_learning_fidelity} we can, in a sense, optimize \textit{both} the fidelity and the waiting time, and we illustrate this explicitly in Sec.~\ref{sec-opt_pol_fidelity}. We remark that this subtlety in the choice of the reward when considering the fidelity has been pointed out already in the case of elementary links in Ref.~\cite{Khatri2021policieselementary}.

\section{Improved policies for the waiting time}\label{sec:opt_pol_waiting_time}

In this section, we apply the Q-learning reinforcement learning algorithm to find policies that reduce the expected waiting time for end-to-end entanglement in a linear quantum repeater chain. We consider four-node and five-node repeater chains, and we compare the policy obtained via Q-learning to the \textsc{swap-asap} (swap-as-soon-as-possible) policy~\cite{coopmans2021thesis,kamin2022exactrateswapasap}. We refer to Appendix~\ref{sec-Q_learning} for a brief description of the Q-learning algorithm and the details of our implementation of it.

\subsection{Evolution of the state and calculation of the waiting time}

As described at the beginning of Sec.~\ref{sec:MDP_model}, the state of the repeater chain evolves in discrete time steps. Initially, all elementary links in the repeater chain are inactive. In every time step, we have a combination of elementary link requests and requests for entanglement swapping. Specifically, we have the following:
\begin{enumerate}
    \item Check if there are any elementary link requests. If so, then increase the age of all existing elementary and virtual links by one; see Appendix~\ref{sec-MDP_details} for the precise update rule. Otherwise, leave the state unchanged.
    
    \item Perform all requests for elementary links. 
   
    \item Perform all entanglement swapping requests.
\end{enumerate}
Since we consider LOCC to be free, any time step in which the action consists only of entanglement swapping requests does not increase the age of existing links and also does not count towards the waiting time. Time steps in which only elementary links are requested or both swaps and elementary links are requested contribute one time step towards the waiting time and also to the age of all links. This is a reasonable assumption for short chains, in which the classical communication time between different nodes of the chain is small compared to the time required to create the elementary links by storing the shared entangled qubits in quantum memories. For longer chains/larger networks, classical communication times become more relevant. Even though we neglect the time taken for entanglement swapping, some reasonable restrictions are put on the allowed actions at any given time step in order to mimic the finite-time requirements of entanglement generation, processing, and storage in memories. Specifically,
%if $A_{i,i}^t = 1$, then $A_{i-1,i}^t = 0$ and $A_{i,i+1}^t = 0$, i.e.,
nodes that request entanglement swapping cannot request elementary links in the same time step, and vice versa. Another restriction is that virtual links cannot cross. For example, consider a seven-link repeater chain. If nodes 1 and 5 are connected, then nodes 2 and 6 cannot be connected; similarly, nodes 3 and 7 cannot be connected, and so on. However, nodes 2, 3, and 4 can form links amongst themselves. We refer to Appendix~\ref{sec-MDP_details} for further details regarding the constraints on the states and actions.

\begin{figure}
    \centering
    \includegraphics[width=0.9\columnwidth]{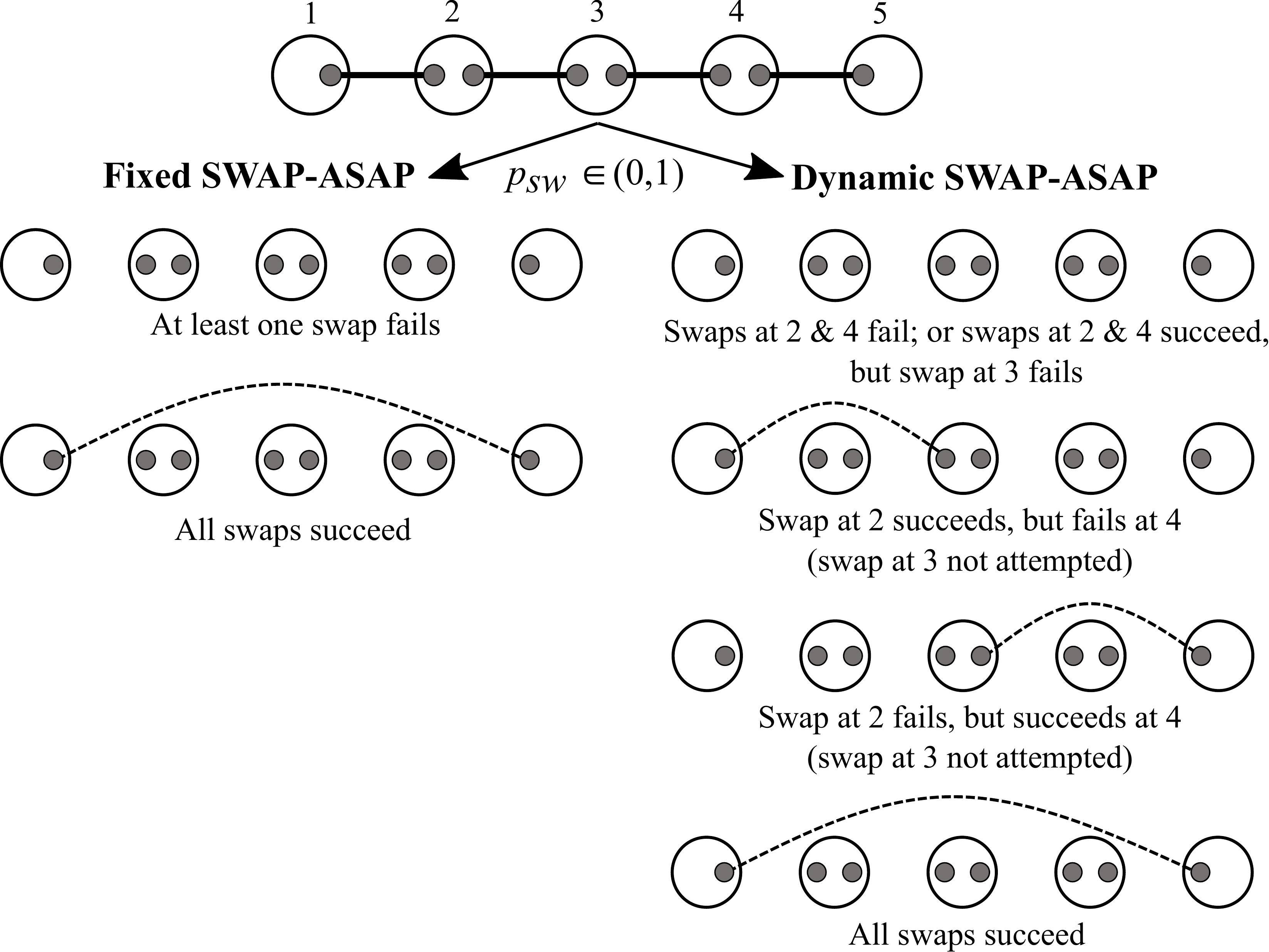}
    \caption{Fixed and dynamic versions of the \textsc{swap-asap} policy. Consider a five-node repeater chain in which all elementary links are active at the same time and the entanglement swapping is non-deterministic. In the fixed \textsc{swap-asap} policy, all Bell measurements are performed at once, leading to two possibilities: either one of the swaps fails and consequently all elementary links become inactive, or all of them succeed and we obtain end-to-end entanglement. On the other hand, for the dynamic \textsc{swap-asap} policy, there are four possibilities illustrated on the right-hand side. These possibilities are allowed within the framework of \textsc{swap-asap} when LOCC are considered free.}
    \label{fig:swap-asap}
\end{figure}

\subsection{Fixed and dynamic \textsc{swap-asap} policies} \label{subsec:swap_asap}

In this work, we consider two versions of the \textsc{swap-asap} policy, one that we refer to as the \textit{dynamic \textsc{swap-asap}} policy (which has been considered before in Ref.~\cite{shchukin2022optimal}) and the other that we refer to as the \textit{fixed \textsc{swap-asap}} policy (which has been considered before in Ref.~\cite{inesta2022optimal}). See Fig.~\ref{fig:swap-asap} for an illustration of the difference between the two versions. 

The dynamic \textsc{swap-asap} policy is defined as follows.
\begin{itemize}
    \item The action $R_\ell$ of requesting an elementary link between two neighboring nodes is performed whenever an elementary link is inactive. 
    
    \item An elementary link is discarded if and only if its age is equal to the maximum cutoff value $m^\star$. 
    
    \item A virtual link is discarded when its age is equal to $m^\star$.
    
    \item Entanglement swaps are performed ``as soon as possible'', i.e., as soon as two elementary links sharing a common node are active.
    
    \item If more than one entanglement swap can be performed sequentially, then we are free to choose the order of the swaps, in a \emph{dynamic} way; see Fig.~\ref{fig:swap-asap} for an example. When LOCC is considered free, all such different orderings correspond to a single time step.
\end{itemize}

The policy described above is called the ``dynamic'' \textsc{swap-asap} policy because the entanglement swapping is performed sequentially when multiple entanglement swapping operations can be done at the same time. On the other hand, the \emph{fixed \textsc{swap-asap}} policy is a strictly local policy in which the entanglement swapping operations are done in parallel rather than sequentially. Doing the entanglement swapping operations in parallel means that they can only succeed together or fail together. For example, in Fig.~\ref{fig:swap-asap}, we see that if one of the entanglement swapping operations fails, then this causes all of the other elementary links to become inactive, even if the swapping succeeds at the other nodes. Instead, in the dynamic \textsc{swap-asap} policy, because the swapping operations are done sequentially, it is possible to communicate the failure of a swapping operation to the adjacent nodes, such the adjacent nodes can avoid doing the entanglement swapping. In fact, a better strategy in the four-segment case is thus to first try to perform the swaps at nodes 2 and 4 and only if they succeed attempt swapping at node 3 (see Fig.~\ref{fig:swap-asap}~(right)). The extra classical communication involved in the dynamic \textsc{swap-asap} policy does not add to the waiting time, because in our model LOCC is considered free. As a result of this freedom in choosing the order of the entanglement swapping, and because LOCC is assumed free, the dynamic \textsc{swap-asap} policy can, in principle, outperform the fixed \textsc{swap-asap} policy.

Despite their differences, from the point of view of actions on elementary links (namely, waiting and requesting on an elementary link), both the fixed and dynamic versions of the \textsc{swap-asap} policy are considered to be local, because in these policies the elementary link actions of waiting and requesting depend only on the state of the nodes performing those actions, and not on the state of the other elementary links or of the network as a whole.

\subsection{Homogeneous repeater chains}

We now analyze the performance of policies obtained via Q-learning, henceforth referred to as ``Q-learning policies''. Specifically, we compare our Q-learning policies to the dynamic \textsc{swap-asap} policy using the improvement factor~\cite{inesta2022optimal}, which is defined as $(T_{RL} - T_{SA})/T_{RL}$, where $T_{RL}$ is the average waiting time using the Q-learning policy and $T_{SA}$ is the average waiting time using the dynamic \textsc{swap-asap} policy.

\begin{figure}
    \centering
    \includegraphics[width=0.48\columnwidth]{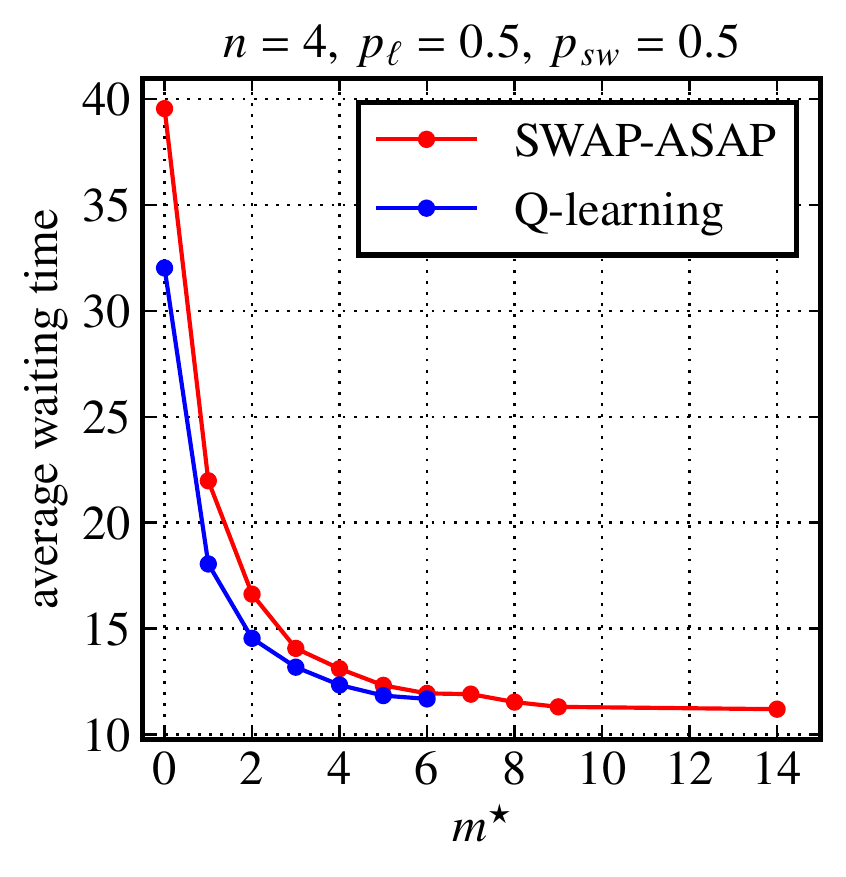}
    \includegraphics[width=0.48\columnwidth]{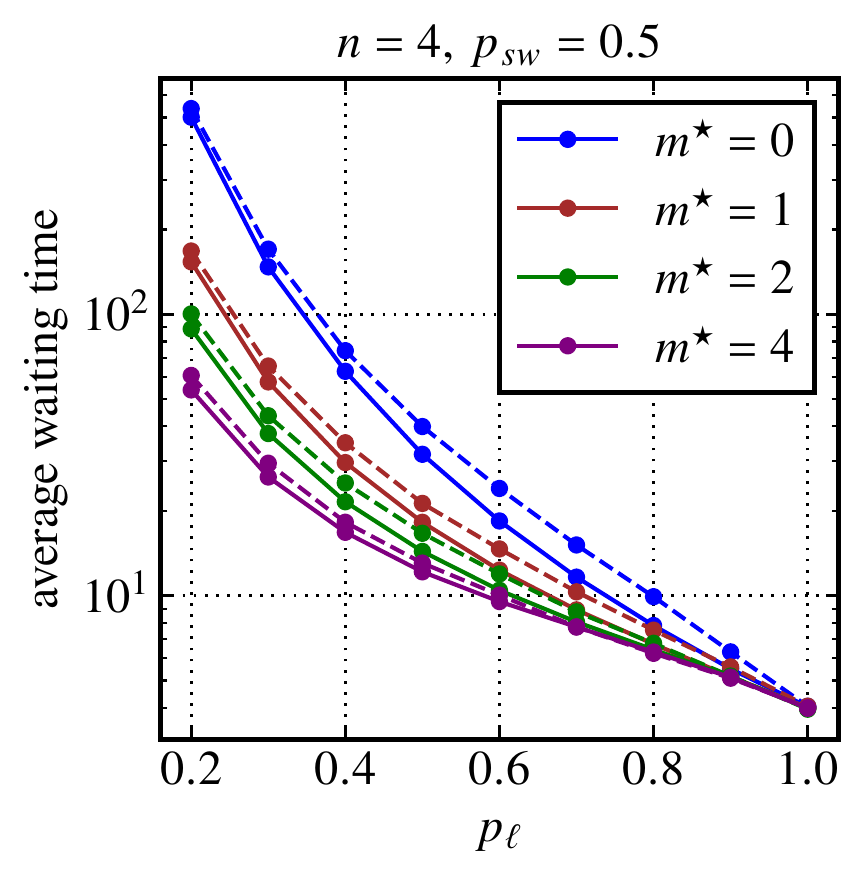}
    \includegraphics[width=0.48\columnwidth]{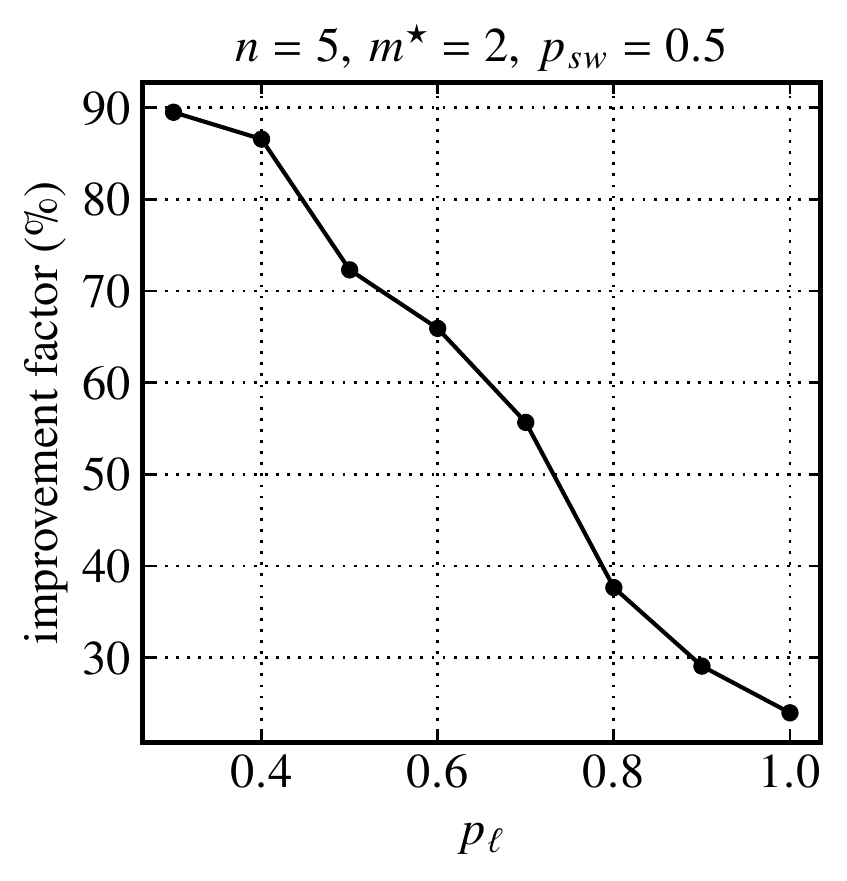}
    \caption{Average waiting time for a repeater chain with four and five nodes and non-deterministic entanglement swapping success probability $p_{sw}=0.5$. (Top left) Average waiting time as a function of the maximum memory cutoff $m^{\star}$, with elementary link success probability $p_{\ell}=0.5$. (Top right) Average waiting time as a function of $p_{\ell}$, for different values of $m^{\star}$. The solid line is the Q-learning policy and the dashed line is the dynamic \textsc{swap-asap} policy. (Bottom) Improvement factor of our Q-learning policy compared to the dynamic \textsc{swap-asap} policy as a function of $p_{\ell}$.}
    \label{fig:t_star_benchmark}
\end{figure}

\begin{figure}
    \centering
    \includegraphics[width=0.48\columnwidth]{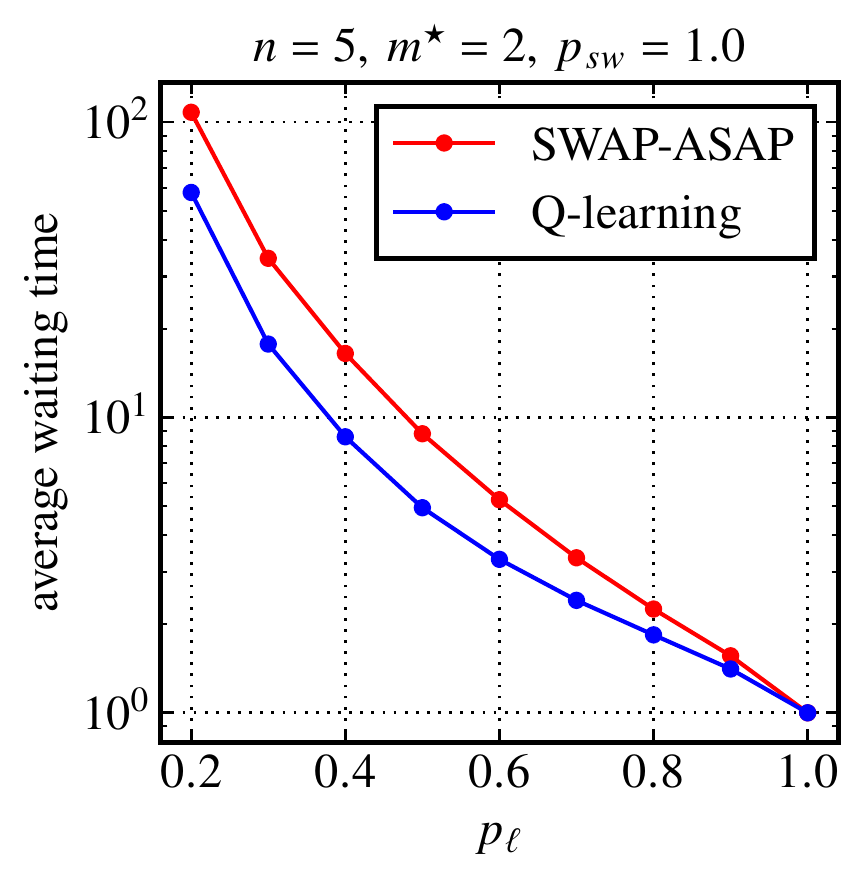}
    \includegraphics[width=0.48\columnwidth]{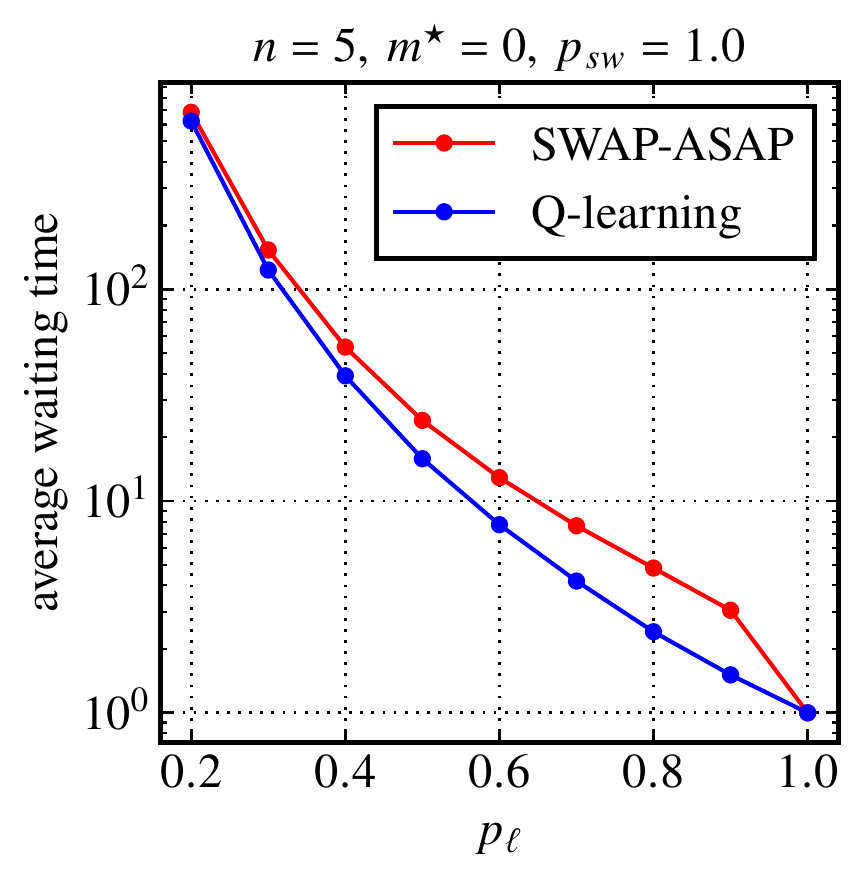}
    \caption{Average waiting time as a function of the elementary link success probability $p_{\ell}$ for a repeater chain with five nodes, deterministic entanglement swapping, $m^{\star}=2$ (left), and $m^{\star}=0$ (right).}
    \label{fig:deterministic}
\end{figure}

Our main results are as follows.
\begin{enumerate}
    \item For a fixed elementary link and entanglement swapping success probability, the Q-learning policy's waiting time curve has a larger gap with the \textsc{swap-asap} curve for smaller values of the maximum memory cutoff $m^{\star}$. In Fig.~\ref{fig:t_star_benchmark}~(top left), we show this trend for $p_{\ell} = 0.5$ and $p_{sw} = 0.5$. As $m^{\star}$ increases, the Q-learning policy approaches the dynamic \textsc{swap-asap} policy, and the improvement factor falls. It is also interesting to note that the waiting time for the dynamic \textsc{swap-asap} policy approaches a fixed value for large $m^{\star}$. In Fig.~\ref{fig:t_star_benchmark}~(top right), we show that for various values of $m^\star$ and $p_{sw}$, the waiting time falls with increasing values of $p_\ell$. We also see that Q-learning policies always yield a lower waiting time than the dynamic \textsc{swap-asap} policy, with the gap reducing with increasing $p_\ell$; see also Fig.~\ref{fig:t_star_benchmark}~(bottom).

    It is also notable that even for $m^{\star}=0$, which is when links have to be used as soon as they are created, the Q-learning policy does not coincide with the \textsc{swap-asap} policy. Instead, the Q-learning policy coincides with the policy of waiting for all of the elementary links to become active at the same time, and then doing all of the entanglement swapping operations at once. The waiting time of this policy has the known analytical expression~\cite{das2018robust} $1/(p_{\ell}^{n-1}p_{sw}^{n-2})$, and this indeed coincides with the $m^{\star}=0$ Q-learning data shown in Fig.~\ref{fig:t_star_benchmark} (top right). This result proves that the \textsc{swap-asap} policy is not optimal even in the simple case of $m^{\star}=0$. On the other hand, when $p_{\ell}=1$, the Q-learning and \textsc{swap-asap} policies have the same average waiting time, which is equal to $1/p_{sw}^{n-2}$. Although the waiting times coincide, we note that the policies are different, a point that we elaborate upon in Sec.~\ref{sec-swap_asap_vs_Q_learning}. 
    
    \item For deterministic entanglement swapping, in Fig.~\ref{fig:deterministic}, we observe similar trends, with the Q-learning policies providing a greater advantage for low values of $p_{\ell}$, and the advantage diminishing with increasing $p_{\ell}$. We note that, in this case, both the fixed and dynamic \textsc{swap-asap} policies have the same waiting times. Also, as in the case of non-deterministic entanglement swapping in Fig.~\ref{fig:t_star_benchmark}, it is interesting to see that the \textsc{swap-asap} policy is again not optimal for $m^{\star}=0$ (see Fig.~\ref{fig:deterministic}~(right)), while the Q-learning policy has the waiting time given by $1/p_{\ell}^{n-1}$, corresponding to waiting for all of the elementary links to become active at the same time and then performing all of the entanglement swapping operations at once.
    
\end{enumerate}

The trends we obtain for the improvement in waiting time in the deterministic swapping case are qualitatively and quantitatively in close agreement with the results in Ref.~\cite{inesta2022optimal}, in which the comparison of their policies was done against the fixed \textsc{swap-asap} policy. This agreement is to be expected, because for deterministic entanglement swapping both the fixed and dynamic \textsc{swap-asap} policies yield the same waiting time. However, in the non-deterministic case, our trends do not agree with Ref.~\cite{inesta2022optimal}, in which the improvement factor increases monotonically with increasing $p_\ell, m^\star$, while in our case the improvement factor decreases. This difference arises because of some of the differences in our assumptions. First, unlike Ref.~\cite{inesta2022optimal}, our MDP formulation allows the memory cutoff to be dynamic, which leads to greater improvement over the policies found in Ref.~\cite{inesta2022optimal}. Furthermore, since we do not allow requests and swaps at the same nodes to take place in the same time step, our waiting times are slightly different from those found in Ref.~\cite{inesta2022optimal}. The improvement factors for a five-node repeater chain in Ref.~\cite{inesta2022optimal} vary between $\approx$ 6-13\% for $p_{sw} = 0.5$, $p_\ell\in(0.3,1.0)$, and $m^\star\in(2,6)$. In our work, for $p_{sw} = 0.5$ and $m^\star = 2$, the improvement factors against the fixed \textsc{swap-asap} and dynamic \textsc{swap-asap} policies vary between $\approx$ 35-50\% and $\approx$ 10-50\%, respectively. We have thus shown that by using our more general MDP formulation, a greater improvement in waiting times, even over the dynamic \textsc{swap-asap} policy (which is itself better than the fixed \textsc{swap-asap} policy), can be obtained.

\subsection{Inhomogeneous repeater chains}\label{sec-waiting_time_inhomogeneous}

\begin{figure}
    \centering
    \includegraphics[width=0.48\columnwidth]{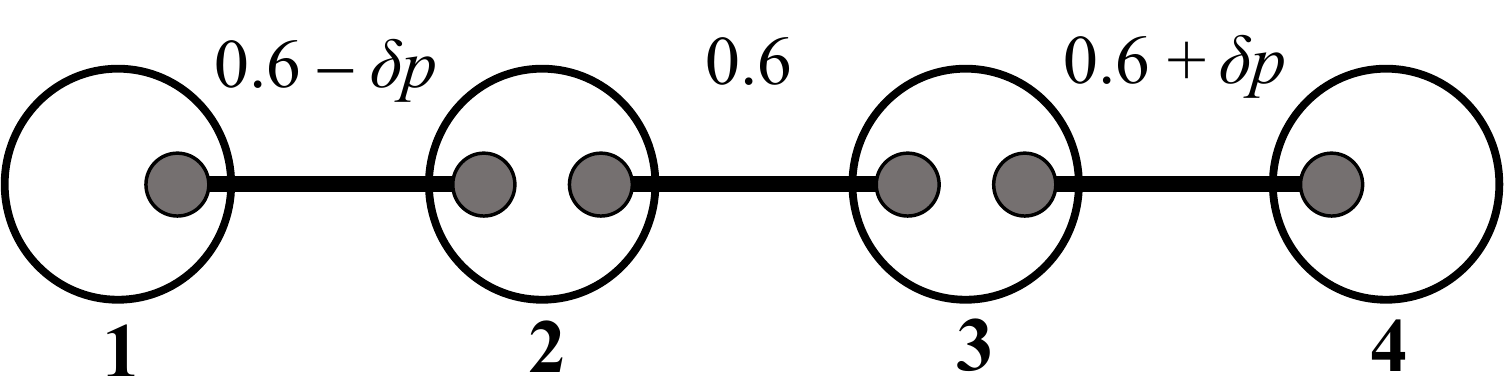}\\ \medskip
    \includegraphics[width=0.70\columnwidth]{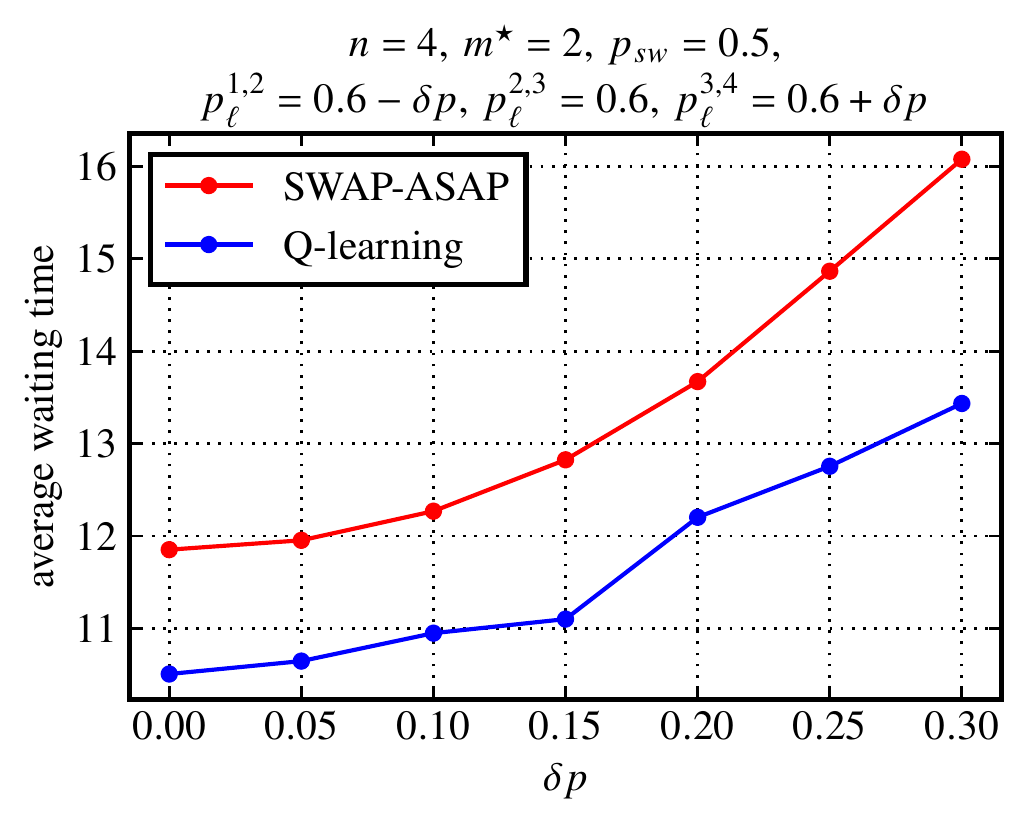}
    \caption{Average waiting time of our Q-learning policy for a four-node inhomogeneous repeater chain compared to the dynamic \textsc{swap-asap} policy as a function of the inhomogeneity $\delta p$ in the elementary link success probabilities, where $\delta p = p_{\ell}^{2,3} - p_{\ell}^{1,2} = p_{\ell}^{3,4} - p_{\ell}^{2,3}$. Here, $p_{\ell}^{i,j}$ is the elementary link success probability between nodes $i$ and $j$. Our Q-learning policies provide an advantage over the \textsc{swap-asap} policy even for short inhomogeneous chains.} \label{fig:asymmetry_comparison} 
\end{figure}

\begin{figure}
    \centering
    \includegraphics[width=0.6\columnwidth]{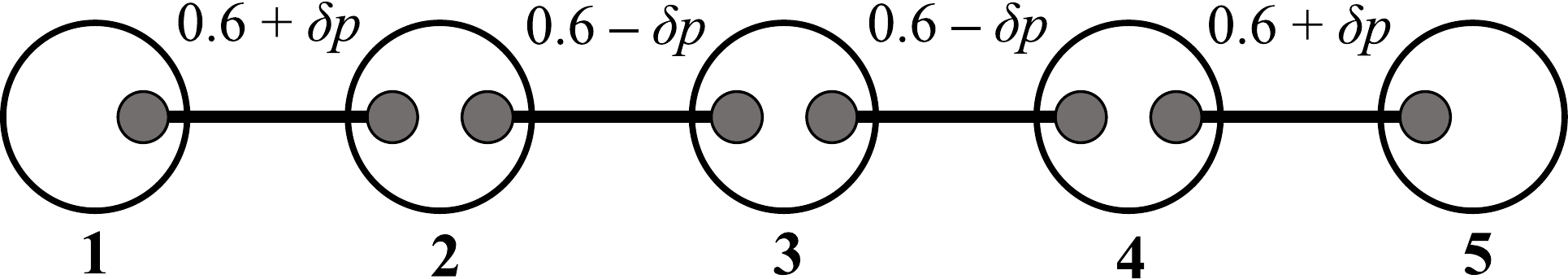}\\ \medskip
    \includegraphics[width=0.70\columnwidth]{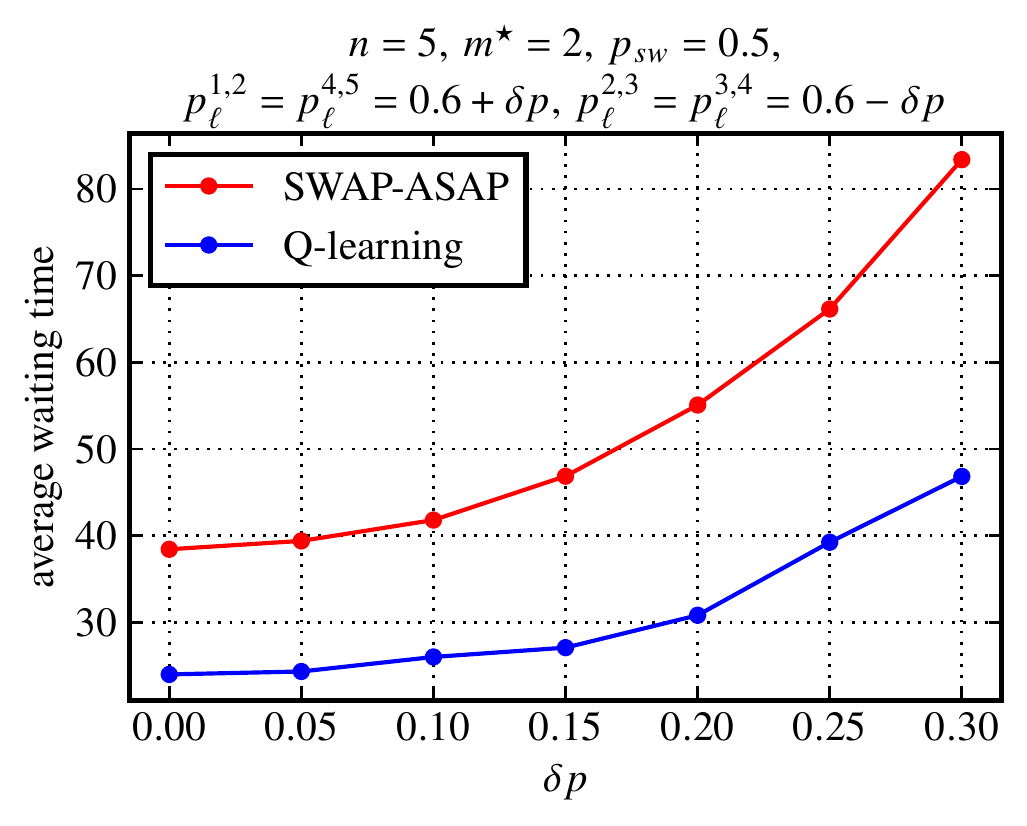}
    \caption{Average waiting time of our Q-learning policy compared to the dynamic \textsc{swap-asap} policy as a function of the inhomogeneity $\delta p$ in the elementary link success probabilities for a five-node inhomogeneous repeater chain. The repeater chain has longer central elementary links ($p_{\ell}^{2,3} = p_{\ell}^{3,4} = 0.6-\delta p$) and shorter outer elementary links ($p_{\ell}^{1,2} = p_{\ell}^{4,5} = 0.6+\delta p$), where $p_{\ell}^{i,j}$ is the elementary link success probability for nodes $i$ and $j$. Such a repeater chain is a toy model for a multi-layered quantum network in which small-scale networks between nearby nodes are connected via long links to a central node.}
    \label{fig:dumbell_comparison}
\end{figure}

We now consider inhomogeneous quantum repeater chains, in which the elementary link success probabilities are different. We provide the following two sets of results in this setting.
\begin{enumerate}
    \item A four-node inhomogeneous repeater chain with all elementary links having different success probabilities $p_{\ell}^{i,j}$, where $p_{\ell}^{i,j}$ is the elementary link success probability for pair $(i,j)$ of adjacent nodes; see Fig.~\ref{fig:asymmetry_comparison}. We fix the central elementary link (between nodes 2 and 3) to have $p_{\ell}^{2,3} = 0.6$, and we vary the other elementary link success probabilities $p_{\ell}^{1,2}$ and $p_{\ell}^{3,4}$. We define the inhomogeneity of the chain as $\delta p = p_{\ell}^{2,3} - p_{\ell}^{1,2} = p_{\ell}^{3,4} - p_{\ell}^{2,3}$. In Fig.~\ref{fig:asymmetry_comparison}, we show that our Q-learning policies can improve the average waiting times compared to the dynamic \textsc{swap-asap} policy. The dependence of the improvement on $\delta p$ does not have a monotonic trend, but overall, more advantage can be gained in the more asymmetric case. The improvement factor varies between 2\% and 8\% in this specific case.
    
    \item Next, we consider a five-node inhomogeneous repeater chain that includes central links (elementary links emanating from the central node of the chain) that have low success probability, and outer elementary links that have high success probability; see Fig.~\ref{fig:dumbell_comparison}. In particular, the first and fourth elementary links have high success probability compared to the interior and central links: $p_{\ell}^{1,2} = p_{\ell}^{4,5} = 0.6 +\delta p$ and $p_{\ell}^{2,3} = p_{\ell}^{3,4} = 0.6 - \delta p$, where as before $\delta p$ quantifies the inhomogeneity of the chain. This configuration mimics a quantum network built out of short (high $p_{\ell}$) local elementary links amongst client nodes and long (low $p_{\ell}$) elementary links to a central hub that connects to several clusters of nodes or small local networks. We consider non-deterministic entanglement swapping, with $p_{sw} = 0.5$. In Fig.~\ref{fig:dumbell_comparison}, we show that even in this inhomogeneous setting our Q-learning policies improve upon the dynamic \textsc{swap-asap} policy. The average waiting time improvement factor varies between 26\% and 38\% for this case. Therefore, as in the case of homogeneous repeater chains, a greater advantage can be gained for longer chains by using our Q-learning policies.
\end{enumerate}

\section{Improved policies for the fidelity}\label{sec-opt_pol_fidelity}

We now look at another figure of merit for the performance of quantum repeater chains, namely, the average age of the end-to-end entangled state, which allows us to determine the fidelity of the end-to-end entangled state with respect to the desired maximally entangled state via \eqref{eq-elem_link_fidelity} and \eqref{eq-virtual_link_fidelity}. Fidelity-based policy optimization using MDPs was previously presented in Ref.~\cite{Khatri2021policieselementary} for elementary links (see also Ref.~\cite[Sec.~II]{Kha22}).

We present the results of using the Q-learning algorithm, with the fidelity-based reward as defined in \eqref{eq-reward_Q_learning_fidelity}. The average age of the end-to-end link is calculated by simulating the learned policy, and the average is performed over multiple runs, each run starting from the state of no active elementary links and letting the network evolve until the end-nodes are connected, just as in the case of average waiting time. We emphasize that the policies obtained in this section, using the reward in \eqref{eq-reward_Q_learning_fidelity}, will in general be different from the policies obtained in Sec.~\ref{sec:opt_pol_waiting_time} based on the waiting time reward in \eqref{eq-opt_waiting_time_reward}, because when using \eqref{eq-reward_Q_learning_fidelity} as the reward the agent, while learning, considers not only the total time taken to reach the terminal state (end-to-end entangled state) but also the age/fidelity of the terminal state, once it is reached. This allows us to explore the trade-off between the waiting time and the fidelity, namely, that links that are stored longer will decrease the waiting time for an end-to-end link, at the cost of increasing the average age of the end-to-end entangled state (decreasing its fidelity); conversely, storing the links for less time will decrease the average age (increase the fidelity), but it will increase the waiting time.

\begin{figure}
    \centering
    \includegraphics[width=0.48\columnwidth]{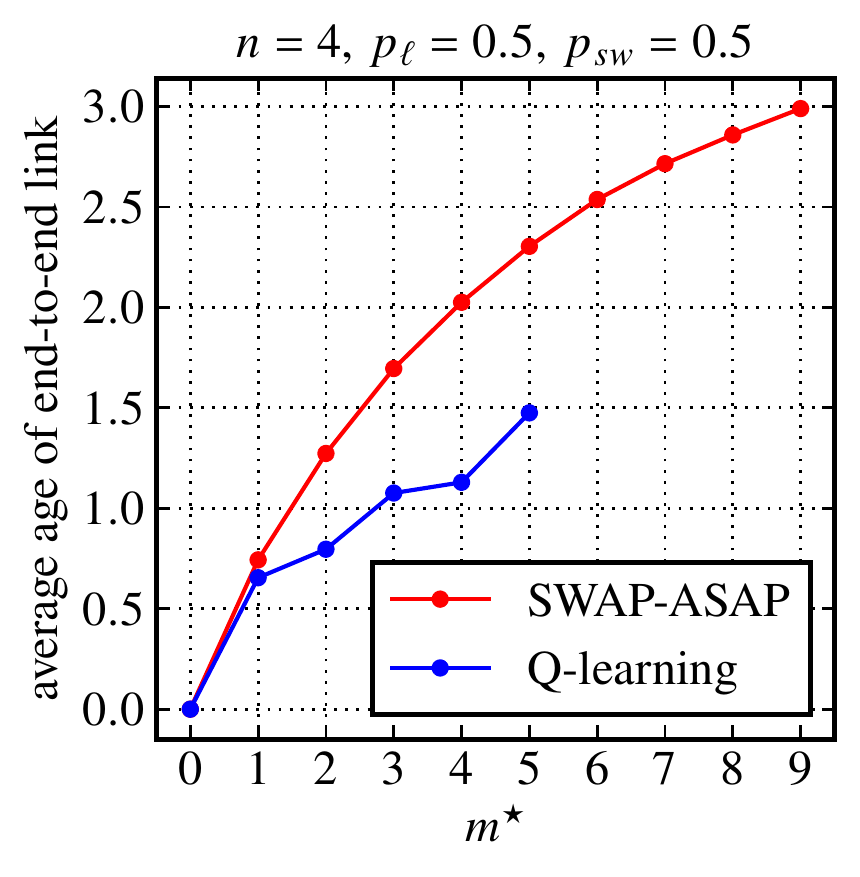}
    \includegraphics[width=0.48\columnwidth]{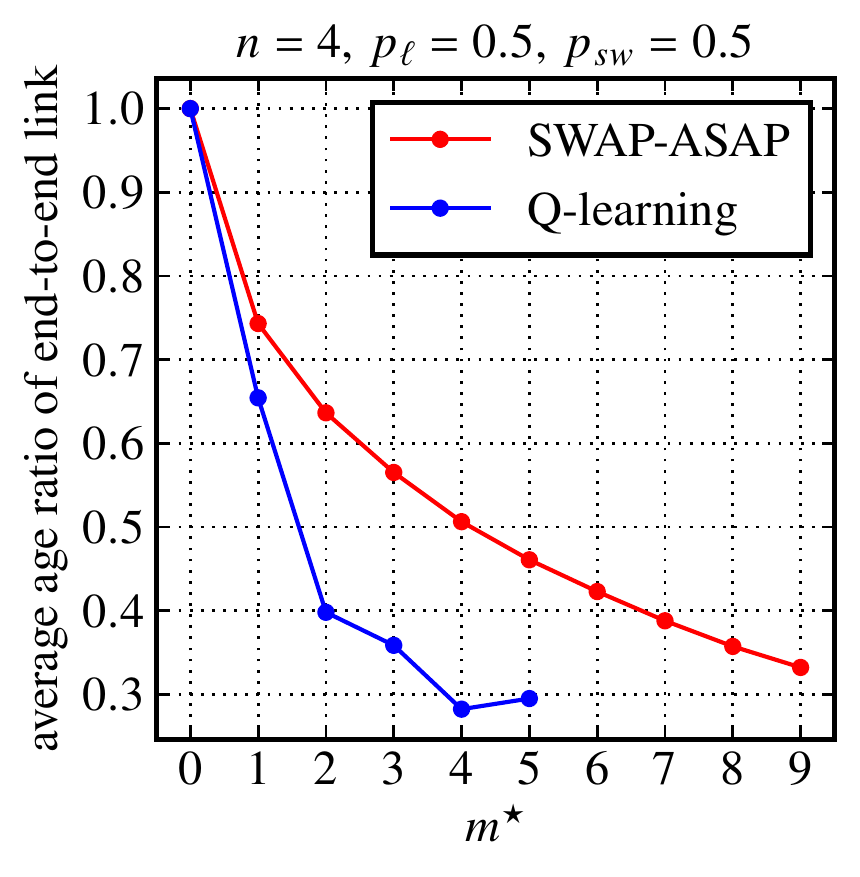}
    \caption{Average age of the end-to-end entangled state (left) and its ratio with respect to the maximum cutoff $m^{\star}$ (right) as a function of $m^{\star}$ for a four-node repeater chain, elementary link success probability $p_{\ell}=0.5$, and entanglement swapping success probability $p_{sw}=0.5$.}
    \label{fig:avg_age}
\end{figure}

\begin{figure}
    \centering
    \includegraphics[width=0.48\columnwidth]{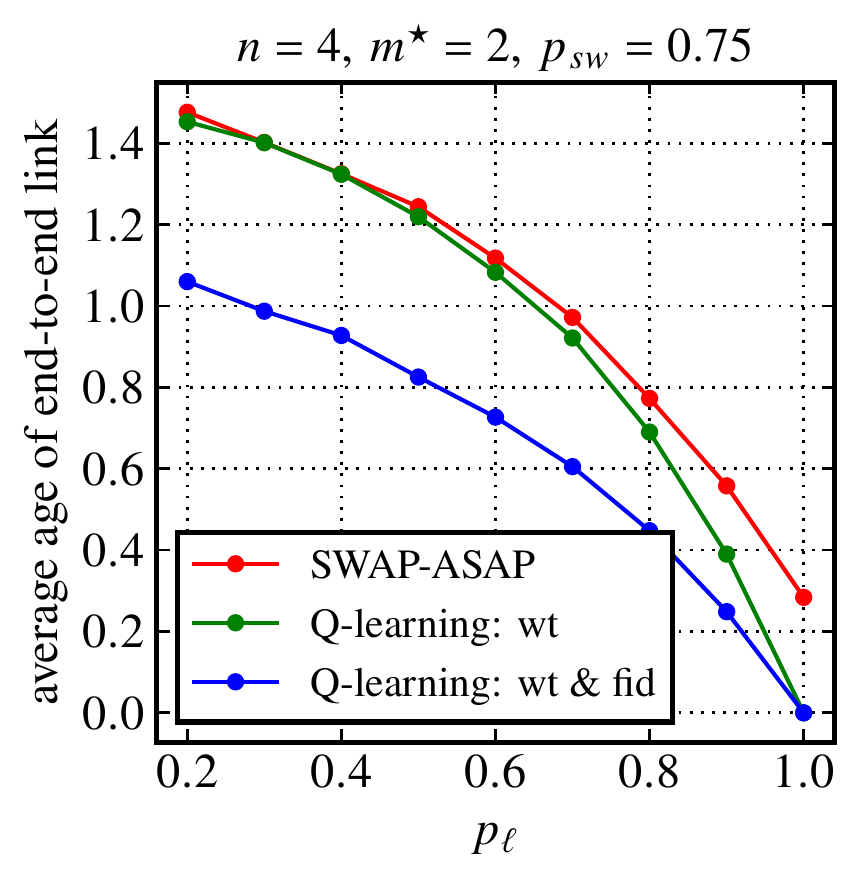}
    \includegraphics[width=0.48\columnwidth]{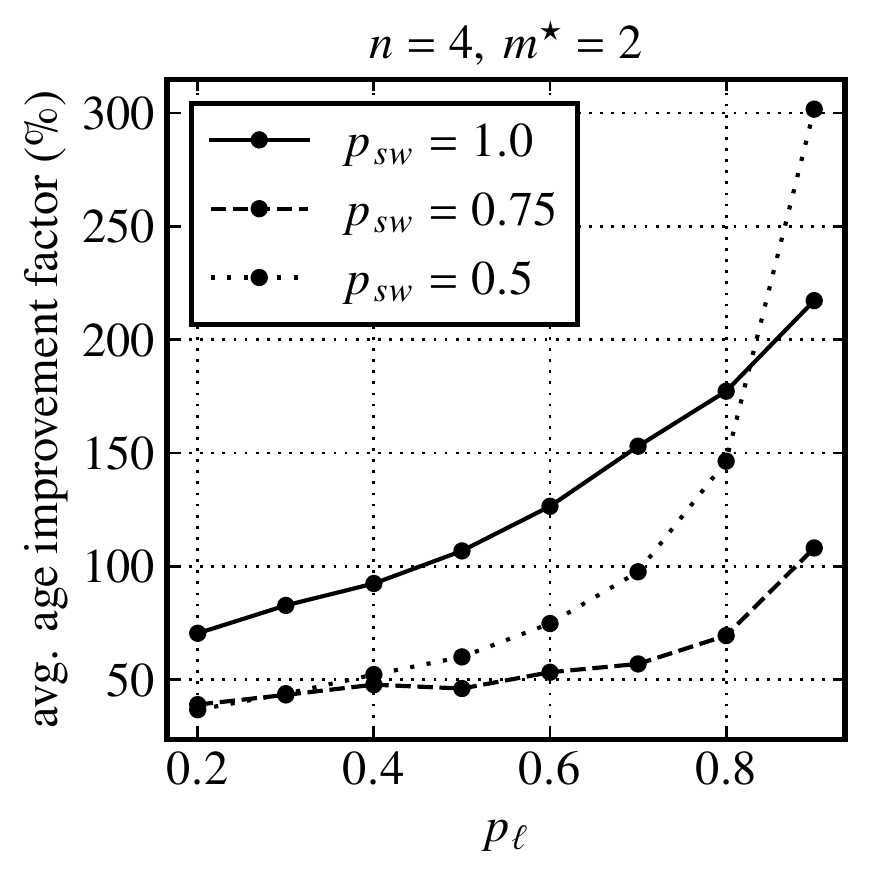}
    \caption{Average age of the end-to-end entangled state as a function of the elementary link success probability $p_{\ell}$ for a four-node repeater chain with maximum cutoff $m^{\star}=2$ and various entanglement swapping success probabilities $p_{sw}$. (Left) Comparison between the dynamic \textsc{swap-asap} policy, the policy based on the waiting time reward in \eqref{eq-opt_waiting_time_reward}, as presented in Sec.~\ref{sec:opt_pol_waiting_time} (``Q-learning: wt''), and the policy based on the waiting time and fidelity based reward in \eqref{eq-reward_Q_learning_fidelity} (``Q-learning: wt \& fid''). Here, $p_{sw}=0.75$. (Right) Average age improvement factor, as a function of $p_{\ell}$ and different choices of $p_{sw}$, for the Q-learning policy based on the reward in \eqref{eq-reward_Q_learning_fidelity}, when compared to the dynamic \textsc{swap-asap} policy. Here, unlike the average waiting time, improvement increases with increasing $p_\ell$.}
    \label{fig:avg_age_improvement}
\end{figure}

\begin{figure}
    \centering
    \includegraphics[width=0.48\columnwidth]{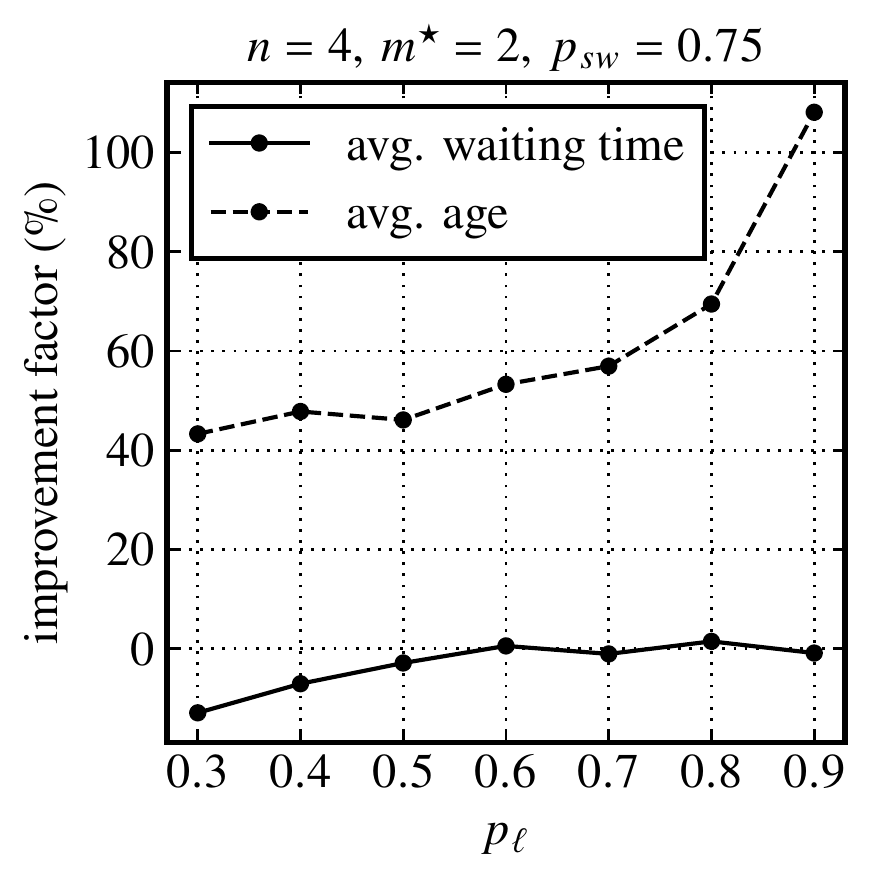}
    \includegraphics[width=0.48\columnwidth]{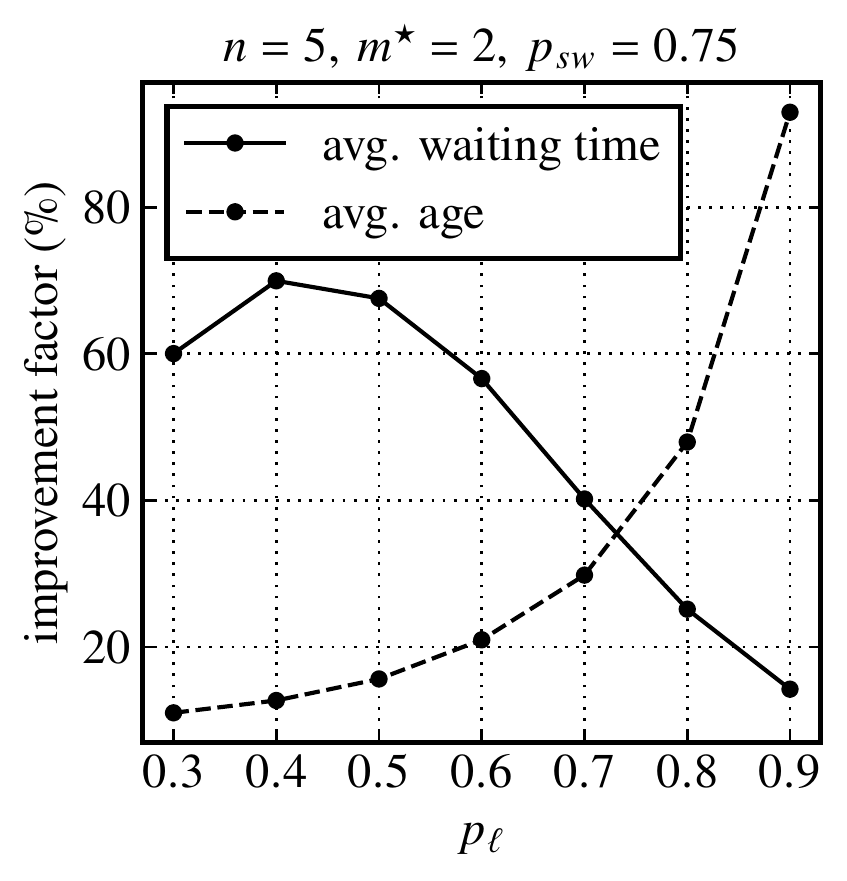}
    \caption{Improvement factor for the average waiting time and the average age of the end-to-end link when our Q-learning policy is based on the fidelity-based reward defined in \eqref{eq-reward_Q_learning_fidelity}. (Left) Four-node homogeneous repeater chain. Here we see that the decrease in the average age for our Q-learning policy comes at the expense of an increase in average waiting time (negative improvement factor) compared to the \textsc{swap-asap} policy. (Right) Five-node homogeneous chain. Interestingly, here, unlike the four-node case, a positive improvement can be obtained simultaneously for both quantities. We still see a trade-off between the average waiting time and the average age improvement. In fact, the advantage in the low-probability regime is on average much higher for the five-node case compared to the four-node case, indicating again that} our Q-learning policies become more advantageous for larger chains.
    \label{fig:trade-off}
\end{figure}

\begin{figure}
    \centering
    \includegraphics[width=0.70\columnwidth]{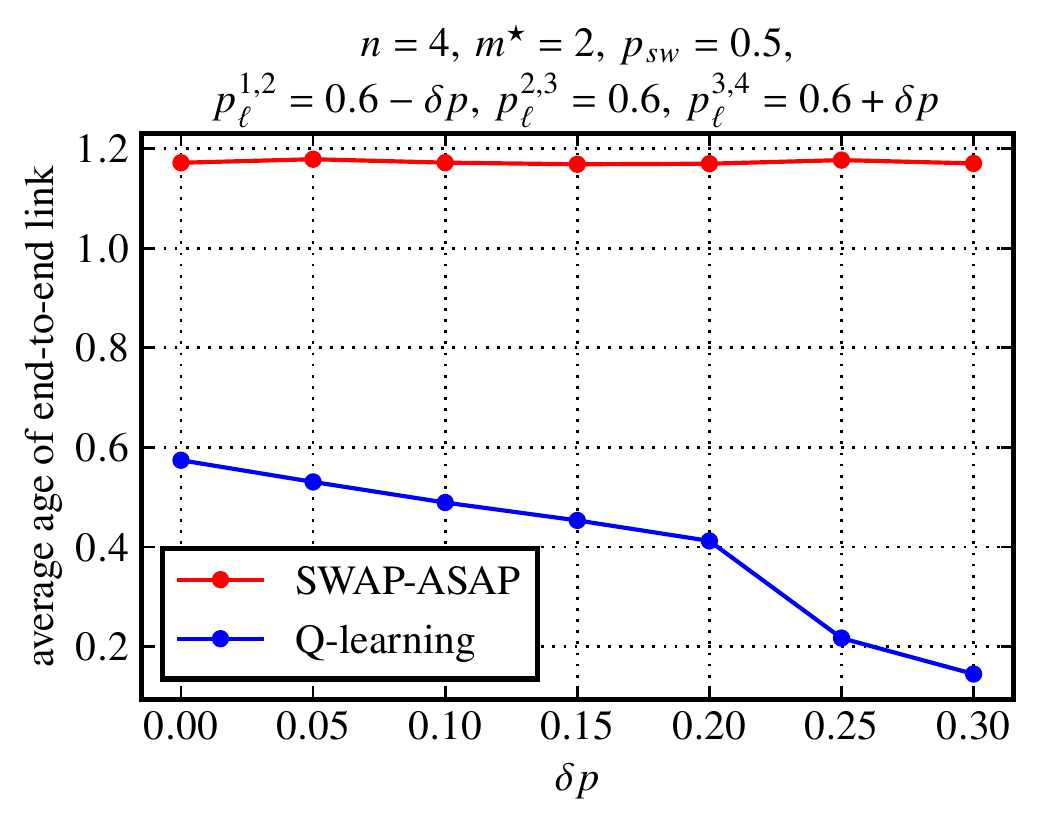}
    \caption{Average age of the end-to-end link for a four-node inhomogeneous repeater chain as shown in Fig.~\ref{fig:asymmetry_comparison}~(top).
    }
    \label{fig:trade-off_asymmetry}
\end{figure}

Our results are shown in Fig.~\ref{fig:avg_age}, Fig.~\ref{fig:avg_age_improvement}, Fig.~\ref{fig:trade-off}, and Fig.~\ref{fig:trade-off_asymmetry}. Similar to the case of waiting time, we compare our Q-learning policies to the fixed and dynamic \textsc{swap-asap} policies.

\begin{enumerate}
\item In homogeneous repeater chains, we see in Fig.~\ref{fig:avg_age}~(left) that as the maximum cutoff $m^{\star}$ increases, so too does the average age of the end-to-end link, for both our Q-learning policy and the \textsc{swap-asap} policy; however, the average age remains lower for the Q-learning policy. In particular, as shown in Fig.~\ref{fig:avg_age}~(right), the ratio between the average age and $m^{\star}$ falls faster for our Q-learning policy than for the \textsc{swap-asap} policy. As we show in Sec.~\ref{sec-practical}, the ratio is the key quantity that determines the fidelity of the end-to-end link.

\item We also find that the average age of the end-to-end entangled state falls with increasing elementary link success probability $p_{\ell}$ and increasing swapping probability $p_{sw}$, as expected. The Q-learning policy reduces the average age of the end-to-end entangled state when compared to both the \textsc{swap-asap} policy and when compared to Q-learning policies based on the waiting time reward in \eqref{eq-opt_waiting_time_reward}; see Fig.~\ref{fig:avg_age_improvement}~(left). Also, in Fig.~\ref{fig:avg_age_improvement}~(left), it is notable that with $p_{\ell}=1$, our Q-learning policy can obtain a perfect fidelity for the end-to-end-state (average age equal to zero), while with the \textsc{swap-asap} policy this is not possible. We emphasize that this improvement over \textsc{swap-asap} is true for \textit{both} the waiting time-based Q-learning policy and the fidelity-based Q-learning policy. In particular, even though the waiting time-based Q-learning policy and the \textsc{swap-asap} policy achieve the same average waiting time at $p_{\ell}=1$, as demonstrated in Fig.~\ref{fig:t_star_benchmark}~(top right), they do not achieve the same average age for the end-to-end link, meaning that the two policies must differ. We elucidate and elaborate on some of these differences in Sec.~\ref{sec-swap_asap_vs_Q_learning} below.

Furthermore, the improvement factor of the average age (compared to the dynamic \textsc{swap-asap} policy), calculated analogously to the waiting time improvement factor, decreases with increasing $p_{\ell}$ for deterministic entanglement swapping, whereas it increases with increasing $p_{\ell}$ for non-deterministic entanglement swapping; see Fig.~\ref{fig:avg_age_improvement}~(right). It is interesting to see that this trend is different from what we obtain when considering the waiting time reward (see Fig.~\ref{fig:t_star_benchmark}). In particular, as shown in Fig.~\ref{fig:trade-off}, even though there exists a trade-off between the average waiting time and the average age of the end-to-end link, as described at the beginning of this section, both can be simultaneously improved (in certain parameter regimes) when the Q-learning agent is trained according to the fidelity-based reward in \eqref{eq-reward_Q_learning_fidelity}. 

\item Fig.~\ref{fig:trade-off_asymmetry} shows that for inhomogeneous repeater chains, a similar improvement in average age as in homogeneous repeater chains can be obtained from our Q-learning policies. In this case, for some parameter regimes, the average age is enhanced at the expense of larger waiting time (negative improvement), unlike the homogeneous case (Fig.~\ref{fig:trade-off}) in which both the average age and waiting time could be simultaneously improved. %While the waiting time improvement increases with increasing inhomogeneity, the fidelity improvements reduce. This result reinforces the trade-off between maximizing fidelity and minimizing waiting time. 
\end{enumerate}
In Appendix~\ref{sec-opt_pol_example}, we provide an explicit example of a Q-learning policy obtained from our fidelity-based reward in \eqref{eq-reward_Q_learning_fidelity}, providing the action for every possible state. We compare this policy to the dynamic \textsc{swap-asap} policy and a Q-learning policy obtained from the waiting time reward in \eqref{eq-opt_waiting_time_reward}.

\subsection{Examples}\label{sec-practical}

Let us now demonstrate how our Q-learning policies from this section, based on the reward in \eqref{eq-reward_Q_learning_fidelity} and formulated abstractly in terms of the age of the end-to-end link, can be used to generate fidelity thresholds for the end-to-end link for given noise models. The primary purpose here is to demonstrate how our abstract theory can be used in practice to analyze the fidelity improvement of our Q-learning policies compared to the \textsc{swap-asap} policy for actual quantum repeater chains.

\paragraph*{Noise model.} Our results can be applied to arbitrary Pauli noise models for the decoherence of the memory qubits. For the purposes of this section, we consider the following Pauli noise model:
\begin{equation}
    \mathcal{N}_{m_1^{\star},m_2^{\star}}(\rho)=p_I\rho+p_X X\rho X+p_Y Y\rho Y+p_Z Z\rho Z,
\end{equation}
where $X,Y,Z$ are the single-qubit Pauli operators, defined as
\begin{equation}\label{eq-Pauli_operators}
    X\coloneqq\begin{pmatrix} 0 & 1 \\ 1 & 0 \end{pmatrix},\quad Y\coloneqq\begin{pmatrix} 0 & -\I \\ \I & 0 \end{pmatrix},\quad Z\coloneqq\begin{pmatrix} 1 & 0 \\ 0 & -1 \end{pmatrix},
\end{equation}
and the probabilities $p_I,p_X,p_Y,p_Z$ are defined as~\cite{sarvepalli2009asymmetriccodes,ghosh2012surfacecodedecoherence}
\begin{align}
    p_I&=\frac{1+\e^{-\frac{1}{m_2^{\star}}}}{2}-\frac{1-\e^{-\frac{1}{m_1^{\star}}}}{4},\label{eq-decoherence_channel_pI}\\
    p_X&=\frac{1-\e^{-\frac{1}{m_1^{\star}}}}{4},\label{eq-decoherence_channel_pX}\\
    p_Y&=\frac{1-\e^{-\frac{1}{m_1^{\star}}}}{4},\label{eq-decoherence_channel_pY}\\
    p_Z&=\frac{1-\e^{-\frac{1}{m_2^{\star}}}}{2}-\frac{1-\e^{-\frac{1}{m_1^{\star}}}}{4}.\label{eq-decoherence_channel_pZ}
\end{align}
This channel represents the Pauli-twirled version of the concatenated amplitude damping and dephasing channels~\cite{sarvepalli2009asymmetriccodes,ghosh2012surfacecodedecoherence}. In this work, because we consider discrete time steps, the quantities $m_1^{\star},m_2^{\star}\in\{1,2,\dotsc\}$ represent the discrete analogues of the continuous $T_1$ and $T_2$ times, respectively. Specifically, the quantity $m_1^{\star}$ represents the coherence time of amplitude damping noise, and $m_2^{\star}$ represents the coherence time of dephasing noise. While the connection between $m_1^{\star}$ and the coherence time of amplitude damping noise is not exact, due to the Pauli-twirled channel that we consider, we can use it as a toy model, especially because our learned policies (and the age-update rule) are based on Pauli noise; refer back to Sec.~\ref{sec-virtual_links} for details. We note that, to the best of our knowledge, most prior works consider only dephasing noise~\cite{RPL09,rozpedek2018parameterregimesrepeater,rozpedek2019neartermrepeaterNV,reiss2022deep,kamin2022exactrateswapasap}, while our model is suited to handle arbitrary Pauli noise models, with the specific noise model presented here being only a particular example.

Let us also define the two-qubit Bell states as follows:
\begin{align}
    \ket{\Phi^{\pm}}&\coloneqq\frac{1}{\sqrt{2}}(\ket{0,0}\pm\ket{1,1}),\quad \Phi^{\pm}\coloneqq\ketbra{\Phi^{\pm}}{\Phi^{\pm}},\label{eq-Bell_states_1}\\
    \ket{\Psi^{\pm}}&\coloneqq\frac{1}{\sqrt{2}}(\ket{0,1}\pm\ket{1,0}),\quad \Psi^{\pm}\coloneqq\ketbra{\Psi^{\pm}}{\Psi^{\pm}}.\label{eq-Bell_states_2}
\end{align}

\paragraph*{Decoherence of an entangled qubit pair.} Now, let us suppose that the initial state of an entangled qubit pair is the perfect maximally-entangled Bell state $\Phi^+$. We do this for simplicity and illustrative purposes. The following development can be straightforwardly generalized to arbitrary initial entangled qubit pairs whose states have non-unit fidelity; see, e.g., Ref.~\cite[Appendix~C]{Kha22}.

After $m\in\{1,2,3,\dotsc\}$ time steps, it is straightforward to show that the decohered entangled state, defined as in \eqref{eq-elem_link_state}, is equal to
\begin{align}
    &(\mathcal{N}_{m_1^{\star},m_2^{\star}}\otimes\mathcal{N}_{m_1^{\star},m_2^{\star}})^{\circ m}(\Phi^+)\nonumber\\
    &\quad=\frac{1}{4}(1+\e^{-2m/m_1^{\star}}+2\e^{-2m/m_2^{\star}})\Phi^+\nonumber\\
    &\qquad +\frac{1}{4}(1+\e^{-2m/m_1^{\star}}-2\e^{-2m/m_2^{\star}})\Phi^-\nonumber\\
    &\qquad +\frac{1}{4}(1-\e^{-2m/m_1^{\star}})\Psi^+\nonumber\\
    &\qquad +\frac{1}{4}(1-\e^{-2m/m_1^{\star}})\Psi^-. \label{eq-decohered_entangled_state}
\end{align}
We provide a proof of this in Appendix~\ref{sec-decohered_entangled_state_pf}. This implies that the fidelity of the state after $m$ time steps is $\frac{1}{4}(1+\e^{-2m/m_1^{\star}}+2\e^{-2m/m_2^{\star}})$.

Let us now connect the noise model with our MDP model of entanglement distribution, so that we can apply this noise model to the results presented in the previous sections. For a given value of $m^{\star}$, as used in our MDP model, let us take
\begin{equation}\label{eq-m1_star_m2_star_assumptions}
    m_1^{\star}=\infty,\quad m_2^{\star}=5m^{\star}.
\end{equation}
In other words, we consider dephasing only for the examples here, in order to make connections to prior works~\cite{RPL09,rozpedek2018parameterregimesrepeater,rozpedek2019neartermrepeaterNV,reiss2022deep,kamin2022exactrateswapasap} and because dephasing is often the dominant source of noise; see, e.g., Ref.~\cite{ourari2023telecomphotonsErion}. The fidelity after $m$ time steps, given by the function defined in \eqref{eq-elem_link_fidelity}, is then equal to
\begin{equation}\label{eq-fidelity_function}
    f(m)=\frac{1}{2}(1+\e^{-\frac{2m}{5m^{\star}}}).
\end{equation}
Note that we have chosen the value of $m_2^{\star}$ in \eqref{eq-m1_star_m2_star_assumptions} such that, at the cutoff time $m^{\star}$, the fidelity of the entangled qubit pair is $f(m^{\star})=\frac{1}{2}(1+\e^{-2/5})\approx 0.8351$. We emphasize that our results can be applied to other choices of $m_1^{\star}$ and $m_2^{\star}$---in particular, choices that could take amplitude damping into account---in order to obtain a more refined performance analysis.

We can invert the fidelity function in \eqref{eq-fidelity_function}, such that for a given fidelity $F$, the corresponding age of the qubits is given by
\begin{equation}\label{eq-fidelity_to_age}
    f^{-1}(F)=\left\lceil-\frac{5m^{\star}}{2}\log(2F-1)\right\rceil,
\end{equation}
for all $F\in(f(m^{\star}),1)$, where we take the ceiling $\ceil{\cdot}$ because we want an integer for the age.

We also remark that the decohered entangled state in \eqref{eq-decohered_entangled_state} is a Bell-diagonal state of the form
\begin{equation}
    (a+b)\Phi^++(a-b)\Phi^-+c\Psi^++c\Psi^-.
\end{equation}
As such, its entanglement can be characterized entirely by the quantity $a+b$, namely, the fidelity of the state with respect to $\Phi^+$. In particular, the state is entangled if and only if this fidelity is greater than or equal to $\frac{1}{2}$; see, e.g.,~\cite[Chapter~2]{Kha21}. Therefore, under the assumptions in \eqref{eq-m1_star_m2_star_assumptions}, such that the fidelity is given by \eqref{eq-fidelity_function}, we find that the state in \eqref{eq-decohered_entangled_state} is entangled for all time steps $m\in\{0,1,2,\dotsc,m^{\star}\}$.

\subsubsection{Fidelity thresholds}

\begin{figure}
        \centering
        \includegraphics[width=0.7\columnwidth]{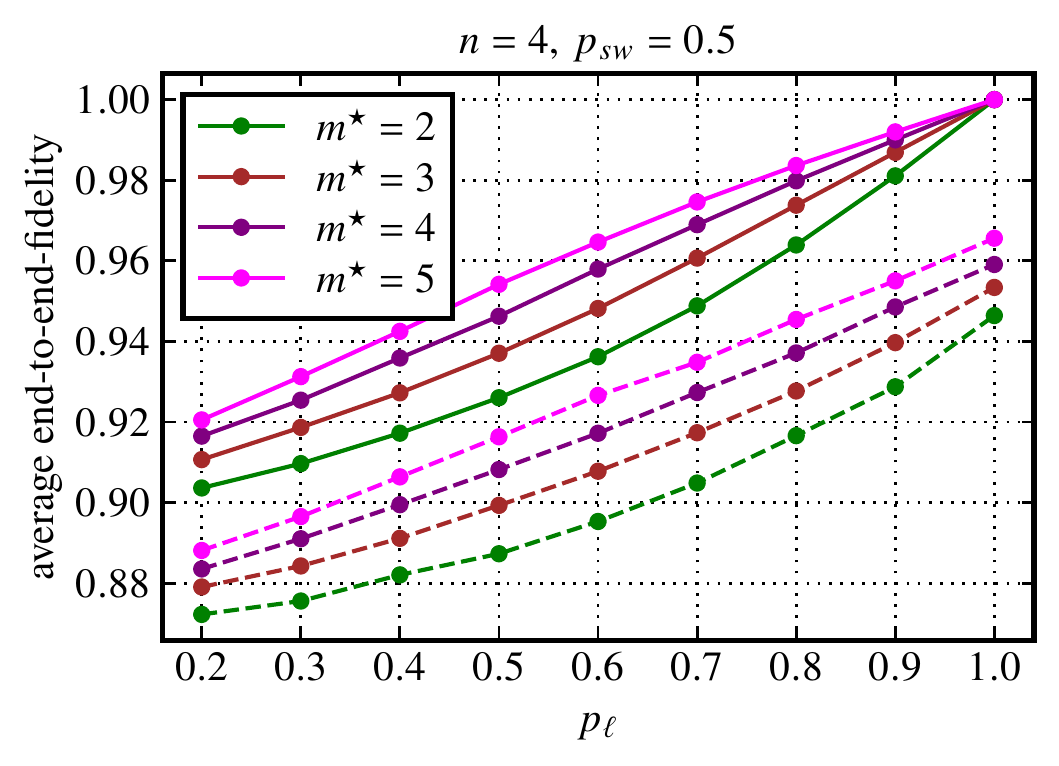}
        \caption{Average fidelities of the end-to-end link for a four-node repeater chain with entanglement swapping success probability $p_{sw}=0.5$ and various values of the maximum cutoff time $m^{\star}$. The solid line is the Q-learning policy, and the dashed line is the \textsc{swap-asap} policy. We make use of the abstract results on the average age of the end-to-end link from the beginning of Sec.~\ref{sec-opt_pol_fidelity} (see, e.g., Fig.~\ref{fig:avg_age_improvement}), and then use the formula in \eqref{eq-fidelity_function} for the dephasing noise model to obtain the fidelity from the age.}
        \label{fig:fidelity_dephasing}
    \end{figure}

    Our Q-learning policies give us lower bounds on the maximum end-to-end fidelity that can be achieved in a repeater chain with a given number $n$ of nodes and a given value $m^{\star}$ for the maximum cutoff of the links. For the dephasing model presented above, we use the formula in \eqref{eq-fidelity_function} in order to obtain the fidelity from the age of the end-to-end link; the results are shown in Fig.~\ref{fig:fidelity_dephasing}. As already pointed out earlier in this section, we can see that with our Q-learning policies we can achieve perfect fidelity for $p_{\ell}=1$, while with the \textsc{swap-asap} policy this is not possible.

    \begin{figure}
        \centering
        \includegraphics[width=0.7\columnwidth]{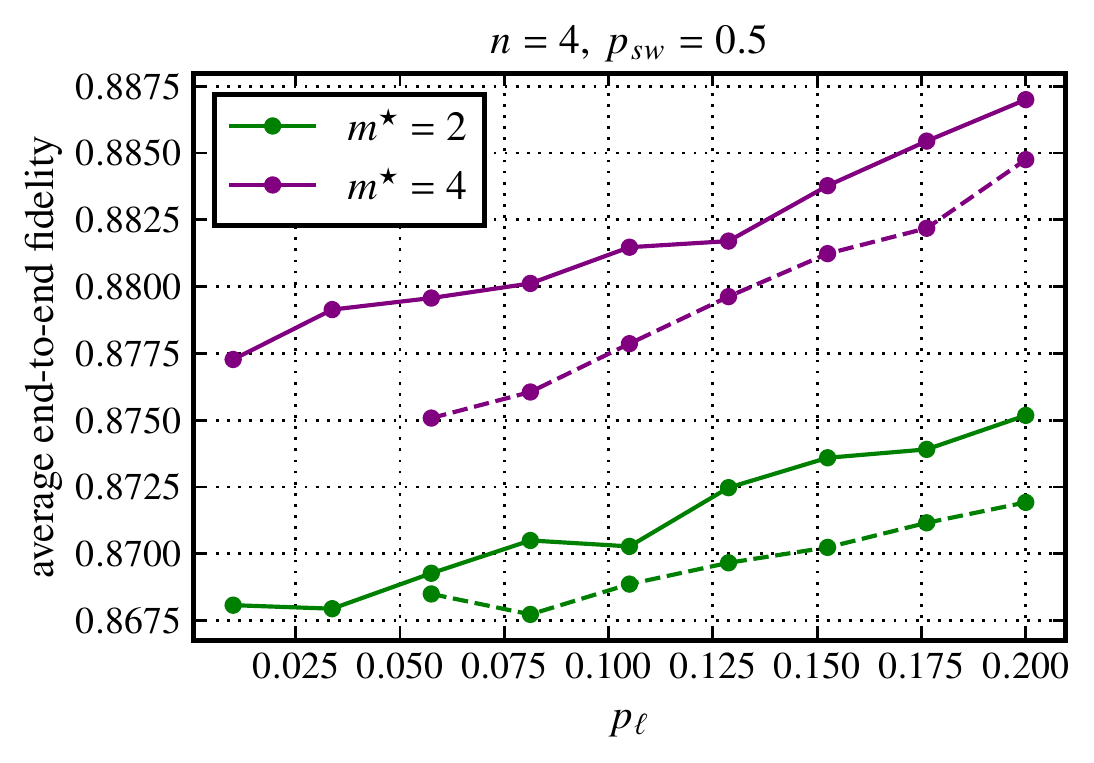}
        \caption{Average fidelities of the end-to-end link for a four-node repeater chain with low values of the elementary link success probability $p_{\ell}$, entanglement swapping success probability $p_{sw}=0.5$, and different values of the maximum cutoff time $m^{\star}$. The solid line is the Q-learning policy, obtained via training at $p_{\ell}=0.3$, and the dashed line is the \textsc{swap-asap} policy.}
        \label{fig:results_low_pl}
    \end{figure}

\subsubsection{Performance for low success probabilities}

    In practice, the elementary link success probability $p_{\ell}$ can be quite low---in fact, considerably lower than the values we have considered so far in this work. However, obtaining good policies via reinforcement learning for quantum repeater chains with low values of $p_{\ell}$ is generally challenging, requiring many episodes of training. On the other hand, it is of course possible to use the policies obtained through training quantum repeater chains with larger values of $p_{\ell}$ and apply them to repeater chains with lower values of $p_{\ell}$. Interestingly, as we show in Fig.~\ref{fig:results_low_pl}, applying these policies still allows us to obtain an advantage over the \textsc{swap-asap} policy. These results demonstrate the utility of our methods even in the practically relevant setting of low $p_{\ell}$. As an interesting direction for future work, in order to obtain even better fidelity thresholds for low values of $p_{\ell}$, one can train explicitly for these values of $p_{\ell}$ by using the policy for a large value of $p_{\ell}$ as a ``warm start'', in order to potentially make the training easier.

\begin{figure*}
    \centering
    \includegraphics[width=0.90\textwidth]{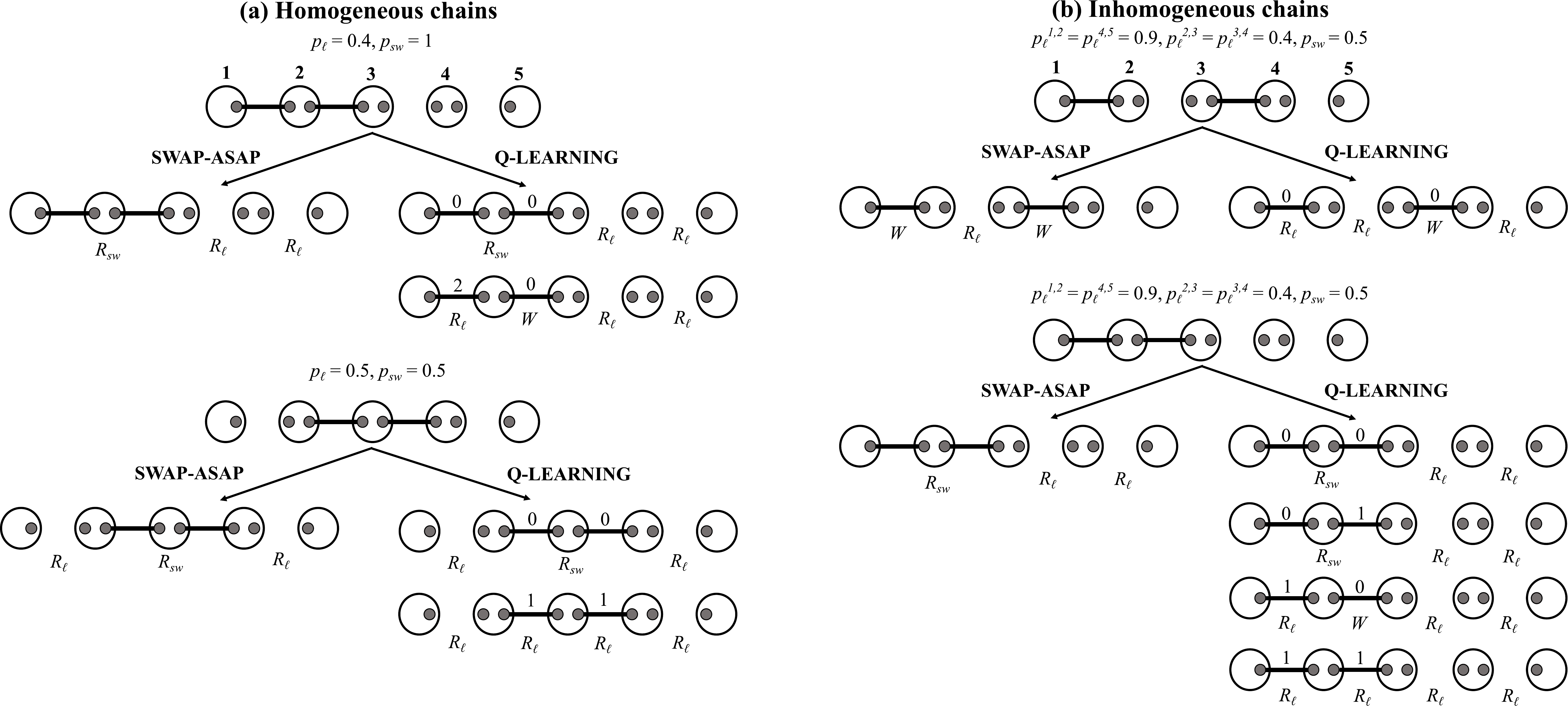}
    \caption{Some examples of differences in the dynamic \textsc{swap-asap} policy and improved Q-learning policies obtained using waiting time-based rewards. We consider $m^\star = 2$ for these illustrations. The ages of active links are written on top of the link. (a)~Homogeneous chains. (Top)~Deterministic swaps when links 1 and 2 are active. The dynamic \textsc{swap-asap} policy (left) always requests swaps for active adjacent links, whereas the Q-learning policy determines the action based on the ages of links and also the status of links in other parts of the chain. (Bottom) Non-deterministic swaps. Here, the Q-learning policy sometimes chooses to discard active links, even before they have reached the maximum cutoff time, in order to provide more chances for elementary link generation, if the chain is still far away from the terminal state. This \emph{dynamic} cutoff becomes even more pronounced for inhomogeneous chains. (b)~Inhomogeneous chains. (Top) The Q-learning policy chooses to discard the link between nodes 1 and 2 before it reaches the maximum cutoff $m^{\star}$, because it has a success probability of $p_{\ell}^{1,2}=0.9$ and is therefore easy to produce. (Bottom) Similarly, the Q-learning policy can sometimes delay swapping depending on the relative ages of the involved links in order to preserve elementary links that are more difficult to produce.}
    \label{fig:swap-asap_vs_optimal}
\end{figure*}

\section{Discussion of results}\label{sec-discussion}

In this work, we have considered entanglement distribution in a linear chain of quantum repeaters in the setting of non-deterministic elementary link creation and entanglement swapping. We have done so by modeling the repeater chain using a Markov decision process. Then, using the Q-learning reinforcement learning algorithm, we have obtained policies that improve upon the \textsc{swap-asap} policy, both in terms of the average waiting time and the average age of the end-to-end link, in certain parameter regimes. In addition, by using the Q-learning algorithm, we can explicitly identify the best action for every state, and we provide an example of this in Appendix~\ref{sec-opt_pol_example}. In this section, we provide some details on the high-level insights about our results. In Sec.~\ref{sec-swap_asap_vs_Q_learning}, we identify some of the key differences between our Q-learning policies and the fixed and dynamic \textsc{swap-asap} policies, which lead to the improvements in waiting time and fidelity shown in Sec.~\ref{sec:opt_pol_waiting_time} and Sec.~\ref{sec-opt_pol_fidelity}. Then, in Sec.~\ref{sec-compare_prior_work}, we outline the differences between our work and prior related work on reinforcement learning for entanglement distribution in quantum repeater chains.

\subsection{Differences between the \textsc{swap-asap} and Q-learning policies}\label{sec-swap_asap_vs_Q_learning}

We outline below of the some key differences that we observed between our Q-learning policies and the \textsc{swap-asap} policies. These differences are also illustrated in Fig.~\ref{fig:swap-asap_vs_optimal}.

\begin{enumerate}
    \item For homogeneous repeater chains with deterministic entanglement swapping, it is clear that most of the advantage comes from adopting a ``global'' policy, in which all of the nodes have knowledge of each other's states. Specifically, this means that entanglement swapping is not always performed right away, but instead some links will wait until a large number of adjacent nodes are all active. (We quantitatively explore this concept of global knowledge in Sec.~\ref{sec:advantage} below.) For an illustration, see Fig.~\ref{fig:swap-asap_vs_optimal}(a)(top), where two different age configurations are shown for a five-node chain with maximum cutoff $m^{\star}=2$. If both links are fresh, a swap is attempted, but if one of the links is at the maximum cutoff, a swap is not attempted; instead, a new link is requested, because the agent knows that links 3 and 4 are still inactive and will need at least one more time step to become active, by which time the swapped link would have crossed its maximum cutoff time of $2$, according to the rule $m'= m_1 + m_2$ (see Sec.~\ref{sec:MDP_model}). Thus, the second link is saved for one more time step by avoiding an unnecessary swap attempt. More generally, waiting can be advantageous when both the elementary link success probability $p_{\ell}$ and the maximum cutoff $m^{\star}$ are low. Indeed, when $p_{\ell}$ is low, discarding a link becomes ``risky'', because the new elementary link attempt is not likely to succeed. Similarly, when $m^{\star}$ is low, then due to our update rule for the age of a virtual link (see Sec.~\ref{sec:MDP_model}), intermediate swapped states will not survive for too long if $m^{\star}$ is low. This also tells us why the advantage of our Q-learning policy is reduced for increasing elementary link success probability and increasing memory cutoff (see Fig.~\ref{fig:deterministic}). 
    
    \item For homogeneous repeater chains with non-deterministic entanglement swapping, as in Fig.~\ref{fig:swap-asap_vs_optimal}(a)(bottom), the Q-learning policy often chooses to discard links even before they have reached the maximum cutoff time, in contrast to the dynamic \textsc{swap-asap} policy. For example, in Fig.~\ref{fig:swap-asap_vs_optimal}(a)(bottom), segments (2, 3) and (3, 4) are active and (1, 2) and (4, 5) are inactive. The \textsc{swap-asap} policy would attempt a swap at node 3 and request links at (1, 2) and (4, 5). A doubling policy (see Ref.~\cite{shchukin2019waiting} for details) would not perform the swap and just request at (1, 2) and (4, 5). On the other hand, the Q-learning policy chooses several different strategies depending on the absolute and relative ages of the two links and also the values $p_{\ell}$, $p_{sw}$, $m^{\star}$, two of which are shown in the figure. 
    
    \item For inhomogeneous repeater chains, the key differences in policy occur with respect to the swapping of links with different elementary link success probabilities. This also explains why the improvement factor increases with increasing asymmetry of the chain (see that the gap between the RL and \textsc{swap-asap} curves increases with increasing $\delta p$ in Fig.~\ref{fig:trade-off_asymmetry}). For example, in Fig.~\ref{fig:swap-asap_vs_optimal}(b)(top), because the elementary link success probability between the outer nodes 1 and 2 is high, whenever a request for a link between central nodes 2 and 3 is made, the adjacent outer link is also requested even when it is already active. In contrast, in the \textsc{swap-asap} policy (either fixed or dynamic), unless the link is at the cutoff time, it would not be requested. Our Q-learning policy adopts this strategy because it helps keep the outer link fresh, so that whenever the central link is created and a successful swap occurs between it and the adjacent outer link, the age of the virtual link remains low. Otherwise, if the created virtual link has a high age, because the easy-to-produce outer link was old, the precious central link's lifetime would be effectively shortened. A similar choice of holding off swaps if the effective age of central links would increase due to swapping with an old but easy-to-produce outer link can be seen in Fig.~\ref{fig:swap-asap_vs_optimal}(b)(bottom).
\end{enumerate}

\subsection{Comparison to prior work}\label{sec-compare_prior_work}

The work presented here is most similar to Refs.~\cite{inesta2022optimal,reiss2022deep}, in which linear, homogeneous repeater chains are studied. One key difference between those works and ours is that our MDP is different from those in Refs.~\cite{inesta2022optimal,reiss2022deep}. In particular, in Ref.~\cite{inesta2022optimal}, the elementary link policy was fixed to the memory-cutoff policy, and only the entanglement swapping policy was optimized. Then, using policy/value iteration, the authors of Ref.~\cite{inesta2022optimal} obtain policies that improve the average waiting time. On the other hand, although optimization of the elementary link policy is considered in Ref.~\cite{reiss2022deep}, the authors of that work fix the swapping policy to \textsc{swap-asap}, and they also assume that the entanglement swapping is perfect and deterministic. In this work, we optimize both the policy of the elementary links as well as the policy for the entanglement swapping, and we take as our figure of merit both the average waiting time as well as the fidelity of the end-to-end entangled state. One advantage of our more general MDP is that cutoffs less than the maximum cutoff $m^{\star}$ can be chosen. This then allows implicitly for an optimization of the cutoffs simultaneously with an optimization of the entanglement swapping strategy, in contrast to prior works~\cite{li2020efficient,reiss2022deep} that perform an optimization of memory cutoffs while keeping the entanglement swapping strategy fixed.

We also emphasize that in this work we use the ``addition'' update rule $m'=m_1+m_2$ for the age of a link after entanglement swapping, while in Ref.~\cite{inesta2022optimal} the ``max'' update rule $m'=\max\{m_1,m_2\}$ is used. It is known that the addition update rule is the exact update rule for Pauli noise acting on the memory qubits (see Appendix~\ref{sec:ent_swap_Pauli_update}), while it is unclear whether the max update rule faithfully corresponds to realistic noise models. Our results in Sec.~\ref{sec-opt_pol_fidelity} therefore provide benchmarks that are meaningful for physically realistic noise models; see specifically Sec.~\ref{sec-practical} above.

Furthermore, in reality, quantum networks will in all likelihood resemble inhomogeneous repeater chains, with elementary links of different lengths, or using different physical platforms/channels, and hence the elementary links would have different success probabilities; see, e.g., Refs.~\cite{khatri2021spooky,wallnofer2022sats}. Thus, in this work we also consider examples of four- and five-node inhomogeneous repeater chains, and we find improved policies for such networks. Finally, the use of the Q-learning algorithm allows us to explicitly extract the learned policy, i.e., a list containing the best action for every possible state (see Appendix~\ref{sec-opt_pol_example} for an example), while for deep reinforcement learning techniques, as used in Ref.~\cite{reiss2022deep}, extracting the policy is generally difficult.

\section{How do improved policies derive their advantage?}\label{sec:advantage}

In this section, we quantify the various sources of advantage of our Q-learning policies, particularly the advantages that we outlined in the previous section. Broadly, there are three key differences in our improved policies when compared to a fixed-cutoff local \textsc{swap-asap} policy, which are: different cutoffs for the elementary links, global knowledge and coordination between non-adjacent nodes, and decision making with foresight.

\subsection{Non-uniform and state-dependent cutoffs}

By analyzing the policies obtained using the Q-learning algorithm, we find that these improved policies have a dynamic and global-state-dependent cutoff for the different elementary links of the network. These cutoffs are, in general, less than the maximum cutoff $m^{\star}$, which is allowed within our MDP framework; see Sec.~\ref{sec:MDP_model} and Appendix~\ref{sec-MDP_details}.

In order to see that the cutoffs are not only state dependent but that different links can have different cutoffs, we calculate an average cutoff $\langle t^\star \rangle_{i,j}$ for every node pair $(i,j)$. We do this by averaging the age of a link over instances when it is discarded, i.e., when a link request $R_\ell$ is performed on that link whenever it is active. Results for four-node and five-node repeater chains are shown in Table~\ref{tab:dynamic_cut-off}. We find that, even for homogeneous repeater chains, interior links (closer to the center of the chain) have a higher average cutoff than outer links (closer to the ends of the chain). It appears that the Q-learning agent views the interior links as more ``precious'' than the outer links. Intuitively, this could be because interior links have the ability to become joined with two links, one on each side, which is not the case for outer links. For inhomogeneous repeater chains, the effect of the difference in the elementary link success probabilities on the choice of the cutoffs can also be seen by looking at the quantities $\langle t^\star \rangle_{i,j}$. For a fixed $m^\star$, the $t^\star$ choice made by the Q-learning agent is on average smaller for elementary links with higher probability, and vice versa.

The above heuristic can in fact be utilized in a practical setting in the following way. Although our optimized policies choose cutoffs for different links in a dynamic, state-dependent way, it is in general advantageous to choose higher cutoffs for low-probability links and low cutoffs for high-probability links. We can see this just by choosing different but still fixed cutoffs at the elementary link level in the \emph{local \textsc{swap-asap} policy}. Indeed, waiting times can be reduced and average ages improved, which might seem counter-intuitive to some extent. However, consider for example a four-link chain with elementary link probabilities [0.9, 0.3, 0.3, 0.9]. If we choose a fixed cut-off $m^\star=2$ for all of the links, then we get an average waiting time of $51.9 \pm 1.2$ and an end-to-end link age of $3.17 \pm 0.02$. On the other hand, by reducing the cutoff at links $1$ and $4$ to $m^\star=0$, we get a lower waiting time of $43.8 \pm 0.6$ ($\approx 25\%$ decrease) and an average age of $2.2 \pm 0.02$ ($\approx 33\%$ decrease). This is a clear example of how the qualitative and quantitative trends that we have identified in our global, optimized policies can be used to modify and improve even local, sub-optimal policies such as the \textsc{swap-asap} policy, without adding very many experimental/engineering challenges.

\begin{table}
\renewcommand{\arraystretch}{1.2}
\begin{center}
\begin{tabular}{|p{1.5cm}|p{1.5cm}|p{1.5cm}|p{1.5cm}|}
\hline
\multicolumn{4}{|c|}{$n=4, m^\star=2, p_{sw}=0.5$} \\
\hline\hline
  & link 1 & link 2 & link 3\\
\hline
$p_{\ell}$  & 0.6 & 0.6 & 0.6 \\
$\langle t^\star \rangle_{ij}$ & 1.3 & 2.0 & 1.4\\
\hline
$p_{\ell}$  & 0.3 & 0.6 & 0.9\\
$\langle t^\star \rangle_{i,j}$ & 2.0 & 1.0 & 0.15 \\
\hline
\end{tabular}\\[0.4cm]
\begin{tabular}{|p{1.5cm}|p{1.5cm}|p{1.5cm}|p{1.5cm}|p{1.5cm}|}
\hline
\multicolumn{5}{|c|}{$n=5, m^\star=2, p_{sw}=0.5$} \\
\hline\hline
  & link 1 & link 2 & link 3 & link 4\\
\hline
$p_{\ell}$  & 0.6 & 0.6 & 0.6 & 0.6\\
$\langle t^\star \rangle_{i,j}$ & 0.3 & 0.9 & 1.0 & 0.2\\
\hline
$p_{\ell}$  & 0.9 & 0.3 & 0.3 & 0.9\\
$\langle t^\star \rangle_{i,j}$ & 0.6 & 2.0 & 2.0 & 0.4\\
\hline
\end{tabular}
\caption{Average values $\langle t^\star \rangle_{i,j}$ of the cutoffs for the links between nodes pairs $(i,j)$ in homogeneous and inhomogeneous repeater chains. In general, interior links are retained for longer compared to outer links, and low probability links are retained for longer than high probability links. In contrast, for a fixed memory cutoff policy, such as \textsc{swap-asap}, the maximum possible cutoff $m^\star$ is chosen for all links irrespective of its position in the chain or its elementary link success probability.}
\label{tab:dynamic_cut-off}
\end{center}
\end{table}

\subsection{Global knowledge and coordination between nodes}

Next, we come to global knowledge and the coordination between non-adjacent nodes. It was noted in Sec.~\ref{subsec:swap_asap} that \textsc{swap-asap} policies use local knowledge of the network to decide on what actions to take. Therefore, only adjacent nodes can coordinate on elementary link requests, waiting, and entanglement swapping requests. On the other hand, in the Q-learning framework, the agent is all-knowing: it can decide on actions based on complete global knowledge of the state of the network. This is one of the main reasons for the improvement over \textsc{swap-asap}, as we demonstrate quantitatively in this section. It is worth mentioning that this source of improvement is already present even within the framework of \textsc{swap-asap}: the fixed version, which is completely local and does not allow for any collaboration, even among adjacent nodes, performs worse than the dynamic version, which allows for some collaboration/flexibility in terms of the decision to do Bell measurements and the order in which swaps are performed.

The extent of global knowledge used in our Q-learning policies can be quantified using correlation functions between actions and states at different links and nodes of the network. Specifically, we use the absolute value of the Pearson correlation coefficient. For two random variables $X$ and $Y$, it is defined as follows:
\begin{equation}
    r[X,Y] = \frac{\vert \langle XY \rangle - \langle X \rangle\langle Y \rangle \vert}{\sigma_X \sigma_Y},
    \label{eq:pearson_coeff}
\end{equation}
where $\vert \cdot \vert$ represents the absolute value, $\langle \cdot \rangle$ stands for the mean value, and $\sigma_X$, $\sigma_Y$ are the standard deviations of $X$ and $Y$, respectively. This quantity varies between 0 and 1, with 0 indicating no correlation between the random variables and 1 representing maximum correlation. We calculate this quantity for our Q-learning policies, for the \textsc{swap-asap} policy, and (to provide a baseline) for a random policy, i.e., a policy that consists of a random choice of action for every state. We consider the following cases. We use the notation $A_{i,j}^t$ to denote an action corresponding to the link between nodes $i$ and $j$ at time $t$, and we use $A_{i,i}^t$ to refer to an action at node $i$ at time $t$. Similarly, we use the notation $S_{i,j}^t\in\{-1,0,1,\dotsc,m^{\star}\}$ to refer to the state of the link connecting nodes $i$ and $j$ at time $t$. We refer to Appendix~\ref{sec-MDP_details} for more details.

\begin{table}
\renewcommand{\arraystretch}{1.2}
\centering
\resizebox{0.95\columnwidth}{!}{
\begin{tabular}{|p{2.8cm}|p{1.5cm}|p{1.5cm}|p{1.5cm}|p{1.5cm}|}
\hline
\multicolumn{5}{|c|}{$n=5, m^\star=2, p_{sw}=0.5, p_{\ell}=0.6$} \\
\hline\hline
Correlator & $\vert i-k \vert = 0$ & $\vert i-k \vert = 1$ & $\vert i-k \vert = 2$ & $\vert i-k \vert = 3$ \\
\hline
$r_{\text{RND}}[A_{i,i}^t,A_{k,k}^t]$ & N/A  & $0.11 \pm 0.03$  & $0.05 \pm 0.02$ & N/A \\
$r_{\text{SA}}[A_{i,i}^t,A_{k,k}^t]$ & N/A & $0.19 \pm 0.04$  & $0.03 \pm 0.02$ & N/A \\
$r_{\text{RL}}[A_{i,i}^t,A_{k,k}^t]$ & N/A  & $0.27 \pm 0.03$  & $0.27 \pm 0.03$ & N/A \\
\hline
$r_{\text{RND}}[A_{i,i+1}^t,A_{k,k+1}^t]$ & N/A & $0.17 \pm 0.04$ & $0.03 \pm 0.03$ & $0.01 \pm 0.01$\\
$r_{\text{SA}}[A_{i,i+1}^t,A_{k,k+1}^t]$ & N/A & $0.22 \pm 0.03$ & $0.03 \pm 0.02$ & $0.04 \pm 0.02$\\
$r_{\text{RL}}[A_{i,i+1}^t,A_{k,k+1}^t]$ & N/A & $0.59 \pm 0.01$ & $0.28 \pm 0.03$ & $0.28 \pm 0.02$\\
\hline
$r_{\text{RND}}[A_{i,i}^t,X_{k,k+1}^t]$ & $0.03 \pm 0.02$  & $0.02 \pm 0.01$  & $0.03 \pm 0.02$ & N/A \\
$r_{\text{SA}}[A_{i,i}^t,X_{k,k+1}^t]$ & $0.81 \pm 0.01$  & $0.29 \pm 0.01$ & $0.03 \pm 0.02$ & N/A \\
$r_{\text{RL}}[A_{i,i}^t,X_{k,k+1}^t]$ & $0.20 \pm 0.01$ & $0.04 \pm 0.01$ & $0.07 \pm 0.01$ & N/A \\
\hline
$r_{\text{RND}}[A_{i,i+1}^t,X_{k,k+1}^t]$ & $0.02 \pm 0.01$ & $0.02 \pm 0.01$ & $0.02 \pm 0.01$ & $0.03 \pm 0.01$ \\
$r_{\text{SA}}[A_{i,i+1}^t,X_{k,k+1}^t]$ & $0.39 \pm 0.01$ & $0.07 \pm 0.01$ & $0.04 \pm 0.01$ & $0.04 \pm 0.01$ \\
$r_{\text{RL}}[A_{i,i+1}^t,X_{k,k+1}^t]$ & $0.09 \pm 0.01$ & $0.03 \pm 0.01$ & $0.03 \pm 0.01$ & $0.12 \pm 0.01$ \\
\hline
\end{tabular}}
\caption{Equal-time correlation coefficients of states and actions (see \eqref{eq:pearson_coeff}) for our Q-learning policy ($r_{\text{RL}}$), the \textsc{swap-asap} policy ($r_{\text{SA}}$), and a random policy ($r_{\text{RND}}$). The values range between 0 and 1 (0 represents no correlation and 1 represents maximum correlation or anti-correlation). Averages of correlators are calculated over randomly generated states and actions determined for those states by the respective policies.
}
\label{tab:equal_time_corr}
\end{table}

\begin{enumerate}
    \item If the actions $A_{i,j}^t$ and $A_{k,l}^t$ at some fixed time $t$ at two faraway links/nodes are coordinated, then the covariance between them should be strictly greater than zero, i.e.,
        \begin{equation}
            r[A_{i,j}^t,A_{k,l}^t] > 0 \quad \text{(collaboration).}
        \end{equation}
        From the first two groups of rows of Table~\ref{tab:equal_time_corr}, it is clear that for the \textsc{swap-asap} policy the correlation between action pairs $A_{i,i}^t,\, A_{k,k}^t$ and $A_{i,i+1}^t,\, A_{k,k+1}^t$ falls off very quickly with increasing distance $\vert i-k \vert$, whereas the correlations for our Q-learning policy survive even between the two ends of the chain. The random policy provides a base or offset level for the correlations. The correlations that exist only because of state and action constraints (see Appendix~\ref{sec-MDP_details}) are reflected by this random policy. Also see Appendix~\ref{sec-apndx_advantage}.
    
    \item Similarly, if the action $A_{i,j}^t$ at fixed time $t$ at a link or node is correlated with the status $X_{i,j}^t$ of a faraway link, i.e., 
        \begin{equation}
            r[A_{i,j}^t,X_{k,l}^t] > 0\quad\text{(global knowledge)},
        \end{equation}
        then this would indicate the use of global knowledge of the state by the policy. Here, the status $X_{i,j}^t$ is defined as $X_{i,j}^t=1$ if $S_{i,j}^t\in\{0,1,\dotsc,m^{\star}\}$, and $X_{i,j}^t=0$ if $S_{i,j}^t=-1$. From the final two groups of rows of Table~\ref{tab:equal_time_corr}, we see that in the case of \textsc{swap-asap} all the correlation is concentrated between the states and actions at the same site, i.e., only local knowledge of the state is being used. On the other hand, for our Q-learning policies, the correlations are well distributed: the actions at a node/link are not only influenced by the state at that link, but by the state of other links. This indicates global knowledge of the state. Again, correlations for a random policy are provided as a base level. For random policies, actions and states have no correlations beyond those arising from the built-in constraints on them in the MDP formulation, as explained in detail in Appendix~\ref{sec-MDP_details}.
\end{enumerate}

For both of the correlation functions considered here, we refer the reader to Appendix~\ref{sec-apndx_advantage} for a detailed discussion on their meaning. Also we note here that for equal time correlators, the time $t$ at which the correlator is calculated is irrelevant, since, averages of these correlators are calculated over randomly generated states and actions determined for those states by the respective policies.

\subsection{Decision making with foresight}

Our Q-learning policies also take future rewards into account, unlike local policies such as \textsc{swap-asap}, since they sometimes decide to take actions that might lead to a worse state in the very next time step, but over a longer time scale, and on average, lead to more frequent and larger network connectivity. This maximization of future rewards on larger time scales, discounting immediate rewards, is quantified in the Bellman equations (see Appendix~\ref{sec-Q_learning}) as the discount factor $\gamma$.

\begin{figure}
    \centering
    \includegraphics[width=0.7\columnwidth]{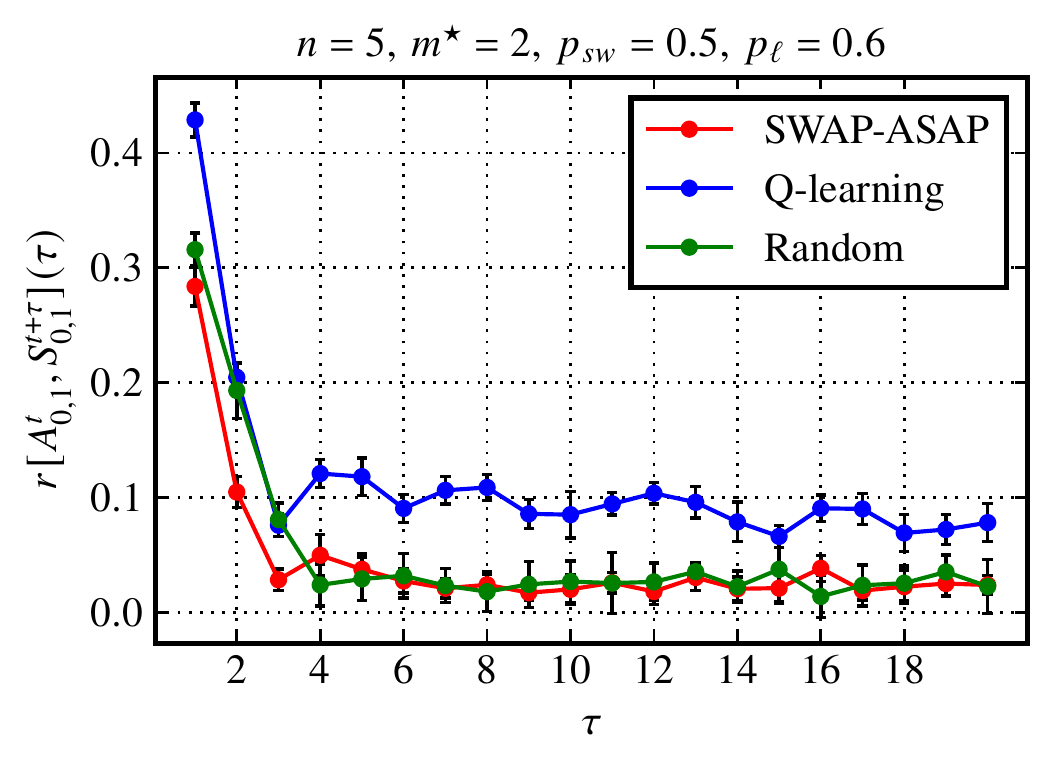}\\
    \includegraphics[width=0.7\columnwidth]{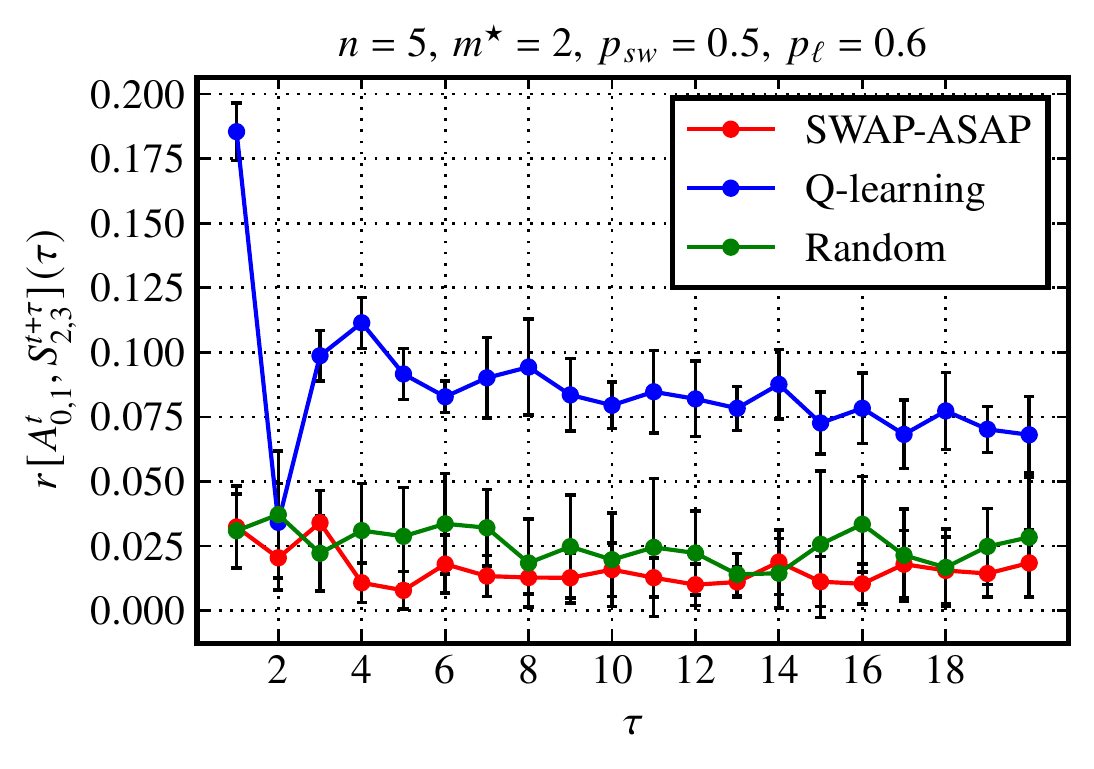}
    \caption{(Top) Unequal-time correlation coefficients for states and actions at the same link. For these plots, we chose $t = 5$. We see that for the \textsc{swap-asap} policy, actions at a time $t$ are influenced only by the states at the next time step, because the correlation falls off exponentially with increasing $\tau$ and reaches very close to 0 very quickly.
    On the other hand, for our Q-learning policies, while the unequal-time correlators also drop with increasing $\tau$, the drop is slower than \textsc{swap-asap}, and the correlations are are sustained for much longer time scales. (Bottom) Unequal-time correlation functions at different links. Here also we chose $t=5$. For the \textsc{swap-asap} policy, these correlations are close to $0$ to begin with, and stay so with increasing $\tau$. For our Q-learning policies, the non-zero equal-time correlations between states and actions at faraway sites are sustained, even for large $\tau$. Error bars represent standard deviation over 50 batches of 1000 independent runs of the repeater chain state evolution.
    }
    \label{fig:unequal_correlators}
\end{figure}

The discounting for future rewards can be examined by looking at unequal-time correlators at a given site or between different sites. In particular, if
\begin{equation}\label{eq-correlator_3}
    r[A_{i,j}^t,X_{k,l}^{t+\tau}] > 0,
\end{equation}
for large $\tau$, then this indicates that actions at time $t$ are correlated with states at time $t+\tau$. Similar conclusions can be drawn if
\begin{equation}
    r[A_{i,j}^t,A_{k,l}^{t+\tau}] > 0.
\end{equation}
In Fig.~\ref{fig:unequal_correlators}, we see that for the \textsc{swap-asap} policy, actions at a time $t$ are influenced only by the state outcomes at the next time step, and the correlator falls off exponentially with increasing $\tau$. This indicates that it maximizes only short-term rewards. On the other hand, our Q-learning policy maximizes both short- and long-term rewards through discounting. The unequal-time correlators fall with increasing $\tau$ but are sustained for much longer time scales. The same is true for correlations between different sites. For \textsc{swap-asap}, these correlations are close to zero to begin with (equal-time), and stay so with increasing $\tau$. In stark contrast, for our Q-learning policies, the non-zero equal-time correlations between states and actions at faraway sites are sustained, even for large $\tau$. For reference, we provide the correlation coefficient values also for a random policy, i.e., a policy in which an action is chosen randomly for every state. The fact that the initial drop in the correlation function exists for all policies (even the random one) shows that it arises simply due to the MDP rules. The state at $t=t+1$ always has some correlation with state at time $t$, regardless of the policy, because the MDP rules remain the same. For instance, if one waits at a link that is below its cutoff age, no matter what policy is chosen, the state can never become inactive in the next time step. This correlation dies off quickly for the random and \textsc{swap-asap} policies because the MDP branches off very quickly with time, except for the RL policies, in which some correlation is retained due to foresight. This also explains why this initial high correlation is absent for random and \textsc{swap-asap} policies in Fig.~\ref{fig:unequal_correlators}~(bottom), because the MDP rules act independently for different links (except for adjacent links involved in a swapping operation). We discuss this and the meaning of foresight in more detail in Appendix~\ref{sec-apndx_advantage}.

As a side remark, we note that the correlation measure in \eqref{eq-correlator_3} is similar in spirit to the ``empowerment'' measure presented in Ref.~\cite{klyubin05empowerment}, which considers, instead of the covariance, the mutual information between the random variables $A_{i,j}^t$ and $S_{k,l}^{t+\tau}$ in order to assess a learning agent's ability to influence its environment.

\section{Improved policies for large repeater chains via nesting} \label{sec:nested}

\begin{figure}
    \centering
    \includegraphics[width=0.9\columnwidth]{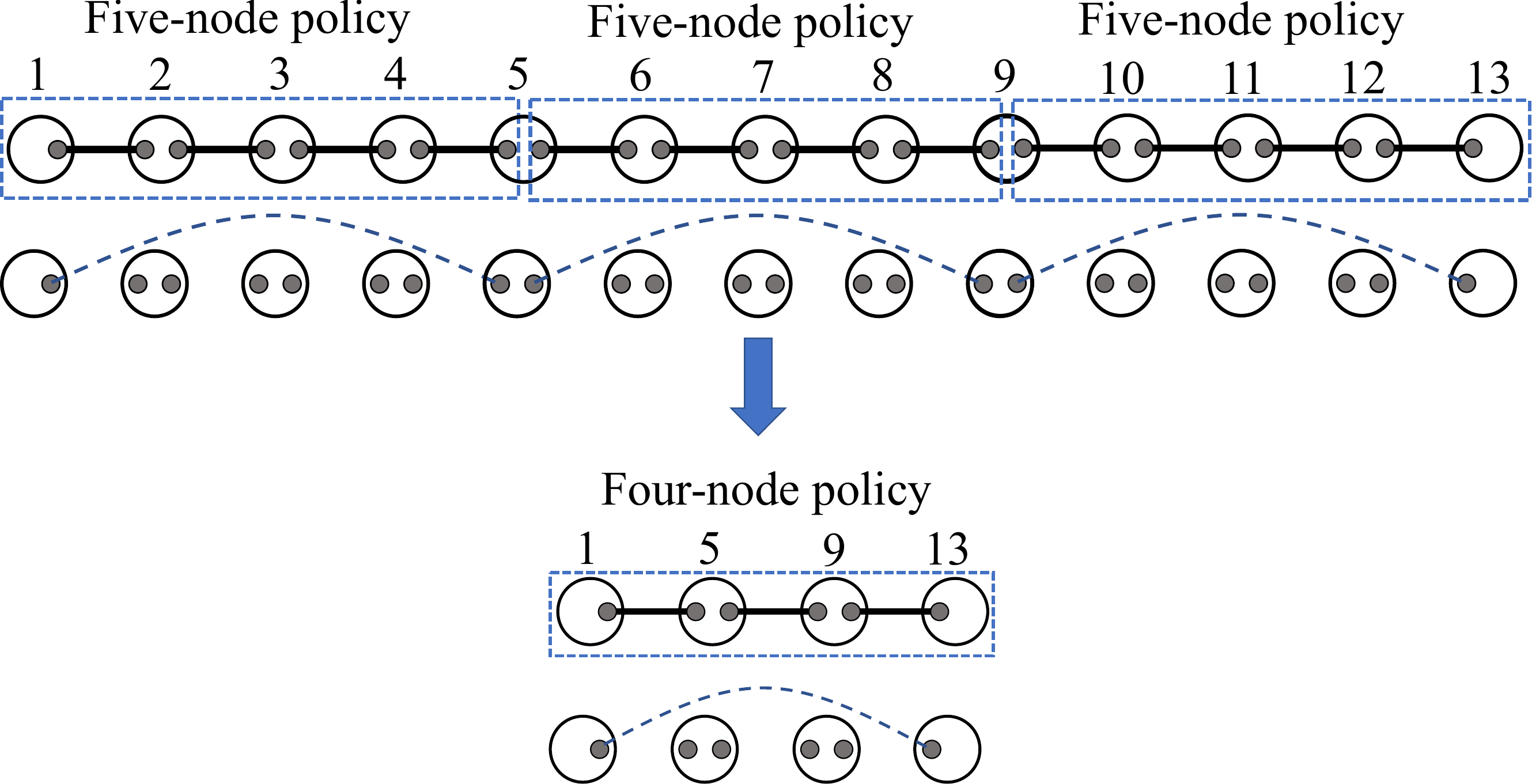}
    \caption{An example of the policy nesting strategy presented in Sec.~\ref{sec:nested}. In this example, the policy for distributing entanglement to the end nodes of a quantum repeater chain with 13 nodes consists of performing a fixed policy for the three five-node segments indicated by the dashed boxes in the ``first'' nesting level. Then, in the ``second'' nesting level, a policy for four nodes is executed.}
    \label{fig:nesting}
\end{figure}

So far, we have considered relatively small repeater chains, up to five nodes, with small values of the maximum cutoff $m^{\star}$. Remarkably, this has been enough for us to extract general features of policies that outperform the \textsc{swap-asap} policy in certain parameter regimes. Now, we would of course like to obtain policies for larger repeater chains. Doing so, however, becomes quickly intractable, even using reinforcement learning, because the number of states and actions in our MDP grows exponentially with the number of nodes. In this section, we present a method for using policies for small repeater chains in order to construct policies for large repeater chains. This allows us to extend our Q-learning policies to larger repeater chains without having to train over the exponentially large state-action space of such chains. To illustrate the method, we present a specific example with two nesting levels. This makes it clear how to generalize the method to an arbitrary number of nesting levels.

Consider the repeater chain shown in Fig.~\ref{fig:nesting}, containing a large number of nodes, and let us split it up into $k$ smaller chains, each with $n$ nodes. (In Fig.~\ref{fig:nesting}, $k=3$ and $n=5$.) Therefore, the full repeater chain has $k(n-1) + 1$ nodes. Now, our ``nesting'' policy refers to the use of a given policy for the smaller chains with $n$ nodes---this is the ``first''  nesting level---and then for the ``second'' nesting level using a policy for $k+1$ nodes. More specifically, the following rules are followed for the two nesting levels.
\begin{enumerate}
    \item An elementary link request $R_\ell$ at the second nesting level is considered as a request to create an end-to-end link for the smaller $n$-node chain in the first nesting level. For example, referring to Fig.~\ref{fig:nesting}, we regard the chain at the second nesting level to be a four-node chain among the nodes 1, 5, 9, and 13, such that a request for an ``elementary link'' between nodes 1 and 5 is interpreted as a request for the end-to-end link in the five-node chain, with end nodes 1 and 5, at the first nesting level.
    
    \item All non-terminal states of the first nesting level are considered inactive states for the second nesting level, and once the terminal state in the first/lower level is achieved, its age becomes the age of the elementary link of the second nesting level.
    
    \item Once a request is performed for elementary link generation at the second level, another request is not allowed for that link until it becomes active. The waiting time keeps on increasing in such a case, and any other active links at the second level keep getting older. Lower level links keep evolving as per the policy at that level.
    
    \item Also, like before, an active link at both levels can be discarded anytime as determined by the policy.
    
    \item A new action is performed at the second level only when the state of at least one link at this level changes. 
\end{enumerate}

\begin{figure}
    \centering
    \includegraphics[width=0.7\columnwidth]{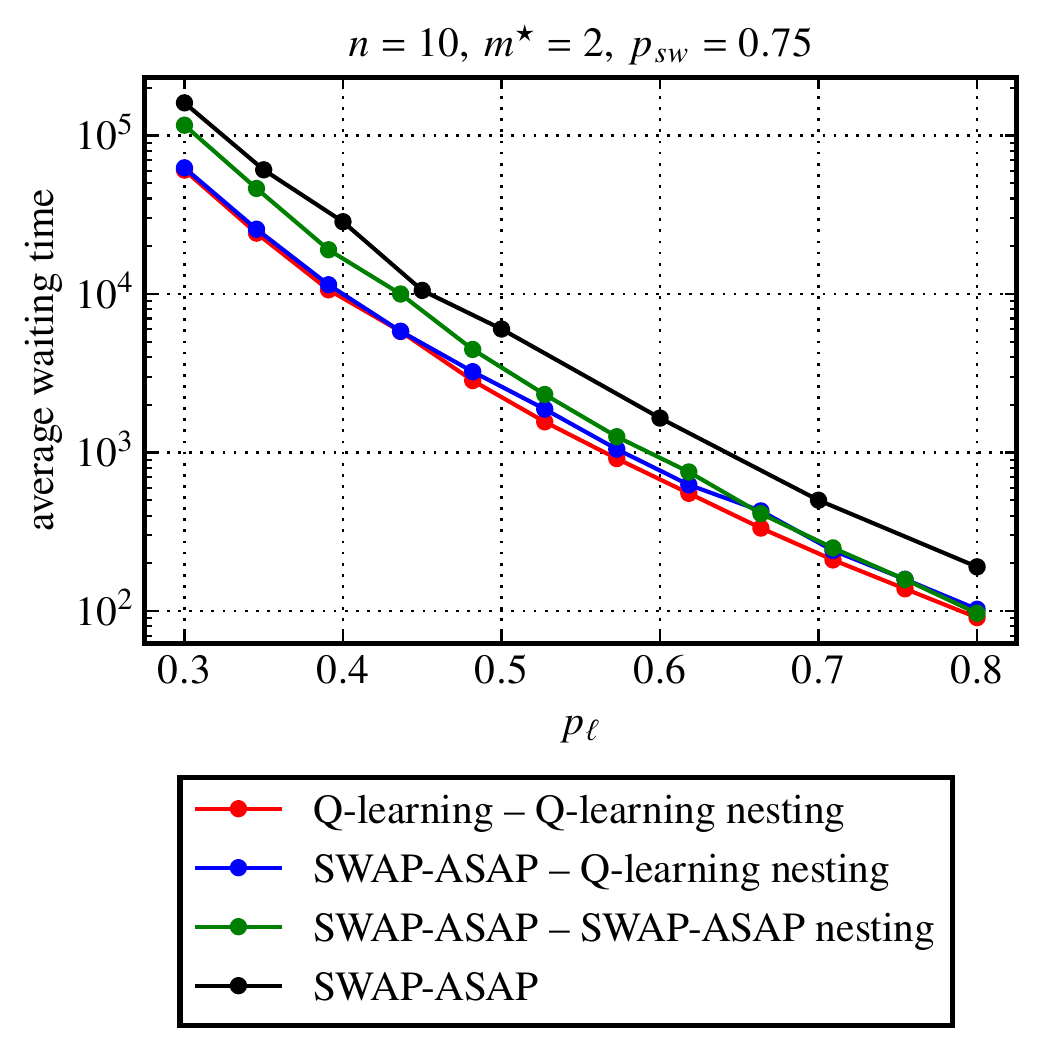}\\
    \includegraphics[width=0.7\columnwidth]{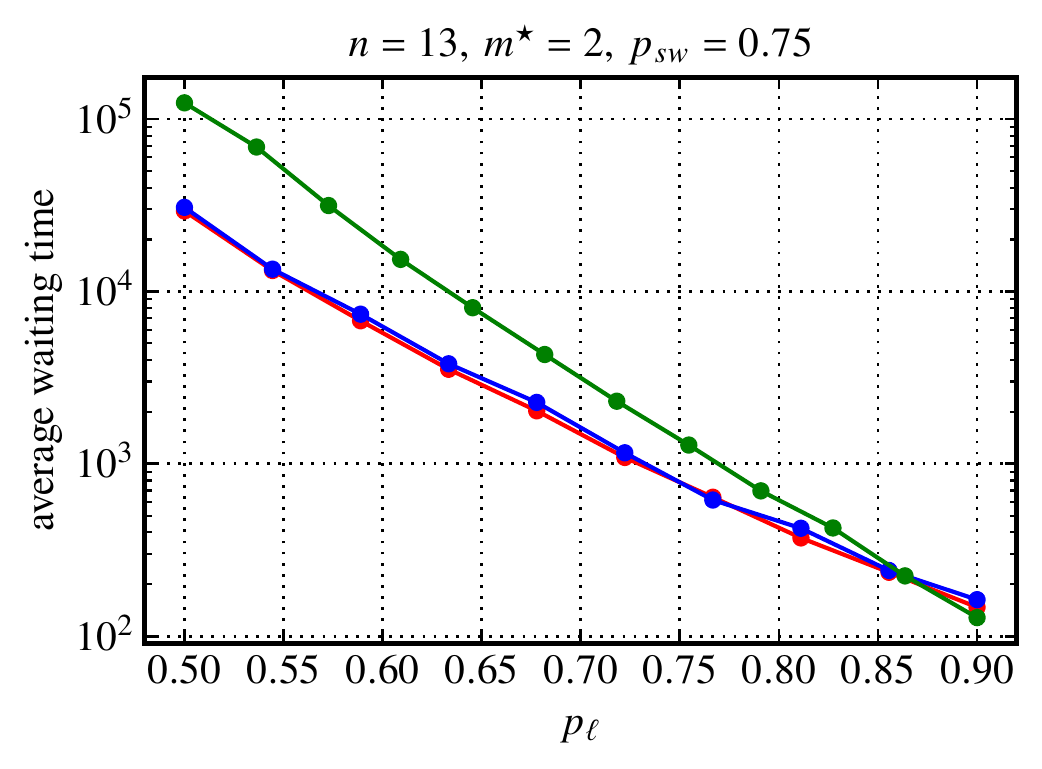}
    \caption{Examples of the application our nesting strategy, as presented in Sec.~\ref{sec:nested}, which uses our Q-learning policies in combination with the local \textsc{swap-asap} policy, in order to improve waiting times for large quantum repeater chains consisting of 10 nodes (three chains with three links each) (top) and 13 nodes (three chains with four links each) (bottom). See the main text for a description of the various nested polices shown. We also provide the waiting time for the simple, non-nested \textsc{swap-asap} policy for the 10-node chain. For the 13-node chain, the exceedingly long simulation times make it difficult to reasonably estimate the waiting time for the simple \textsc{swap-asap} policy.
    }
    \label{fig:concat}
\end{figure}

For the budgeting of classical communication time, it is intuitive that any policy for the second nesting level should ideally be local, because nodes at the second nesting level will, in general, be far away from each other, and because global policies require long-distance classical communication, which increases the waiting time. We therefore consider three kinds of nesting of policies for two nesting levels, and we provide examples of the waiting times for these policies for large chains of 10 and 13 nodes in Fig.~\ref{fig:concat}.
\begin{enumerate}
    \item \textsc{swap-asap} -- \textsc{swap-asap} nesting: This is a ``local-local'' nesting of policies, i.e., the fixed \textsc{swap-asap} policy is used at both levels. This is similar to the ``doubling'' policy studied in previous literature and is known to be better (in terms of the average waiting time) than a fully local \textsc{swap-asap} policy (without any nesting) for large chains with low elementary link generation success probability and low entanglement swapping success probability~\cite{shchukin2022optimal}. 
    
    \item \textsc{swap-asap} -- Q-learning nesting: This we refer to as a ``quasi-local'' or ``local-global'' policy, in which the fixed \textsc{swap-asap} policy is used at the highest (second) nesting level and a Q-learning policy is used at the lowest (first) nesting level. This type of nesting makes sense because, for many repeater chains and network scenarios more generally, it is possible that small sections of the chain (at lower nesting levels) can collaborate amongst themselves and thus use a global policy, without heavily impacting waiting times due to slow classical communication. At the higher (second) level, however, because the corresponding nodes are generally far away, a local policy such as \textsc{swap-asap} should be utilized in order to keep waiting times low. Overall, this type of nesting can reduce waiting times compared to using \textsc{swap-asap} for the full repeater chain. Intuitively, this can also be thought of as increasing the effective link probability of the elementary link of the next nesting level, in turn, reducing the total waiting time. In Fig.~\ref{fig:concat}, we can see that this policy offers a considerable advantage compared to the fully local \textsc{swap-asap} -- \textsc{swap-asap} nested policy.
    
    \item Q-learning -- Q-learning nesting: This refers to the use of a Q-learning policy at both nesting levels. This is not necessarily a practical policy from the point of view of classical communication time, but it serves as a benchmark to see how advantageous the \textsc{swap-asap} -- \textsc{swap-asap} and the \textsc{swap-asap} -- Q-learning nested policies can be. In Fig.~\ref{fig:concat}, we see that, in fact, this policy only slightly outperforms the \textsc{swap-asap} -- Q-learning nested policies, indicating that the latter is quite efficient in a practical setting.
\end{enumerate}
It is also important to note that all of the nesting policies outperform the simple, non-nested version of dynamic \textsc{swap-asap}; see Fig.~\ref{fig:concat}~(top) for the waiting times of this policy for a 10-node chain. For the 13-node chain (Fig.~\ref{fig:concat}~(bottom)), this simple \textsc{swap-asap} policy performs so poorly that we were unable to obtain any reasonable estimate for the average waiting time, due to the exceedingly long simulation times.

We remark that ideas similar in spirit to our nesting idea can be found in Refs.~\cite{jiang2007optimal,vanMeter2011recursiverepeaters,wallnofer2020MLQComm,azuma2023networksaggregation}. Furthermore, from the perspective of Markov decision processes, our nesting idea can be thought of as an example of a multi-agent strategy (as opposed to the single-agent strategy that we have considered throughout this work so far), in which each agent is in charge of a portion of the full chain at the lowest nesting level and acts independently of the other agents. These agents coordinate via another agent at the highest level, which executes a policy that joins the individual portions of the full chain in order to obtain the full end-to-end entanglement.

While we have illustrated our nesting method only for a specific example with two nesting levels, the method can be straightforwardly generalized to an arbitrary number of nesting levels. Also, unlike the example shown here, the repeater chain segments at each nesting level do not have to have the same size. Examining various nesting scenarios for a given repeater chain, in order to determine optimal nesting strategies, is an interesting direction for future work.

\section{Summary and outlook}

In this work, we have presented policies for entanglement distribution for linear quantum repeater chains, both homogeneous (with equal elementary link probabilities throughout) and inhomogeneous (in which the elementary link probabilities can be different). We have shown that these policies improve upon the fixed and dynamic versions of the ``swap-as-soon-as-possible'' (\textsc{swap-asap}) policy, both with respect to the waiting time and the fidelity of the end-to-end entanglement; see Sec.~\ref{sec:opt_pol_waiting_time} and Sec.~\ref{sec-opt_pol_fidelity}. While it might not be surprising that we would obtain improvements, understanding \textit{how} this improvement comes about is certainly of interest, and it is this question that we have aimed to address in this work. Specifically, we have identified some of the key differences between our policies and the \textsc{swap-asap} policies that lead to these improvements; see Sec.~\ref{sec-discussion}. We have also quantified three of these key features, namely: state-dependent cutoff times for the elementary and virtual links, collaboration between the nodes, and knowledge of how current actions will affect the state of the repeater chain far in to the future; see Sec.~\ref{sec:advantage}. Although we have considered relatively small chains of up to five nodes, and limited values of the maximum cutoff time $m^{\star}$, this was already enough for us to extract general principles about how the improvement in waiting time and fidelity comes about. We then applied these principles in Sec.~\ref{sec:nested} in order to obtain waiting time improvements for larger repeater chains with 10 and 13 nodes.

We used the Q-learning reinforcement learning algorithm to obtain our improved policies. Q-learning is a model-free reinforcement learning algorithm, unlike model-dependent methods such as value and policy iteration. Drawing an analogy to a game of chess, the agent in Q-learning not only learns the best strategy to win, but does so while learning the rules of the game on the go. This method is thus well suited to situations in which the exact transition probabilities between the different states of a quantum network are either unknown or very difficult to establish, for example in the case of large networks. Furthermore, unlike deep reinforcement learning techniques, the Q-learning algorithm automatically provides us explicitly with the learned policy. 

One noteworthy aspect of our results is that the improvement due to our policies is the largest in the most non-ideal cases, that is, when the maximum memory cutoff is small, the elementary link success probabilities are low, and the asymmetries in the repeater chain are the highest. This shows that collaboration between the nodes, one of the central features of our policies, is crucial when dealing with realistic, small-to-medium-scale networks with noisy, imperfect quantum devices. 

Our work opens up several interesting avenues for future work. One direction for future work is to add the possibility of doing entanglement distillation, which is a form of quantum error correction, to the MDP formalism presented here. As entanglement distillation is now within the reach of experiments~\cite{PSBZ01,Kalb+17,RWC+21}, this question is especially pertinent, and much remains to be explored about what policies are optimal in this scenario, building on prior works that have considered specific policies~\cite{duer1999repeaterspurification,RPL09,VLMN09,AME11,bratzik2013repeatersdistill,PWD18,PD19,MDV19}; see also Refs.~\cite{CKD+20,wallnofer2022ReQuSim}, which numerically investigate several of these scenarios. We therefore expect our insights on the features of improved policies to be a valuable starting point when designing better protocols for quantum repeater chains with multiple memories and entanglement distillation. In particular, it would be interesting to compare a policy that incorporates entanglement distillation to one without entanglement distillation that simply does the memory-cutoff policy with a single memory~\cite{wallnofer2022ReQuSim}. It is not obvious that entanglement distillation will always give a better overall fidelity, and understanding in what parameter regimes this is the case will help guide us towards a concept of a noise and loss threshold for quantum communication, similar to the fault-tolerance thresholds for quantum computation that are defined by the point at which the logical error rate of an error-corrected quantum computation is less than the raw, physical error rate.

The analysis in this work considers only the number of nodes in the chain as the input to the problem, and not the physical distances between the nodes. Consequently, the policies we obtain do not explicitly take the actual classical communication time (in physical units of time) into account. While our work still provides an important first step towards understanding what types of policies improve waiting times and fidelities, ultimately, in order for these policies to be useful in practice, we have to take classical communication explicitly into account. One way to do this, albeit indirectly, would be to incorporate the quantifiers considered in Sec.~\ref{sec:advantage} into the reward, such that the policies obtained have locality constraints based on the fact that classical communication times between distant nodes can reduce the end-to-end entanglement distribution rate. It would then be interesting to see whether any global-knowledge policy can outperform the local \textsc{swap-asap} policy.

Finally, the ideas in Sec.~\ref{sec:nested} provide an example of using a multi-agent strategy for discovering improved policies in quantum networks. It would be interesting to explore how this nesting strategy, and multi-agent strategies more generally, could be extended to networks with arbitrary topologies, not just linear topologies as we have considered here. Doing so would help guide efforts to develop good policies for arbitrary networks, with the ultimate goal of providing policies for large-scale entanglement distribution, towards a global-scale quantum internet.

\begin{acknowledgments}
   The authors thank Paras Regmi and Roy Pace for helpful discussions. This work was supported by the Army Research Office Multidisciplinary University Research Initiative (ARO MURI) through the grant number W911NF2120214. PB and HL also acknowledge the support of the U.S. Air Force Office of Scientific Research as well as the US-Israel Binational Science Foundation. In addition, SK is supported by the German BMBF (Hybrid).
\end{acknowledgments}

\onecolumngrid
\appendix

\section{Details of the Markov decision process}\label{sec-MDP_details}

Here, we provide further details on our Markov decision process (MDP) framework for quantum repeater chains, which is summarized in Fig.~\ref{fig:mdp} of the main text.

A Markov decision process (MDP)~\cite{Puterman2014book} (see also Ref.~\cite[Appendix~A]{Kha22}) is a mathematical model of an agent that has the ability to interact with a system by taking different actions on it. The system consists of a set $\mathbf{S}$ of (classical) states, and similarly the actions form a set $\mathbf{A}$. The transition probabilities between states are given by transition matrices $\{T_A\}_{A\in\mathbf{A}}$, such that the matrix element $T_A(S';S)$ is equal to the probability of going to the state $S' \in \mathbf{S}$, when the current state is $S \in \mathbf{S}$ and the action taken is $A \in \mathbf{A}$. We now present these defining elements of an MDP within the context of quantum repeater chains. We also note that our definitions of the states and actions in a quantum repeater chain are similar to, but not exactly the same as, those in Refs.~\cite{inesta2022optimal,reiss2022deep}; see Sec.~\ref{sec-compare_prior_work} for details on the differences.

\subsection{States}

The state space consists of all possible configurations of the repeater chain. In particular, we represent the state of the repeater chain at any given time $t$ by an $n\times n$ matrix $S^t$, where $n$ is the number of nodes in the chain. The state of the link between two nodes $i,j\in\{1,2,\dotsc,n\}$ is represented by the entry $S^t_{i,j}$, which can lie between $-1$ and $m^{\star}$, with $S^t_{i,j}\geq 0$ indicating that the link is active and $S_{i,j}^{t}=-1$ indicating that the link is inactive.
    
    For example, consider a four-node repeater chain, with an elementary link between node 1 and 2 that is active for one time step and a virtual link between 2 and 4 that is active for two time steps. Such a state is represented as
    \begin{equation}
    S^t = \begin{bmatrix}
    \ast & 1 & \ast & \ast\\
    1 & \ast & \ast & 2\\
    \ast & \ast & \ast & \ast\\
    \ast & 2 & \ast & \ast
    \end{bmatrix},
    \end{equation}
    where the asterisks `$\ast$' indicate that the corresponding entries could be arbitrary elements in the set $\{-1,0,1,\dotsc,m^{\star}\}$. The \textit{terminal/absorbing states} are states in which the first and the last node are connected by a virtual link. Such states have the form $S^t_{1,n}=m$, for $m\in\{0,1,\dotsc,m^{\star}\}$.
   
    We remark that not all $n\times n$ matrices are possible states of the repeater chain. First, the states are symmetric and nodes cannot link to themselves, so that diagonal elements of the form $S^t_{i,i}$ are not allowed. Furthermore, because we are only dealing with a linear repeater chain in which every node has at most two quantum memories, every node can be connected to not more than two more nodes, one ahead it and one behind it. Therefore, every row and column of $S$ can have only two elements that are not equal to $-1$. These constraints mean that the number of possible states scales as $\approx (m^{\star}+1)^{n}$. For example, for a repeater chain with $n=5$ nodes and maximum cutoff $m^{\star} = 2$, the number of allowed states is 562.

\subsection{Actions}

The action space consists of the following.
\begin{itemize}
    \item Request ($R_\ell$):
    This action is a request to create an elementary link between two adjacent nodes. The success probability of elementary link creation is denoted by $p_{\ell}$. If the elementary link is already active between the two nodes, then this action first discards the link and then attempts to create a new one.
    \item Wait ($W$): This action keeps a link, either a virtual link or an elementary link, stored in the quantum memories. It increases the age of active links by one; however, if the current age is equal to the maximum cutoff $m^{\star}$, then the link becomes inactive.
    \item Request Swapping ($R_{sw}$) : This action can take place at any node when it is linked to two other nodes via elementary or virtual links. We assume that the entanglement swapping is error free but in general non-deterministic, succeeding with probability $p_{sw}\in[0,1]$. If it succeeds, a virtual link is established between the corresponding end nodes. In case of failure, both links become inactive.
\end{itemize}
Mathematically, similar to the states, we describe the action taken on a state at any given time $t$ by an $n\times n$ matrix $A^t$. Now, $A^t_{i,j}$ represents the action taken by the nodes $i,j$ together. This is also a symmetric matrix. An elementary link request ($R_{\ell}$) is given by $A^t_{i,j}=1$, with $j=i+1$. The wait action ($W$) is given by $A^t_{i,j}=0$ for $i\neq j$. Entanglement swapping requests ($R_{sw}$) are given by diagonal elements of the matrix, so that $A_{i,i}^t=1$ means a request for entanglement swapping at the $i^{\text{th}}$ node. For example, consider a four-node repeater chain. The action of requesting an elementary link between node 1 and node 2 along with a swapping request at node 3 is represented as
    \begin{equation*}
    A^t = \begin{bmatrix}
    0 & 1 & 0 & 0\\
    1 & 0 & 0 & 0\\
    0 & 0 & 1 & 0\\
    0 & 0 & 0 & 0
    \end{bmatrix}.
    \end{equation*}
    As with the states, the number of allowed actions is limited. Again, the actions are symmetric, but now the elements $A^t_{i,i}$ can be non-zero. Also, no $R_\ell$ action is allowed between non-adjacent nodes, so the only non-diagonal non-zero elements are of the form $A_{i,i+1}^t$. These constraints bring down the number of possible actions to the order of $\approx 2^{(2n-3)}$ for a repeater chain with $n$ nodes. Further, if a swap action $R_{sw}$ is taken at a node, then the same node is not involved in an elementary link request $R_\ell$; therefore, if $A_{i,i}^t = 1$, then the $i^{\text{th}}$ row and column can have no other non-zero entry. These constraints reduce the number of allowed actions even more. For example, for a repeater chain with $n=5$ nodes, the number of allowed actions is only 34 (out of the $2^{(2n-3)} = 128$ possible actions taking all combinations of elementary link generation and entanglement swap requests).

\subsection{Transition rules}

Let us now look at the effect of various actions on the states of the repeater chain.
\begin{itemize}
    \item Effect of request ($R_\ell$): If $A^t_{i,i+1} = 1$, then with probability $p_{\ell}$ we obtain $S^{t+1}_{i,i+1} = 1$, and with probability $1 - p_{\ell}$ we obtain $S^{t+1}_{i,i+1} = -1$. 
    
    \item Effect of wait ($W$): 
    If $A^t_{i,j} = 0$ and $S^t_{i,j} \geq 0$, then
    \begin{equation}\label{eq-state_update_wait}
        S^{t+1}_{i,j} = -1 + ((S^t_{i,j} + 2)\text{ mod } (m^\star+2)).
    \end{equation}
    If the link is already inactive, i.e., if $S^t_{i,j} = -1$, then $S^{t+1}_{i,j} = S^t_{i,j}$. Observe that the action $W$ automatically discards any link (elementary or virtual) that has an age equal to the maximum cutoff time $m^\star$.
    
    \item Effect of request for entanglement swapping ($R_{sw}$):
    If $A^t_{i,i} =1 $, then check if $S^t_{i,k}\geq 0$ and $S^t_{j,i}\geq 0$, i.e., check if the node $i$ has an active link with some nodes $j$ and $k$. If so, then with probability $p_{sw}$, $S^{t+1}_{j,k} = S_{i,k}^t+S_{j,i}^t$, and with probability $1 - p_{sw}$, $S^{t+1}_{j,k} = -1$. For both outcomes, the parent links become inactive, i.e., $S^{t+1}_{i,k} = -1$ and $S^{t+1}_{j,i} = -1$. Note that if an entanglement swap is requested at a node that is connected to only one other node, then that link becomes inactive. Also, if a swap operation is attempted when $S_{i,k}^t+S_{j,i}^t > m^\star$, then $S^{t+1}_{j,k} = -1$, even if the operation succeeds. Therefore, virtual links are always discarded at the maximum cutoff age.

\end{itemize}
The request and wait actions effect individual links independently. The swap request, on the other hand, depends on and effects the states of two links.

The transition rules described above form the transition matrices $\{T_A\}_{A\in\mathbf{A}}$ of our MDP. As a result of having absorbing states, we can partition every transition matrix $T_A$, $A\in\mathbf{A}$, into a block matrix as follows:
\begin{equation}
    T_A=\begin{pmatrix} Q_A & 0 \\ R_A & \mathbbm{1} \end{pmatrix},
\end{equation}
where the top-left block $Q_A$ corresponds to transitions between transient (non-terminal) states, the bottom-left block $R_A$ corresponds to transitions from transient states to terminal states, and the bottom-right block corresponds to transitions between terminal states, which is the identity matrix $\mathbbm{1}$ by our definition of a terminal state.

\subsection{Optimization of waiting time and fidelity} 

A policy is defined by the function $\pi:\mathbf{A}\times\mathbf{S}\to[0,1]$, such that $\pi(A|S)$ is defined as the probability of taking action $A\in\mathbf{A}$ given that the state of the repeater chain is $S\in\mathbf{S}$. In other words, the policy is a collection of conditional probability distributions, one for every state.

For every policy $\pi$, we define the transition matrix $P_{\pi}$ such that, for all $S,S'\in\mathbf{S}$,
\begin{equation}
    P_{\pi}(S';S)=\sum_{A\in\mathbf{A}}\pi(A|S)T_A(S';S).
\end{equation}
Based on the block structure of the transition matrices $T_A$, as described above, we have that for every policy $\pi$ the transition matrix $P_{\pi}$ has the following block structure:
\begin{equation}
    P_{\pi}=\begin{pmatrix} Q_{\pi} & 0 \\ R_{\pi} & \mathbbm{1} \end{pmatrix},
\end{equation}
where
\begin{equation}
    Q_{\pi}(S';S)\coloneqq\sum_{A\in\mathbf{A}}\pi(A|S)Q_A(S';S)
\end{equation}
for all transient states $S,S'$, and 
\begin{equation}
    R_{\pi}(S';S)\coloneqq\sum_{A\in\mathbf{A}}\pi(A|S)R_A(S';S)
\end{equation}
for all transient states $S$ and terminal states $S'$.

\paragraph*{Optimization of waiting time.} For a given policy $\pi$, let
\begin{equation}
    N_{\pi}\coloneqq (\mathbbm{1}-Q_{\pi})^{-1}.
\end{equation}
Then, it follows from the theory of Markov chains (see, e.g., \cite[Theorem~9.6.1]{Stewart09_book}) that the expected time to reach a terminal state (i.e., the expected waiting time for an end-to-end link) is given by
\begin{equation}
    W_{\pi}\coloneqq \sum_{S,S'\in\mathbf{S}} N_{\pi}(S';S)p_1(S),
\end{equation}
where $p_1(S)\equiv\Pr[S^1=S]$ is the initial probability distribution of (transient) states at time $t=1$. The minimum expected waiting time is then given by the following optimization problem:
\begin{equation}\label{eq-waiting_time_opt_problem}
    \begin{array}{l l} \text{minimize} & W_{\pi} \\[1ex] \text{subject to} & \pi:\mathbf{A}\times\mathbf{S}\to[0,1],\, \sum_{A\in\mathbf{A}}\pi(A|S)=1~~\forall~S\in\mathbf{S}.\end{array}
\end{equation}

\paragraph*{Optimization of fidelity.} For every state $S\in\mathbf{S}$ of the repeater chain, we define
\begin{equation}
    \widetilde{f}(S)\coloneqq\left\{\begin{array}{l l} f(S_{1,n}) & \text{if } S \text{ is a terminal state}, \\ 0 & \text{otherwise},  \end{array}\right.
\end{equation}
where we recall the function $f$ defined in \eqref{eq-elem_link_fidelity}. In other words, $\widetilde{f}(S)$ gives the fidelity of the end-to-end entangled state, when it is active. We are then interested in the expected value of this quantity with respect to a policy $\pi$ in the steady-state limit:
\begin{align}
    \widetilde{F}_{\pi}&\coloneqq \lim_{t\to\infty}\widetilde{F}_{\pi}(t),\\
    \widetilde{F}_{\pi}(t)&\coloneqq\langle\widetilde{f}(S^t)\rangle_{\pi}=\sum_{S,S'\in\mathbf{S}} \widetilde{f}(S')P_{\pi}^{t-1}(S';S)p_1(S),
\end{align}
where in the last line $p_1(S)\equiv\Pr[S^1=S]$ is the initial probability distribution of (transient) states at time $t=1$. The maximum fidelity is then given by the following optimization problem:
\begin{equation}\label{eq-fidelity_opt_problem}
    \begin{array}{l l} \text{maximize} & \widetilde{F}_{\pi} \\[1ex] \text{subject to} & \pi:\mathbf{A}\times\mathbf{S}\to[0,1],\, \sum_{A\in\mathbf{A}}\pi(A|S)=1~~\forall~S\in\mathbf{S}. \end{array}
\end{equation}

\section{The Q-learning algorithm}\label{sec-Q_learning}

In this work, we used the Q-learning algorithm~\cite{SuttonBarto2018book} to train the learning agent. Q-learning employs the following update rule: 
\begin{equation}
Q'(S^t,A^t) = Q(S^t,A^t) +\alpha[R^{t+1}+\gamma \max_a Q(S^{t+1},a)-Q(S^t, A^t)],
\label{eqn:bellman}
\end{equation}
where $Q$ is the expected return of the state-action pair ($S^t$ and $A^t$ at time $t$, respectively). The Q-learning algorithm is known to converge to an optimal policy in the limit that all state-action pairs are visited infinitely often~\cite{watkins89_thesis,tsitsiklis94Qlearning,jaakkola94Qlearning}. Convergence rates and sampling complexities for Q-learning have been studied in Refs.~\cite{kearns98Qlearningconvergence,dar03Qlearningconvergence}.

\begin{figure*}
    \centering
    \includegraphics[width=0.85\textwidth]{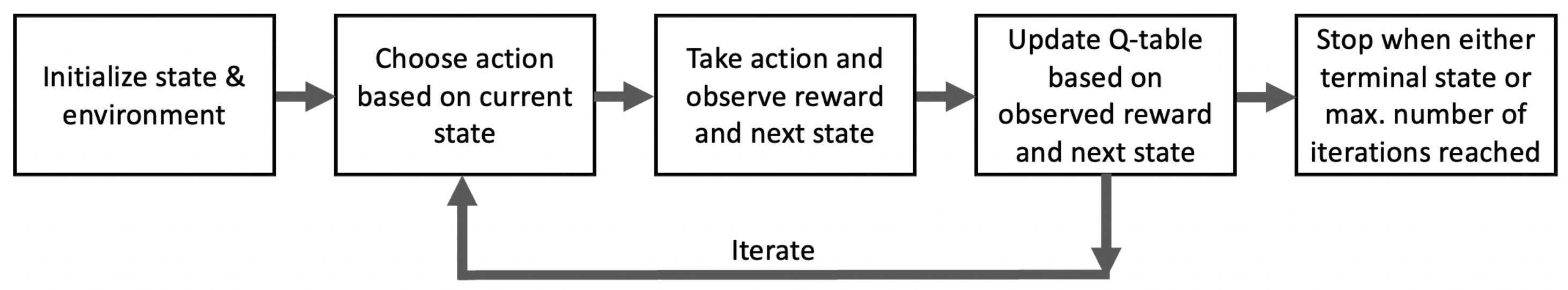}
    \caption{Schematic of the Q-learning algorithm, which is a model-free reinforcement learning technique. The Q-table is the central object in the algorithm and its update through numerous episodes starting from a disconnected state to a terminal state (end-to-end connected via a virtual link) constitutes the learning process. The final Q-table is an instruction on what are the best actions to take in a particular state, with the aim to reach the terminal state and at the same time optimize the desired figure of merit, namely the waiting time and the end-to-end link fidelity. The update of the Q-table is done according to the update rule in \eqref{eqn:bellman}.}
    \label{fig:q_learn}
\end{figure*}

The Q-learning algorithm initializes a Q-matrix of dimension $|\mathbf{S}| \times |\mathbf{A}$|. The initial state is chosen randomly from all possible non-terminal states. Then, the initial action is chosen using an $\epsilon$-greedy algorithm, in which the reward is maximized with probability $\epsilon$, otherwise a random action is taken. The new Q-value for the state-action pair is updated in the Q-matrix according to a pre-established reward. The algorithm repeats this cycle until a terminal state is reached; see Fig.~\ref{fig:q_learn}. This sequence starting from an initial state and reaching the terminal state is defined as an episode. After the algorithm has run over an adequate number of episodes, the column corresponding to the maximum value in the $i^{\text{th}}$ row of the Q-matrix represents the best possible action in the state ``$i$''. Hence, the Q-matrix explicitly determines the improved policy.

\begin{figure*}
    \centering
    \includegraphics[width=0.25\textwidth]{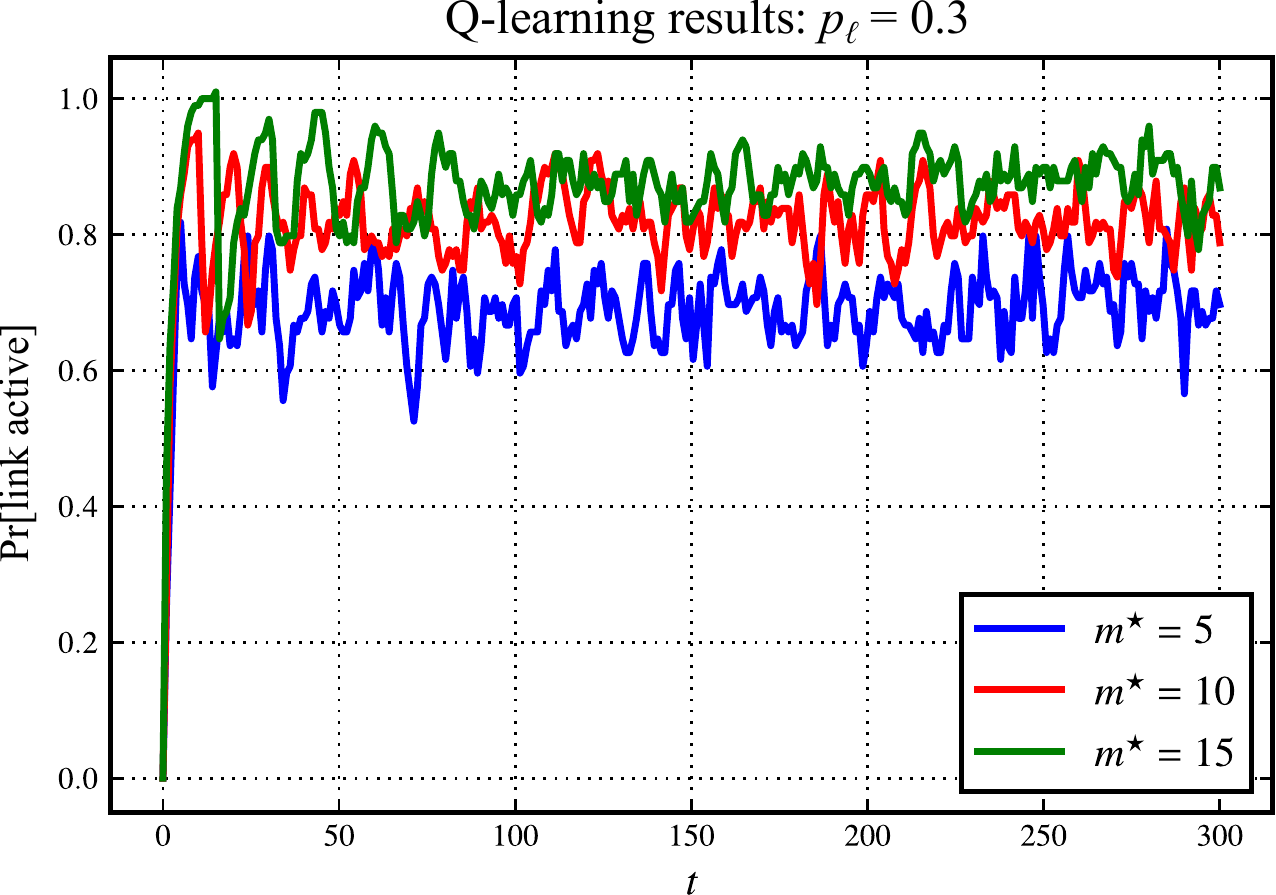}\quad\quad
    \includegraphics[width=0.25\textwidth]{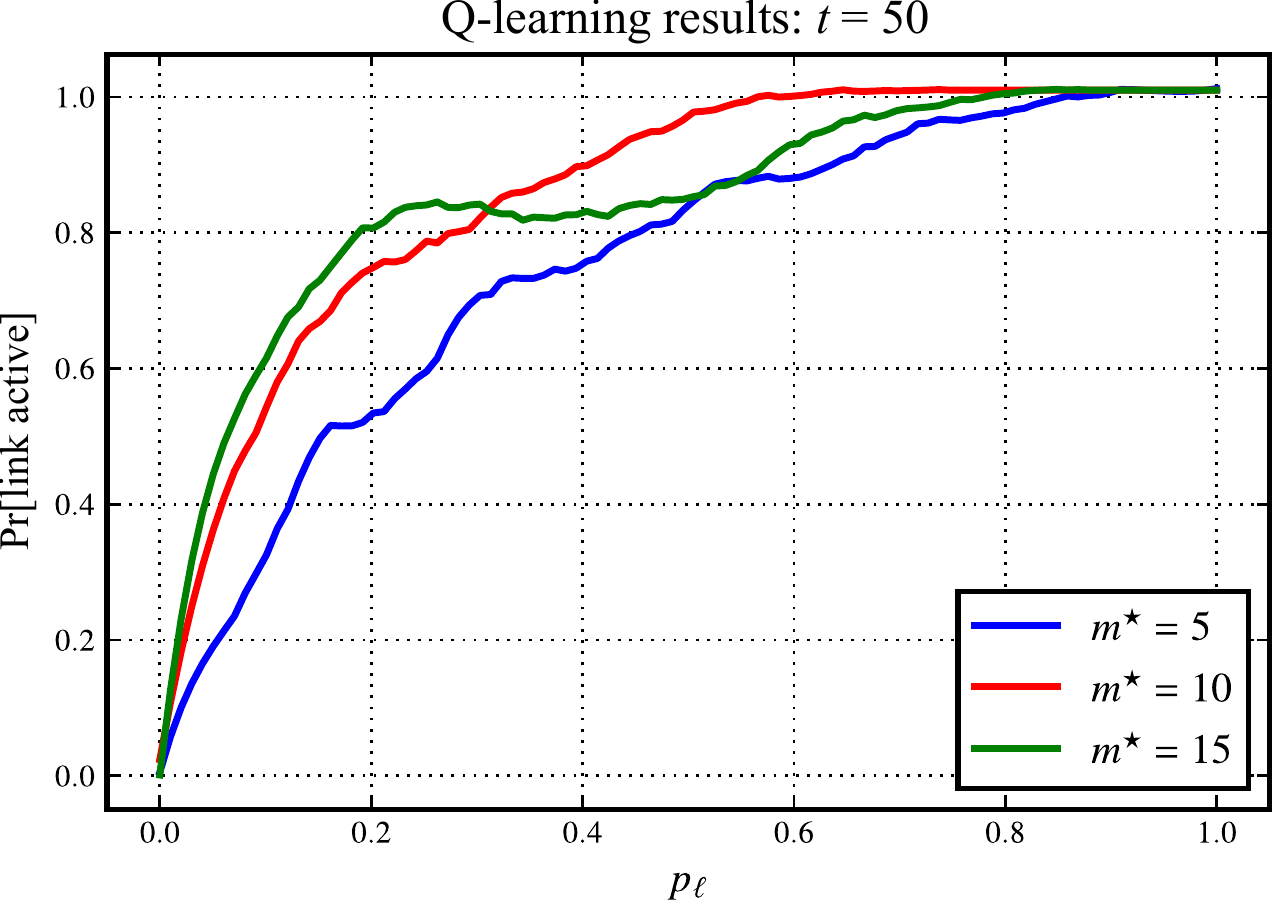}\quad\quad
    
    \includegraphics[width=0.25\textwidth]{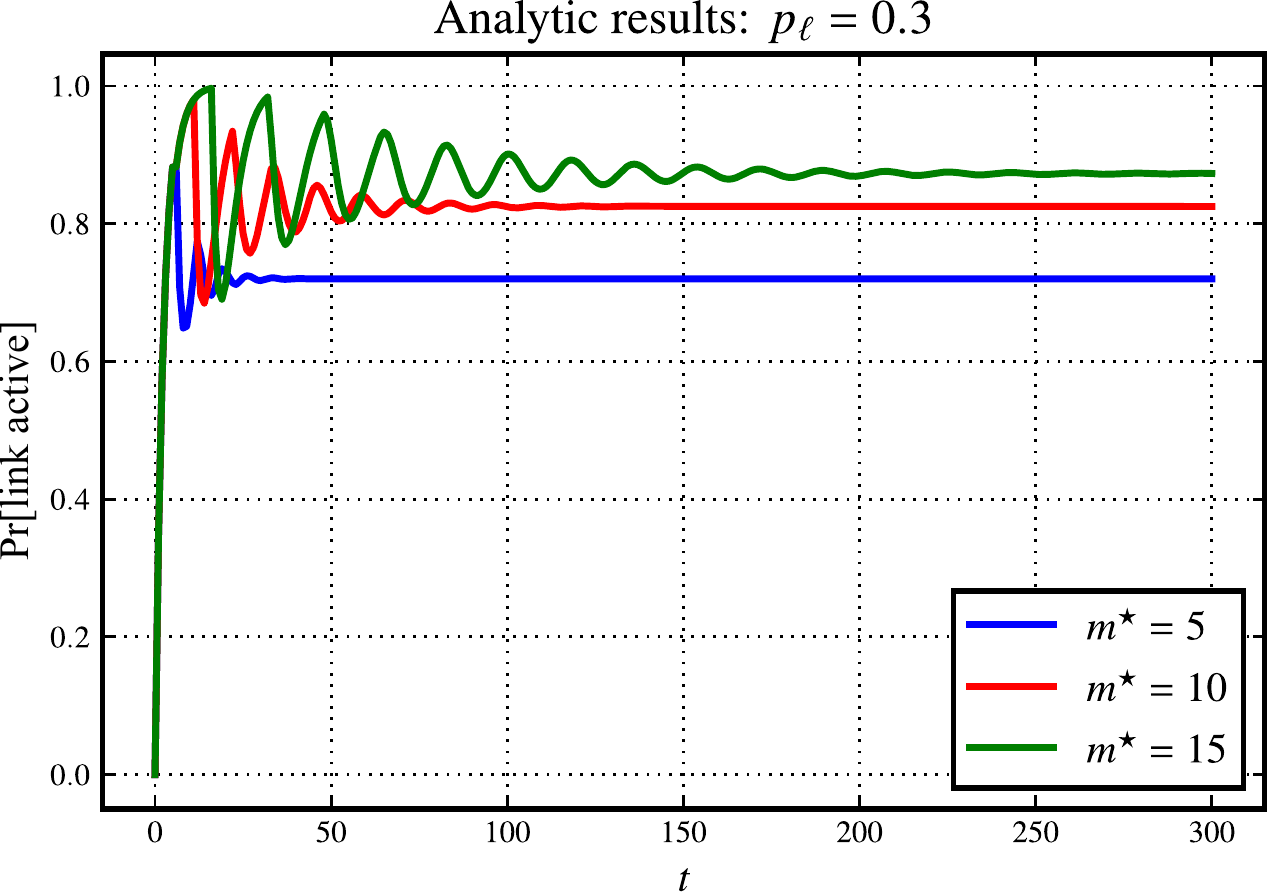}\quad\quad
    \includegraphics[width=0.25\textwidth]{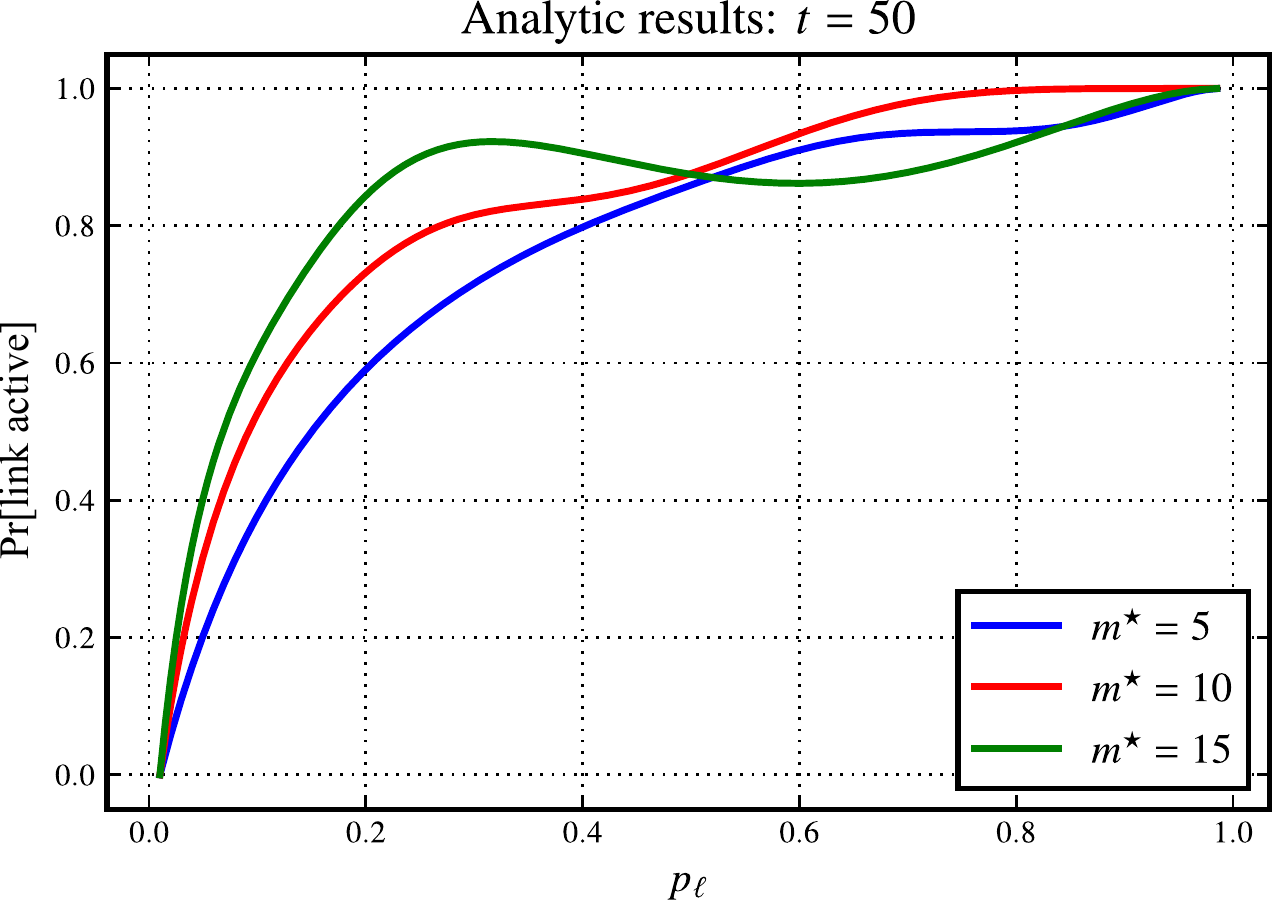}
    \caption{Probability that an elementary link is active for the memory-cutoff policy with cutoff $m^{\star}$, showing the Q-learning results alongside the analytical results from Ref.~\cite[Corollary~4.3]{Khatri2021policieselementary}.}
    \label{fig:elem_link_Q_learning}
\end{figure*}

\paragraph*{Role of training hyperparameters in learning.} Four training hyperparameters, namely $\alpha, \gamma, \epsilon$, and the number of training episodes, need to be tuned in order to find improved policies.
\begin{enumerate}
    \item The role of $\alpha$ is to determine the learning rate. A very low $\alpha$ makes the training too slow; a very large $\alpha$ makes us ``overshoot'' the optimal policy. A value of $\alpha \approx 0.01$ was found to be ideal for training in this work.
    \item The discount factor $\gamma$ determines the discounting for future rewards. This parameter also needs to be tuned in order to find a good balance between prioritizing immediate (next state) rewards versus long-term rewards. We found $\gamma \approx 0.8$ to be ideal.
    \item The hyperparameter $\epsilon$, as mentioned above, controls the trade-off between the exploration of the agent (choosing random actions) and its exploitation (using learned actions). We found $\epsilon$ to be the most relevant hyperparameter, and the one that requires the most fine tuning for finding improved policies. We found $\epsilon \approx 0.15$ to be ideal.
    \item The figures of merit improve with an increasing number of training episodes (at the expense of a higher training time) until they saturate and more training does not help. Furthermore, the larger the state-action space (i.e., the higher the number of nodes and/or the maximum cutoff $m^\star$), and the smaller the probability of creating elementary and virtual links (small $p_\ell$, $p_{sw}$), the larger the number of training episodes required for getting an improved policy. The number of training episodes needed for saturating the figures of merit varied between a few hundred thousand to approximately one million. 
\end{enumerate} 
Here we also note that training need not be done for the entire range of repeater chain parameters that need to be simulated. Our Q-learning policies are different for different values of $p_{\ell}$, $m^{\star}$, etc., but they are not very sensitive to small changes in these parameters. For example, to simulate the repeater chain for $p_\ell = 0.2,0.3,0.4$, we trained and determined the policy only for $p_\ell = 0.3$. Training separately for $p_{\ell}=0.2, 0.4$, and then using those policies, gave no improvement in the waiting time or fidelity compared to using the $p_\ell = 0.3$ policy. We can view this as an example of ``transferable reinforcement learning'', and we could also use policies for one value of $p_{\ell}$ in order to ``warm start'' the training for another value of $p_{\ell}$, in order to improve training times. This is especially relevant to finding policies for chains with very small values of $p_\ell$ and $p_{sw}$, for which training from scratch is extremely resource intensive.

\paragraph*{Verification of the Q-learning algorithm for a single elementary link.} At the elementary link level, if one assumes a finite memory cutoff, the improved policy is to request ($R_\ell$) until the link is created and then wait for $m^{\star}$ time steps before requesting again. For the single link, analytical expressions for figures of merit such as expected state of the link or average wait times can be obtained (see Corollary 4.3 in Ref.~\cite{Khatri2021policieselementary}) and show that the aforementioned policy is optimal. As a sanity check, we see in Fig.~\ref{fig:elem_link_Q_learning} that the Q-learning algorithm also rediscovers this policy and gives qualitative and quantitative agreement with the aforementioned analytical results.

\section{Entanglement swapping update rule for Pauli noise}\label{sec:ent_swap_Pauli_update}

In this section, we prove the formula $f_{\textsc{swap}}(m_1,m_2)=f(m_1+m_2)$ for the fidelity after entanglement swapping of two entangled states impacted by Pauli noise, with ages $m_1$ and $m_2$. This formula can be viewed as a special case of the general formula in Ref.~\cite[Proposition~III.1]{Kha22} or as a consequence of the developments in Ref.~\cite[Appendix~D]{schmidt2020memoryassisted}; however, here we provide a direct proof. To that end, let $\mathcal{N}$ be an arbitrary Pauli channel, so that
\begin{equation}
    \mathcal{N}(\rho)=p_I\rho + p_XX\rho X+p_YY\rho Y+p_ZZ\rho Z,
\end{equation}
where $p_I,p_X,p_Y,p_Z$ are probabilities and $X,Y,Z$ are the single-qubit Pauli operators. Consider noisy bipartite states with ages $m_1$ and $m_2$, as defined in \eqref{eq-elem_link_state}:
\begin{align}
    \sigma_{AR_1}(m_1)&=(\mathcal{N}_A^{\circ m_1}\otimes\mathcal{N}_{R_1}^{\circ m_1})(\Phi_{AR_1}^+),\\
    \sigma_{R_2B}(m_2)&=(\mathcal{N}_{R_2}^{\circ m_2}\otimes\mathcal{N}_B^{\circ m_2})(\Phi_{R_2B}^+),
\end{align}
with the qubits $R_1$ and $R_2$ being involved in the entanglement swapping operation. Using the property that $(M\otimes\mathbbm{1})\ket{\Phi^+}=(\mathbbm{1}\otimes M^{\t})\ket{\Phi^+}$ for every linear operator $M$, we can write these states as
\begin{align}
    \sigma_{AR_1}(m_1)&=(\mathcal{N}_A^{\circ (2m_1)}\otimes\id_{R_1})(\Phi_{AR_1}^+),\\
    \sigma_{R_2B}(m_2)&=(\id_{R_2}\otimes\mathcal{N}_B^{\circ (2m_2)})(\Phi_{R_2B}^+),
\end{align}
such that for the first state the noise acts entirely on qubit $A$ and for the second state the noise acts entirely on qubit $B$. Now, the entanglement swapping operation can be represented as the following quantum channel~\cite[Sec.~III.B.1]{Kha22}:
\begin{equation}
    \mathcal{L}_{AR_1R_2B\to AB}(\rho_{AR_1R_2B})=\sum_{z,x=0}^1 \left(\bra{\Phi_{z,x}}_{R_1R_2}\otimes Z_B^zX_B^x\right)\rho_{AR_1R_2B}\left(\ket{\Phi_{z,x}}_{R_1R_2}\otimes X_B^xZ_B^z\right),
\end{equation}
where $\ket{\Phi_{z,x}}=(Z^zX^z\otimes\mathbbm{1})\ket{\Phi^+}$. Then, because the Pauli corrections $Z^zX^x$ being applied to the qubit $B$ as part of the entanglement swapping protocol are themselves Pauli channels, it holds that these channels commute with the noise channel $\mathcal{N}$, which means that
\begin{align}
    &\rho_{AB}(m_1,m_2)\coloneqq\mathcal{L}_{AR_1R_2B\to AB}(\sigma_{AR_1}(m_1)\otimes\sigma_{R_2B}(m_2))\nonumber\\
    &\quad=\sum_{z,x=0}^1 \left(\bra{\Phi_{z,x}}_{R_1R_2}\otimes Z_B^zX_B^x\right)\left(\sigma_{AR_1}(m_1)\otimes\sigma_{R_2B}(m_2)\right)\left(\ket{\Phi_{z,x}}_{R_1R_2}\otimes X_B^xZ_B^z\right)\\
    &\quad=\left(\mathcal{N}_A^{\circ (2m_1)}\otimes\id_B\right)\left( \sum_{z,x=0}^1 \left(\bra{\Phi_{z,x}}_{R_1R_2}\otimes Z_B^zX_B^x\right)\left(\Phi_{AR_1}^+\otimes(\mathbbm{1}_{R_2}\otimes\mathcal{N}_B^{\circ (2m_2)})(\Phi_{R_2B}^+)\right)\left(\ket{\Phi_{z,x}}_{R_1R_2}\otimes X_B^xZ_B^z\right) \right)\\
    &\quad=\left(\mathcal{N}_A^{\circ (2m_1)}\otimes\mathcal{N}_B^{\circ(2m_2)}\right)\left(\mathcal{L}_{AR_1R_2B\to AB}(\Phi_{AR_1}^+\otimes\Phi_{R_2B}^+)\right)\\
    &\quad=\left(\mathcal{N}_A^{\circ (2m_1)}\otimes\mathcal{N}_B^{\circ(2m_2)}\right)(\Phi_{AB}^+)\\
    &\quad=(\mathcal{N}_A^{\circ(m_1+m_2)}\otimes\mathcal{N}_B^{\circ(m_1+m_2)})(\Phi_{AB}^+)\\
    &\quad=\sigma_{AB}(m_1+m_2).
\end{align}
Therefore, using \eqref{eq-elem_link_fidelity} and \eqref{eq-virtual_link_fidelity}, we obtain $f_{\textsc{swap}}(m_1,m_2)\coloneqq\bra{\Phi^+}\rho_{AB}(m_1,m_2)\ket{\Phi^+}=\bra{\Phi^+}\sigma_{AB}(m_1+m_2)\ket{\Phi^+}=f(m_1+m_2)$, as required.

\section{Comparing the \textsc{swap-asap} policy with Q-learning policies}\label{sec-opt_pol_example}

In Sec.~\ref{sec-discussion}, we compared our Q-learning policies with the fixed and dynamic \textsc{swap-asap} policies. Our Q-learning policies are based on two different rewards. The first reward is based on optimizing the average waiting time alone (see \eqref{eq-opt_waiting_time_reward}), and the second reward is based on optimizing both the average waiting time and the fidelity of the end-to-end entangled state (see \eqref{eq-reward_Q_learning_fidelity}). In Table~\ref{Q_table}, we compare these improved policies with each other and also with the dynamic \textsc{swap-asap} policy. Since the dynamic \textsc{swap-asap} policy allows freedom in the order of swaps (see Sec.~\ref{subsec:swap_asap}), when an action consisting of two swap requests is taken, these swaps are performed sequentially, i.e., if the first swap fails the second one is not performed. On the other hand, for the Q-learning policies, such freedom is not allowed. All swaps that are requested by the policy are performed in one go. This is in keeping with our assumptions that Q-learning is free to find the optimal order of actions, which includes the order of swaps. In the case as shown in Table~\ref{Q_table}, the Q-learning policies do in fact choose to perform the swaps in one go.

\newpage

{\renewcommand{\arraystretch}{1.2}
\begin{longtable}[c]{| p{2.5cm} | p{3cm} | p{3cm} | p{3cm}|}
\hline
\centering State    & \centering Q-learning action (waiting time) & \centering Q-learning action (fidelity \& waiting time) & \centering\arraybackslash \textsc{swap-asap} action \\ 
\hline\hline
($-1$, $-1$, $-1$)     & ($R_\ell$, $R_\ell$, $R_\ell$)                         & ($R_\ell$, $R_\ell$, $R_\ell$)                                    & ($R_\ell$, $R_\ell$, $R_\ell$)            \\
($-1$, $-1$, $0$)    & ($R_\ell$, $R_\ell$, $W$)                         & ($R_\ell$, $R_\ell$, $W$)                                  & ($R_\ell$, $R_\ell$, $W$)            \\
($-1$, $-1$, $1$)     & ($R_\ell$, $R_\ell$, $R_\ell$)                         & ($R_\ell$, $R_\ell$, $R_\ell$)                                    & ($R_\ell$, $R_\ell$, $W$)            \\
($-1$, $-1$, $2$)     & ($R_\ell$, $R_\ell$, $R_\ell$)                         & ($R_\ell$, $R_\ell$, $R_\ell$)                                    & ($R_\ell$, $R_\ell$, $R_\ell$)            \\
($-1$, $0$)       & ($R_\ell$, $W$)                           & ($R_\ell$, $W$)                                      & ($R_\ell$, $W$)              \\
($-1$, $1$)      & ($R_\ell$, $W$)                           & ($R_\ell$, $W$)                                      & ($R_\ell$, $W$)              \\
($-1$, $2$)     & ($R_\ell$, $R_\ell$, $R_\ell$)                         & ($R_\ell$, $R_\ell$, $R_\ell$ )                                   & ($R_\ell$, $R_\ell$, $R_\ell$)            \\
($-1$, 0, $-1$)     & ($R_\ell$, $W$, $R_\ell$)                         & ($R_\ell$, $R_\ell$, $R_\ell$)                                    & ($R_\ell$, $W$, $R_\ell$)            \\
($-1$, 0, 0)     & ($R_\ell$, $R_{sw}$)                         & ($R_\ell$, $R_{sw}$)                                    & ($R_\ell$, $R_{sw}$)            \\
($-1$, 0, 1)     & ($R_\ell$, $R_{sw}$)                         & ($R_\ell$, $R_\ell$, $R_\ell$)                                    & ($R_\ell$, $R_{sw}$)            \\
($-1$, 0, 2)     & ($R_\ell$, $W$, $R_\ell$)                         & ($R_\ell$, $R_\ell$, $R_\ell$)                                    & ($R_\ell$, $R_{sw}$)            \\
($-1$, 1, $-1$)     & ($R_\ell$, $R_\ell$, $R_\ell$)                         & ($R_\ell$, $R_\ell$, $R_\ell$)                                   & ($R_\ell$, $W$, $R_\ell$)            \\
($-1$, 1, 0)     & ($R_\ell$, $R_{sw}$)                         & ($R_\ell$, $R_\ell$, $R_\ell$)                                    & ($R_\ell$, $R_{sw}$)            \\
($-1$, 1, 1)     & ($R_\ell$, $R_\ell$, $R_\ell$)                         & ($R_\ell$, $R_\ell$, $R_\ell$)                                    & ($R_\ell$, $R_{sw}$)            \\
($-1$, 1, 2)    & ($R_\ell$, $R_\ell$, $R_\ell$)                         & ($R_\ell$, $R_\ell$, $R_\ell$)                                    & ($R_\ell$, $R_{sw}$)            \\
($-1$, 2, $-1$)     & ($R_\ell$, $R_\ell$, $R_\ell$)                         & ($R_\ell$, $R_\ell$, $R_\ell$)                                    & ($R_\ell$, $R_\ell$, $R_\ell$)            \\
($-1$, 2, 0)     & ($R_\ell$, $R_\ell$, $W$)                         & ($R_\ell$, $R_\ell$, $R_\ell$)                                    & ($R_\ell$, $R_{sw}$)            \\
($-1$, 2, 1)     & ($R_\ell$, $R_\ell$, $R_\ell$)                         & ($R_\ell$, $R_\ell$, $R_\ell$)                                    & ($R_\ell$, $R_{sw}$)            \\
($-1$, 2, 2)     & ($R_\ell$, $R_\ell$, $R_\ell$)                         & ($R_\ell$, $R_\ell$, $R_\ell$)                                    & ($R_\ell$, $R_{sw}$)            \\
0 - ($-1$, $-1$, $-1$)        & N/A                  & N/A                             & N/A     \\
0 - ($-1$, 0, $-1$) & N/A                  & N/A                             & N/A     \\
0 - ($-1$, 1, $-1$) & N/A                  & N/A                             & N/A     \\
0 - ($-1$, 2, $-1$) & N/A                  & N/A                             & N/A     \\
1 - ($-1$, $-1$, $-1$)        & N/A                  & N/A                             & N/A    \\
1 - ($-1$, 0, $-1$) & N/A                  & N/A                             & N/A     \\
1 - ($-1$, 1, $-1$) & N/A                  & N/A                             & N/A     \\
1 - ($-1$, 2, $-1$) & N/A                  & N/A                             & N/A     \\
2 - ($-1$, $-1$, $-1$)        & N/A                  & N/A                             & N/A     \\
2 - ($-1$, 0, $-1$) & N/A                  & N/A                             & N/A     \\
2 - ($-1$, 1, $-1$) & N/A                  & N/A                             & N/A     \\
2 - ($-1$, 2, $-1$) & N/A                  & N/A                             & N/A    \\
(0, $-1$)       & ($W$, $R_\ell$)                           & ($W$, $R_\ell$)                                      & ($W$, $R_\ell)$              \\
(0, 0)       & $R_{sw}$                           & $R_{sw}$                                      & $R_{sw}$              \\
(0, 1)       & $R_{sw}$                           & $R_{sw}$                                      & $R_{sw}$              \\
(0, 2)       & $R_{sw}$                           & $R_{sw}$                                      & $R_{sw}$              \\
(1, $-1$)       & ($W$, $R_\ell$)                           & ($W$, $R_\ell$)                                    & ($W$, $R_\ell$)              \\
(1, 0)       & $R_{sw}$                           & $R_{sw}$                                      & $R_{sw}$              \\
(1, 1)       & $R_{sw}$                           & $R_{sw}$                                      & $R_{sw}$              \\
(1, 2)       & ($W$, $R_\ell$, $R_\ell$)                           & ($R_\ell$, $R_\ell$, $R_\ell$)                                      & $R_{sw}$              \\
(2, $-1$)       & ($R_\ell$, $R_\ell$, $R_\ell$)                         & ($R_\ell$, $R_\ell$, $R_\ell$)                                   & ($R_\ell$, $R_\ell$, $R_\ell$)            \\
(2, 0)       & $R_{sw}$                           & $R_{sw}$                                      & $R_{sw}$              \\
(2, 1)       & ($R_\ell$, $W$)                           & ($R_\ell$, $W$)                                     & ($R_\ell$, $W$)              \\
(2, 2)       & ($R_\ell$, $R_\ell$, $R_\ell$)                           & ($R_\ell$, $R_\ell$, $R_\ell$)                                      & $R_{sw}$              \\
(0, $-1$, $-1$)     & ($W$, $R_\ell$, $R_\ell$)                         & ($R_\ell$, $R_\ell$, $R_\ell$)                                    & ($W$, $R_\ell$, $R_\ell$)            \\
(0, $-1$, 0)     & ($W$, $R_\ell$, $W$)                         & ($W$, $R_\ell$, $W$)                                    & ($W$, $R_\ell$, $W$)            \\
(0, $-1$, 1)     & ($W$, $R_\ell$, $W$)                         & ($W$, $R_\ell$, $W$)                                    & ($W$, $R_\ell$, $W$)            \\
(0, $-1$, 2)     & ($W$, $R_\ell$, $R_\ell$)                         & ($R_\ell$, $R_\ell$, $R_\ell$)                                    & ($W$, $R_\ell$, $R_\ell$)            \\
(0, 0, $-1$)     & ($R_{sw}$, $R_\ell$)                         & ($R_{sw}$, $R_\ell$)                                    & ($R_{sw}$, $R_\ell$)            \\
(0, 0, 0)     & ($R_{sw}$, $R_{sw}$)                       & ($R_{sw}$, $R_{sw}$)                                  & ($R_{sw}$, $R_{sw}$)          \\
(0, 0, 1)     & ($R_{sw}$, $R_{sw}$)                       & ($R_{sw}$, $R_{sw}$)                                  & ($R_{sw}$, $R_{sw}$)          \\
(0, 0, 2)     & ($R_{sw}$, $R_{sw}$ )                      & ($R_{sw}$, $R_{sw}$ )                                 & ($R_{sw}$, $R_{sw}$)          \\
(0, 1, $-1$)     & ($R_{sw}$, $R_\ell$)                         & ($R_{sw}$, $R_\ell$ )                                   & ($R_{sw}$, $R_\ell$)            \\
(0, 1, 0)     & ($R_{sw}$, $R_{sw}$)                       & ($R_{sw}$, $R_{sw}$)                                  & ($R_{sw}$, $R_{sw}$)          \\
(0, 1, 1)     & ($R_{sw}$, $R_{sw}$)                       & ($R_{sw}$, $R_{sw}$)                                  & ($R_{sw}$, $R_{sw}$)          \\
(0, 1, 2 )    & ($R_{sw}$, $R_\ell$)                       & ($R_{sw}$, $R_{sw}$)                                  & ($R_{sw}$, $R_{sw}$)          \\
(0, 2, $-1$)     & ($W$, $R_\ell$, $R_\ell$)                         & ($W$, $R_\ell$, $R_\ell$)                                  & ($R_{sw}$, $R_\ell$)            \\
(0, 2, 0)     & ($R_{sw}$, $R_{sw}$)                       & ($R_{sw}$, $R_{sw}$)                                  & ($R_{sw}$, $R_{sw}$)          \\
(0, 2, 1)     & ($W$, $R_\ell$, $R_\ell$)                      & ($R_\ell$, $R_\ell$, $R_\ell$)                                  & ($R_{sw}$, $R_{sw}$)          \\
(0, 2, 2)     & ($W$, $R_\ell$, $R_\ell$)                      & ($R_\ell$, $R_\ell$, $R_\ell$)                                  & ($R_{sw}$, $R_{sw}$)          \\
(1, $-1$, $-1$)     & ($W$, $R_\ell$, $R_\ell$)                         & ($R_\ell$, $R_\ell$, $R_\ell$)                                    & ($W$, $R_\ell$ $R_\ell$)            \\
(1, $-1$, 0)     & ($R_\ell$, $R_\ell$, $W$)                         & ($R_\ell$, $R_\ell$, $R_\ell$)                                    & ($W$, $R_\ell$, $W$)            \\
(1, $-1$, 1)     & ($W$, $R_\ell$, $W$)                         & ($R_\ell$, $R_\ell$, $R_\ell$)                                    & ($W$, $R_\ell$, $W$)            \\
(1, $-1$, 2)     & ($R_\ell$, $R_\ell$, $R_\ell$)                         & ($R_\ell$, $R_\ell$, $R_\ell$)                                    & ($W$, $R_\ell$, $R_\ell$)            \\
(1, 0, $-1$)     & ($R_{sw}$, $R_\ell$)                         & ($R_\ell$, $R_\ell$, $R_\ell$)                                    & ($R_{sw}$, $R_\ell$)            \\
(1, 0, 0)     & ($R_{sw}$, $R_{sw}$)                       & ($R_{sw}$, $R_{sw}$)                                  & ($R_{sw}$, $R_{sw}$)          \\
(1, 0, 1)     & ($R_{sw}$, $R_{sw}$)                       & ($R_{sw}$, $R_{sw}$)                                  & ($R_{sw}$, $R_{sw}$)          \\
(1, 0, 2)     & ($R_{sw}$, $R_\ell$)                       & ($R_{sw}$, $R_\ell$)                                  & ($R_{sw}$, $R_{sw}$)          \\
(1, 1, $-1$)     & ($R_{sw}$, $R_\ell$)                         & ($R_\ell$, $R_\ell$, $R_\ell$)                                    & ($R_{sw}$, $R_\ell$)            \\
(1, 1, 0)     & ($R_{sw}$, $R_{sw}$)                       & ($R_{sw}$, $R_{sw}$)                                  & ($R_{sw}$, $R_{sw}$)          \\
(1, 1, 1)     & ($R_\ell$, $R_\ell$, $R_\ell$)                       & ($R_\ell$, $R_\ell$, $R_\ell$)                                  & ($R_{sw}$, $R_{sw}$)          \\
(1, 1, 2)     & ($W$, $R_\ell$, $R_\ell$)                       & ($R_\ell$, $R_\ell$, $R_\ell$)                                  & ($R_{sw}$, $R_{sw}$)          \\
(1, 2, $-1$)     & ($R_\ell$, $R_\ell$, $R_\ell$)                         & ($R_\ell$, $R_\ell$, $R_\ell$)                                    & ($R_{sw}$, $R_\ell$)            \\
(1, 2, 0)     & ($R_\ell$, $R_\ell$, $W$)                       & ($R_\ell$, $R_\ell$, $R_\ell$)                                  & ($R_{sw}$, $R_{sw}$)          \\
(1, 2, 1)     & ($W$, $R_\ell$, $R_\ell$)                       & ($R_\ell$, $R_\ell$, $R_\ell$)                                  & ($R_{sw}$, $R_{sw}$)          \\
(1, 2, 2)     & ($R_\ell$, $R_\ell$, $R_\ell$)                       & ($R_\ell$, $R_\ell$, $R_\ell$)                                  & ($R_{sw}$, $R_{sw}$)          \\
(2, $-1$, $-1$)     & ($R_\ell$, $R_\ell$, $R_\ell$)                         & ($R_\ell$, $R_\ell$, $R_\ell$)                                    & ($R_\ell$, $R_\ell$, $R_\ell$)            \\
(2, $-1$, 0)     & ($R_\ell$, $R_\ell$, $W$)                         & ($R_\ell$, $R_\ell$, $R_\ell$)                                    & ($R_\ell$, $R_\ell$, $W$)            \\
(2, $-1$, 1)     & ($R_\ell$, $R_\ell$, $R_\ell$)                         & ($R_\ell$, $R_\ell$, $R_\ell$)                                    & ($R_\ell$, $R_\ell$, $W$)            \\
(2, 0, $-1$)     & ($R_\ell$, $W$, $R_\ell$)                         & ($R_\ell$, $W$, $R_\ell$)                                    & ($R_\ell$, $W$, $R_\ell$)            \\
(2, 0, 0)     & ($R_{sw}$, $R_{sw}$)                       & ($R_{sw}$, $R_{sw}$)                                  & ($R_{sw}$, $R_{sw}$)          \\
(2, 0, 1)     & ($R_{sw}$, $R_{sw}$)                       & ($R_\ell$, $R_{sw}$)                                  & ($R_{sw}$, $R_{sw}$)          \\
(2, 0, 2)     & ($R_{sw}$, $R_{sw}$)                       & ($R_\ell$, $R_\ell$, $R_\ell$)                                  & ($R_{sw}$, $R_{sw}$)          \\
(2, 1, $-1$)     & ($R_\ell$, $R_\ell$, $R_\ell$)                         & ($R_\ell$, $R_\ell$, $R_\ell$)                                    & ($R_{sw}$, $R_\ell$)            \\
(2, 1, 0)     & ($R_\ell$, $R_{sw}$)                       & ($R_\ell$, $R_{sw}$)                                  & ($R_{sw}$, $R_{sw}$)          \\
(2, 1, 1)     & ($R_\ell$, $R_\ell$, $R_\ell$)                       & ($R_\ell$, $R_\ell$, $R_\ell$)                                  & ($R_{sw}$, $R_{sw}$)          \\
(2, 1, 2)     & ($R_\ell$, $R_\ell$, $R_\ell$)                       & ($R_\ell$, $R_\ell$, $R_\ell$)                                  & ($R_{sw}$, $R_{sw}$)          \\
(2, 2, $-1$)      & ($R_\ell$, $R_\ell$, $R_\ell$)                         & ($R_\ell$, $R_\ell$ $R_\ell$)                                    & ($R_{sw}$, $R_\ell$)            \\
(2, 2, 0)     & ($R_\ell$, $R_\ell$, $W$)                       & ($R_\ell$, $R_\ell$, $R_\ell$)                                  & ($R_{sw}$, $R_{sw}$)          \\
(2, 2, 1)     & ($R_\ell$, $R_\ell$, $R_\ell$)                       & ($R_\ell$, $R_\ell$, $R_\ell$)                                  & ($R_{sw}$, $R_{sw}$)          \\
(2, 2, 2)     & ($R_\ell$, $R_\ell$, $R_\ell$)                       & ($R_\ell$, $R_\ell$, $R_\ell$)                                  & ($R_{sw}$, $R_{sw}$)      \\   
\hline
\caption{Comparison of the dynamic \textsc{swap-asap} policy with our Q-learning policies for a four-node repeater chain with $p_{\ell} = 0.6$, $p_{sw} = 0.5$, and $m^\star = 2$. See Sec.~\ref{sec:MDP_model} for a description of the states and actions. Here, we write the states as $(m_1,m_2,m_3)$, with $m_i$ indicating the age of the $i^{\text{th}}$ elementary link. States like $(m_1, m_2)$ indicate an elementary link with age $m_1$ between nodes $1$ and $2$ and a virtual link of age $m_2$ between nodes $2$ and $4$. It also represents the mirror-image state, i.e., a state with a virtual link with age $m_2$ between nodes $1$ and $3$ and an elementary link with age $m_1$ between nodes $3$ and $4$, because this state of the repeater chain results in identical optimal (mirror image) actions for the homogeneous chain under consideration. Terminal states are written as $m_1$ - $(-1,m_2,-1)$, indicating that the end-to-end link has age $m_1$ and that, in addition, there exists an elementary link between nodes $2$ and $3$ with age $m_2$. The action N/A indicates that no action needs to be taken, because the state is terminal.}\label{Q_table}
\end{longtable}
}

\section{Proof of Eq.~\texorpdfstring{\eqref{eq-decohered_entangled_state}}{(15)}}\label{sec-decohered_entangled_state_pf}

We start by expressing the Bell states in \eqref{eq-Bell_states_1} and \eqref{eq-Bell_states_2} terms of the Pauli operators in \eqref{eq-Pauli_operators} as follows:
\begin{align}
    \Phi^+&=\frac{1}{4}(\mathbbm{1}\otimes\mathbbm{1}+X\otimes X-Y\otimes Y+Z\otimes Z),\\
    \Phi^-&=\frac{1}{4}(\mathbbm{1}\otimes\mathbbm{1}-X\otimes X+Y\otimes Y+Z\otimes Z),\\
    \Psi^+&=\frac{1}{4}(\mathbbm{1}\otimes\mathbbm{1}+X\otimes X+Y\otimes Y-Z\otimes Z),\\
    \Psi^-&=\frac{1}{4}(\mathbbm{1}\otimes\mathbbm{1}-X\otimes X-Y\otimes Y-Z\otimes Z).
\end{align}
These relations can be readily inverted to obtain
\begin{align}
    \mathbbm{1}\otimes\mathbbm{1}&=\Phi^++\Phi^-+\Psi^++\Psi^-,\label{eq-Pauli_to_Bell_1}\\
    Z\otimes Z&=\Phi^++\Phi^--\Psi^+-\Psi^-,\label{eq-Pauli_to_Bell_2}\\
    X\otimes X&=\Phi^+-\Phi^-+\Psi^+-\Psi^-,\label{eq-Pauli_to_Bell_3}\\
    -Y\otimes Y&=\Phi^+-\Phi^--\Psi^++\Psi^-.\label{eq-Pauli_to_Bell_4}
\end{align}

Consider now an arbitrary single-qubit Pauli channel, which has the following action on an arbitrary linear operator $\rho$:
\begin{equation}
    \mathcal{P}(\rho)=p_I\rho+p_X X\rho X+p_Y Y\rho Y+p_Z Z\rho Z.
\end{equation}
Observe that the Pauli operators are ``eigenvectors'' of the Pauli channel, in the following sense:
\begin{align}
    \mathcal{P}(\mathbbm{1})&=(p_I+p_Z+p_X+p_Y)\mathbbm{1},\\
    \mathcal{P}(Z)&=(p_I+p_Z-p_X-p_Y)Z,\\
    \mathcal{P}(X)&=(p_I-p_Z+p_X-p_Y)X,\\
    \mathcal{P}(Y)&=(p_I-p_Z-p_X+p_Y)Y,
\end{align}
which follows from the fact that the Pauli operators mutually anti-commute. Throughout the rest of this proof, we let
\begin{align}
    \lambda_I&\equiv p_I+p_Z+p_X+p_Y,\\
    \lambda_Z&\equiv p_I+p_Z-p_X-p_Y,\\
    \lambda_X&\equiv p_I-p_Z+p_X-p_Y,\\
    \lambda_Y&\equiv p_I-p_Z-p_X+p_Y.
\end{align}
(Note that, because $\mathcal{P}$ is trace preserving by definition, we have that $\lambda_I=1$.) It now immediately follows that
\begin{align}
    \mathcal{P}^{\circ m}(\mathbbm{1})&=\lambda_I^m\mathbbm{1},\\
    \mathcal{P}^{\circ m}(Z)&=\lambda_Z^m Z,\\
    \mathcal{P}^{\circ m}(X)&=\lambda_X^m X,\\
    \mathcal{P}^{\circ m}(Y)&=\lambda_Y^m Y,
\end{align}
for all $m\in\{1,2,\dotsc\}$. Therefore,
\begin{align}
    (\mathcal{P}^{\circ m}\otimes\mathcal{P}^{\circ m})(\Phi^+)&=\frac{1}{4}\mathcal{P}^{\circ m}(\mathbbm{1})\otimes\mathcal{P}^{\circ m}(\mathbbm{1})+\frac{1}{4}\mathcal{P}^{\circ m}(Z)\otimes\mathcal{P}^{\circ m}(Z)\nonumber\\
    &\qquad+\frac{1}{4}\mathcal{P}^{\circ m}(X)\otimes\mathcal{P}^{\circ m}(X)-\frac{1}{4}\mathcal{P}^{\circ m}(Y)\otimes\mathcal{P}^{\circ m}(Y)\\
    &=\frac{1}{4}\lambda_I^{2m}\mathbbm{1}\otimes\mathbbm{1}+\frac{1}{4}\lambda_Z^{2m}Z\otimes Z+\frac{1}{4}\lambda_X^{2m}X\otimes X-\frac{1}{4}\lambda_Y^{2m}Y\otimes Y\\
    &=\frac{1}{4}(1+\lambda_X^{2m}+\lambda_Z^{2m}+\lambda_Y^{2m})\Phi^+\nonumber\\
    &\qquad+\frac{1}{4}(1+\lambda_Z^{2m}-\lambda_X^{2m}-\lambda_Y^{2m})\Phi^-\nonumber\\
    &\qquad+\frac{1}{4}(1-\lambda_Z^{2m}+\lambda_X^{2m}-\lambda_Y^{2m})\Psi^+\nonumber\\
    &\qquad+\frac{1}{4}(1-\lambda_Z^{2m}-\lambda_X^{2m}+\lambda_Y^{2m})\Psi^-.\label{eq-decohered_entangled_state_general}
\end{align}
where the final equality follows from using \eqref{eq-Pauli_to_Bell_1}--\eqref{eq-Pauli_to_Bell_4}, and we recall that $\lambda_I=1$, because $\mathcal{P}$ is trace preserving.

Now, when we take $p_I,p_X,p_Y,p_Z$ as in \eqref{eq-decoherence_channel_pI}--\eqref{eq-decoherence_channel_pZ}, we find that $\lambda_I=1,\,\lambda_X=\e^{-1/m_2^{\star}},\,\lambda_Z=\e^{-1/m_1^{\star}},\,\lambda_Y=\e^{-1/m_2^{\star}}$. Plugging these into \eqref{eq-decohered_entangled_state_general} and simplifying, we obtain
\begin{align}
    (\mathcal{N}_{m_1^{\star},m_2^{\star}}^{\circ m}\otimes\mathcal{N}_{m_1^{\star},m_2^{\star}}^{\circ m})(\Phi^+)&=\frac{1}{4}(1+\e^{-2m/m_1^{\star}}+2\e^{-2m/m_2^{\star}})\Phi^+\nonumber\\
    &\qquad +\frac{1}{4}(1+\e^{-2m/m_1^{\star}}-2\e^{-2m/m_2^{\star}})\Phi^-\nonumber\\
    &\qquad +\frac{1}{4}(1-\e^{-2m/m_1^{\star}})\Psi^+\nonumber\\
    &\qquad +\frac{1}{4}(1-\e^{-2m/m_1^{\star}})\Psi^-,
\end{align}
which is precisely \eqref{eq-decohered_entangled_state}, as required, completing the proof.

\section{Comments about correlation functions for actions and states at different nodes and links} \label{sec-apndx_advantage}

In this section, we elaborate on the equal-time and unequal-time correlation functions that we introduce in Sec.~\ref{sec:advantage}. Let us start with a brief explanation of how to understand these correlation functions for the states and actions in terms of the underlying Markov decision process, as defined in Appendix~\ref{sec-MDP_details}. For any policy $\pi$ (in particular, for the policy determined by the Q-learning algorithm), the joint probability distribution between all of the states and actions up to some time $\tau$ is given by (see, e.g., \cite{Puterman2014book} or \cite[Appendix~A.2]{Kha22})
\begin{equation}\label{eq-MDP_history_distribution}
    \Pr[S^1,A^1,S^2,A^2,\dotsc, A^{\tau-1},S^{\tau}]=\Pr[S^1]\prod_{t=1}^{\tau-1} T_{A^t}(S^{t+1};S^t)\pi(A^t|S^t),
\end{equation}
where $\Pr[S^1]$ refers to the probability distribution of the repeater chain at the initial time. In our case, the repeater chain starts deterministically in the state in which all links are inactive. This joint probability distribution now defines the marginal distribution $\Pr[A^t]$ of actions at any time $t$, which can then we used to obtain joint distributions $\Pr[A_{i,j}^t,A_{k,\ell}^t]$ between actions at different elementary links at time $t$. These probabilities in turn define the expectation values $\langle A_{i,j}^t A_{k,\ell}^t\rangle$, $\langle A_{i,j}^t\rangle$, and $\langle A_{k,\ell}^t\rangle$ that are needed to determine the correlation coefficient $r[A_{i,j}^t,A_{k,\ell}^t]$. We can analogously obtain the unequal-time correlation functions for states an actions by calculating the corresponding marginal probability distributions of the relevant random variables using \eqref{eq-MDP_history_distribution}.

\paragraph*{Equal-time correlation functions.} For the equal-time correlation functions, we highlight the fact that some of the contribution to the correlations comes directly from the constraints introduced on the states and actions, and this contribution to the correlations does not depend on the specifics of the policy. For example, when a swap request is made at a node, elementary link requests are not allowed at that node, i.e., if $A_{i,i}^t = 1$, then we impose the constraint $A_{i,i+1}^t = A_{i-1,i}^t = 0$. Therefore, the actions at a node and its adjoining links are always anti-correlated. As an illustration of how the constraint induced offset emerges, here we show a simple calculation of the correlation coefficient between actions at the (adjacent) links of a two-link chain. The relevant action space is $R_\ell R_\ell, R_\ell W, WR_\ell, W W, W R_{sw} W$, and in terms of action matrices (as defined in Appendix~\ref{sec-MDP_details}) these actions are represented as \begin{equation*}
    \begin{bmatrix}
    0 & 1 & 0 \\
    1 & 0 & 1 \\
    0 & 1 & 0 
    \end{bmatrix}, 
    \begin{bmatrix}
    0 & 1 & 0 \\
    1 & 0 & 0 \\
    0 & 0 & 0
    \end{bmatrix}, 
    \begin{bmatrix}
    0 & 0 & 0 \\
    0 & 0 & 1 \\
    0 & 1 & 0
    \end{bmatrix},
    \begin{bmatrix}
    0 & 0 & 0 \\
    0 & 0 & 0 \\
    0 & 0 & 0
    \end{bmatrix},
    \begin{bmatrix}
    0 & 0 & 0 \\
    0 & 1 & 0 \\
    0 & 0 & 0 
    \end{bmatrix},
    \end{equation*} 
    respectively, from which the values of the relevant random variables (i.e., the actions at the first and second links) can be easily read off. Now, even when each action is chosen randomly (i.e., with equal probability of $1/5$), we have $\langle A_{1,2} A_{2,3} \rangle = 1/5$ (only time they can be 1 together is when no swap occurs, i.e., the first action in the list above). Also, the mean values of the actions at the first link (connecting nodes 1 and 2) and the second link (connecting nodes 2 and 3) are $\langle A_{1,2} \rangle = \langle A_{2,3} \rangle = 2/5$. Again, since we assumed a random distribution of the actions (all 5 actions are equiprobable), the variances of $A_{1,2}$ and $A_{2,3}$ are equal and given by $\langle A_{1,2}^2 \rangle - \langle A_{1,2}\rangle^2=\langle A_{2,3}^2 \rangle - \langle A_{2,3}\rangle^2 = 2/5 - 4/25 = 6/25$. Thus, from the definition of the correlation coefficient in \eqref{eq:pearson_coeff}, we obtain $r[A_{1,2}, A_{2,3}] = \frac{1/5 - 4/25}{6/25} = 1/6\approx 0.167$. In the same way, it is straightforward to see that for a 4-link chain as well, if we sample the action space randomly, this constraint gives an offset value of approximately $0.15$ for the relevant correlation coefficient. This is in fact the value that the random policy has for $r[A_{i,i+1}^t, A_{i+1,i+2}^t]$, within the error bars (specifically, we obtain the value $0.17 \pm 0.04$ in Table \ref{tab:equal_time_corr}). Thus, we have shown that every policy will have an offset level of correlation between some action pairs just due to the MDP constraints.

Next, let us focus on how the correlation rises above the offset value when nodes collaborate on what actions to choose and have global knowledge of the state. To calculate the correlation coefficient $r$, we choose states randomly and then find the correlation between the different parts of the state matrix and action matrix provided to us by the policy for that randomly chosen state. To make the discussion concrete, we first focus on action-action correlations, and let us specifically look at the correlation coefficient $r[A_{i,i+1}^t, A_{k,k+1}^t]$. In this case, the possible actions are $R_\ell R_\ell, R_\ell W, WR_\ell, WW$. Now, for two links that are well separated in the network, if the nodes did not collaborate at all and did not make use of any knowledge each other's states, the requests would be completely independent. Say $m^\star = 0$, since states are selected randomly without bias, all 4 states ($(-1,-1), (0,-1), (-1,0), (0,0)$) for these two links occur with equal probability. Thus, if the nodes did not collaborate, all four actions would be chosen with equal probability and would lead to $r[A_{i,i+1}^t, A_{k,k+1}^t] = 0$. On the other hand, if there is some collaboration between nodes $i$ and $k$, then the policy will choose some action pairs out of these four more frequently than others, thus on average raising the value of $r$ above its offset value. For example, when only one state is active, the RL policy might decide to request on both links rather than wait on the active link and request on the one inactive link. This increases the likelihood of the $R_\ell R_\ell$ action and reduces the likelihood of the $R_\ell W , W R_\ell$ actions compared to the uncorrelated case, since all 4 states are still generated with equal probability. This makes $r[A_{i,i+1}^t, A_{k,k+1}^t] \neq 0$. This is the sense in which we quantify action matrix correlations as a quantifier of collaboration. In the same way, the correlation of some part of the action matrix with some part of the state matrix above the offset value indicates the fact that the states and actions are not agnostic of each other, in particular, that the nodes are using the knowledge of the state of that part of the network. Thus, state-action matrix correlations quantify global knowledge of the agent. We stress here that these correlations are not due to evolution of different parts of a connected chain, but this is the correlation in a predetermined set of states and actions (as determined by the policy), thus we refer to it as a quantifier of collaboration.

\paragraph*{Unequal-time correlation functions.} Let us also briefly discuss the meaning of the unequal-time correlation functions. Initially, the reinforcement learning (RL) agent goes through a large number of training episodes in which it tries to explore the entire state space. Intuitively speaking, it experiences approximately all possible scenarios during a long-enough training phase and builds-up a Q-table that dictates the best action possible for a given state. A well-trained RL agent, equipped with global knowledge of the network (information about the state of the entire repeater chain when executing actions at a particular node), tends to make the most informed decisions, and that is what we mean by ``foresight'' in Sec.~\ref{sec:advantage}. It does not mean that the agent knows the outcomes of the probabilistic process (as dictated by the Markov decision process) in advance. By ``foresight'', we mean the following: in every new training episode, the agent utilizes the experience gained in previous episodes, which involves evolution of the repeater chain over a long period of time, and because of that experience it has some power to make an educated guess about the future evolution of the network. Consequently, its decisions/actions are not just dependent on the next state; they can depend on future states as well. This foresight is quantified by the discount factor in the Bellman equation \eqref{eqn:bellman}. In other words, the policy is predetermined and the best actions are determined by the experience of the agent, which (by virtue of its experience) has seen various future states of the repeater chain based on its initial actions. During evolution of the repeater chain, a non-zero value of the unequal-time correlator implies that the state at some later time is affected by the action in previous time steps. Thus, this does not indicate anything acausal. But the difference in the fall-offs of the correlation with time for different policies is a manifestation of the “training foresight”. For Q-learning derived policies, actions at a certain time impact states at much later times, since the choice of the action is based on agent’s experience of the entire time evolution of the network and not just on what is the next state on application of the action---the latter is what a random or \textsc{swap-asap} policy does, and thus late time-values of the correlators are lower for them compared to the Q-learning policy.
\twocolumngrid
\bibliography{main.bib}

%\clearpage
%\newpage

\end{document}